\documentclass[12pt]{article}
\pdfoutput=1
\usepackage{amsmath,amssymb,amsthm,amsxtra,overpic,bbm,bm,epsfig,subfigure}
\usepackage{hyperref}
\usepackage{graphicx}
\usepackage{color}
\usepackage{comment}
\usepackage{epstopdf}
\usepackage{float}
\usepackage{multirow}
\numberwithin{equation}{section}
\usepackage{cite}
\usepackage{hyperref}
\usepackage{slashed,stmaryrd}

\usepackage{lscape}%
\usepackage{array}
\usepackage{booktabs}%

\def\thefootnote{\fnsymbol{footnote}}

\begin{document}

\vspace{0.2cm}

\begin{center}
{\Large\bf The Minimal Seesaw Model with a Modular $S^{}_4$ Symmetry}
\end{center}

\vspace{0.2cm}

\begin{center}
{\bf Xin Wang}~$^{a,~b}$~\footnote{E-mail: wangx@ihep.ac.cn},
\quad
{\bf Shun Zhou}~$^{a,~b}$~\footnote{E-mail: zhoush@ihep.ac.cn}
\\
\vspace{0.2cm}
{\small
$^a$Institute of High Energy Physics, Chinese Academy of Sciences, Beijing 100049, China\\
$^b$School of Physical Sciences, University of Chinese Academy of Sciences, Beijing 100049, China}
\end{center}

\vspace{1.5cm}

\begin{abstract}
In this paper, we incorporate the modular $S^{}_4$ flavor symmetry into the supersymmetric version of the minimal type-I seesaw model, in which only two right-handed neutrino singlets are introduced to account for tiny Majorana neutrino masses, and explore its implications for the lepton mass spectra, flavor mixing and CP violation. The basic idea is to assign two right-handed neutrino singlets into the unique two-dimensional irreducible representation of the modular $S^{}_4$ symmetry group. Moreover, we show that the matter-antimatter asymmetry in our Universe can be successfully explained via the resonant leptogenesis mechanism working at a relatively-low seesaw scale $\Lambda^{}_{\rm SS} \approx 10^7~{\rm GeV}$, with which the potential problem of the gravitino overproduction can be avoided. In this connection, we emphasize that the observed matter-antimatter asymmetry may lead to a stringent constraint on the parameter space and testable predictions for low-energy observables.
\end{abstract}


\def\thefootnote{\arabic{footnote}}
\setcounter{footnote}{0}

\newpage

\section{Introduction}\label{sec:intro}

Neutrino oscillation experiments in the past two decades have firmly established that neutrinos are massive and lepton flavor mixing is significant~\cite{Xing:2011zza, Tanabashi:2018oca}. To account for tiny neutrino masses, one can naturally extend the standard model (SM) by introducing three right-handed neutrinos $N^{}_{i{\rm R}}$ (for $i = 1, 2, 3$), which are singlets under the ${\rm SU}(2)_{\rm L} \times {\rm U}(1)^{}_{\rm Y}$ gauge group of the SM. Therefore, the gauge-invariant Lagrangian relevant for lepton masses and flavor mixing can be written as
\begin{eqnarray}\label{eq:LagSM}
-{\cal L}^{}_{\rm lepton} = \overline{\ell^{}_{\rm L}} Y^{}_l H E^{}_{\rm R} + \overline{\ell^{}_{\rm L}} Y^{}_\nu \widetilde{H} N^{}_{\rm R} + \frac{1}{2} \overline{N^{\rm C}_{\rm R}} M^{}_{\rm R} N^{}_{\rm R} + {\rm h.c.} \; ,
\end{eqnarray}
where $\ell^{}_{\rm L}$ and $H$ denote the left-handed lepton doublet and the Higgs doublet, $E^{}_{\rm R}$ and $N^{}_{\rm R}$ are the right-handed charged-lepton and neutrino singlets, $M^{}_{\rm R}$ is the Majorana mass matrix of right-handed neutrinos. In Eq.~(\ref{eq:LagSM}), $\widetilde{H} \equiv {\rm i} \sigma^{}_2 H^*$ and $N^{\rm C}_{\rm R} \equiv C\overline{N^{}_{\rm R}}^{\rm T}$ with $C = {\rm i}\gamma^2\gamma^0$ being the charge-conjugation matrix have been defined. After the Higgs doublet acquires its vacuum expectation value (vev), i.e., $\langle H \rangle = (0, v/\sqrt{2})^{\rm T}$ with $v \approx 246~{\rm GeV}$, the gauge symmetry is spontaneously broken down, and the charged-lepton and Dirac neutrino mass matrices are given by $M^{}_l \equiv Y^{}_l v/\sqrt{2}$ and $M^{}_{\rm D} = Y^{}_\nu v/\sqrt{2}$, respectively. At the low-energy scale, the effective Majorana neutrino mass matrix is then obtained via the famous seesaw formula $M^{}_\nu \approx - M^{}_{\rm D} M^{-1}_{\rm R} M^{\rm T}_{\rm D}$.\footnote{In this paper, we shall adopt the left-right convention for all the fermion mass terms. Therefore, the Majorana mass term of ordinary left-handed neutrinos in the Lagrangian reads $-\overline{\nu^{}_{\rm L}} M^{}_\nu \nu^{\rm C}_{\rm L}/2 + {\rm h.c.}$ This convention should be taken care of when one works in the supersymmetric framework, where the superpotential is usually constructed in terms of the left-handed chiral superfields.} In this canonical seesaw model~\cite{Minkowski:1977sc, Yanagida:1979, GellMan1979, Mohapatra:1979ia}, given ${\cal O}(M^{}_{\rm D}) \sim 10^2~{\rm GeV}$ at the electroweak scale, the smallness of light neutrino masses ${\cal O}(M^{}_\nu) \lesssim 0.1~{\rm eV}$ can thus be ascribed to the heaviness of right-handed neutrino masses ${\cal O}(M^{}_{\rm R}) \sim 10^{14}~{\rm GeV}$, which are not subject to the spontaneous symmetry breaking and turn out to be close to the energy scale of grand unified theories (GUT) at $\Lambda^{}_{\rm GUT} \approx 2\times 10^{16}~{\rm GeV}$.

Although tiny neutrino masses can be well accommodated in the canonical seesaw model, the flavor structures of all the relevant lepton mass matrices are left unspecified. As a consequence, the model is in general lacking of predictive power for lepton mass spectra, flavor mixing pattern and CP violation~\cite{Xing:2019vks}. To further reduce the number of free parameters, one can consider the so-called minimal seesaw model (MSM)~\cite{King:1999mb,Frampton:2002qc,Guo:2006qa}, in which only two right-handed neutrino singlets are introduced. In such a minimal scenario, the lightest neutrino appears to be massless, namely, $m^{}_1 = 0$ in the case of normal neutrino mass ordering (NO) with $m^{}_1 < m^{}_2 < m^{}_3$ or $m^{}_3 = 0$ in the case of inverted neutrino mass ordering (IO) with $m^{}_3 < m^{}_1 < m^{}_2$. Given $\Delta m^2_{21} \equiv m^2_2 - m^2_1 \approx 7.39\times 10^{-5}~{\rm eV}^2$ and $|\Delta m^2_{31}| \equiv |m^2_3 - m^2_1| \approx 2.53\times 10^{-3}~{\rm eV}^2$ from the global-fit analysis of current neutrino oscillation data~\cite{Esteban:2018azc}, we have $m^{}_1 = 0$, $m^{}_2 = \sqrt{\Delta m^2_{21}} \approx 8.6~{\rm meV}$ and $m^{}_3 = \sqrt{\Delta m^2_{31}} \approx 50.3~{\rm meV}$ in the MSM with NO or $m^{}_3 = 0$, $m^{}_1 = \sqrt{|\Delta m^2_{31}|} \approx 50.3~{\rm meV}$ and $m^{}_2 = \sqrt{|\Delta m^2_{31}| + \Delta m^2_{21}} \approx 50.4~{\rm meV}$ in the MSM with IO. In addition, the MSM also provides a natural framework to realize realistic and testable neutrino mass models at the TeV scale~\cite{Xing:2006ms, Zhang:2009ac}.

On the other hand, the observed pattern of leptonic flavor mixing can be explained by further imposing non-Abelian discrete flavor symmetries on the generic Lagrangian of the seesaw model. See, e.g., Refs.~\cite{Altarelli:2010gt, Ishimori:2010au, King:2013eh, King:2014nza, King:2017guk}, for recent reviews on this topic. The basic strategy for the model building with discrete flavor symmetries is to suppose that there exists an overall discrete symmetry in the theory at some high-energy scale, which is then broken down into distinct residual symmetries in the charged-lepton and neutrino sectors at low-energy scales~\cite{Lam:2008rs, Lam:2008sh, Ge:2011ih, Ge:2011qn, Hernandez:2012ra, Feruglio:2012cw}. However, the breaking pattern of the discrete symmetries usually requires a number of scalar fields (i.e., flavons), which are singlets under the SM gauge symmetries, and the flavor mixing pattern is then essentially determined by the vev's of those flavons. Therefore, these neutrino mass models with discrete flavor symmetries suffer from the problems how to experimentally pin down the free parameters associated with the flavon fields and how to directly prove the existence of the flavons themselves. In this regard, the recent suggestion of modular symmetries as a solution to the flavor mixing puzzle is attractive and deserves further investigations~\cite{Feruglio:2017spp}. Under one single modular symmetry, the Yukawa couplings in the lepton sector take the modular forms, in which only one complex parameter, i.e., the modulus $\tau$, is involved. Once the value of $\tau$ is fixed, the modular symmetry is broken and the leptonic flavor mixing pattern is determined with a limited number of free coupling constants that are unconstrained by the modular symmetry. In this case, the flavon fields are no longer needed. In the literature, there have been a great number of works on the model building based on the modular group $\Gamma^{}_N$, which for a given value of $N$ is isomorphic to the well-known non-Abelian discrete symmetry groups, e.g., $\Gamma^{}_{2} \simeq S^{}_{3}$~\cite{Kobayashi:2018vbk, Kobayashi:2018wkl, Kobayashi:2019rzp, Okada:2019xqk}, $\Gamma^{}_{3} \simeq A^{}_{4}$~\cite{Kobayashi:2018scp, Criado:2018thu, deAnda:2018ecu, Okada:2018yrn, Nomura:2019jxj, Nomura:2019lnr, Ding:2019zxk, Nomura:2019yft, Okada:2019mjf, Asaka:2019vev, Zhang:2019ngf}, $\Gamma^{}_{4} \simeq S^{}_4$~\cite{Penedo:2018nmg, Novichkov:2018ovf, Okada:2019lzv} and $\Gamma^{}_{5} \simeq A^{}_5$~\cite{Novichkov:2018nkm, Ding:2019xna, Criado:2019tzk}. Moreover, other interesting aspects of modular symmetries have also been studied, such as the combination of modular symmetries and the generalized CP symmetry~\cite{Novichkov:2019sqv}, multiple modular symmetries~\cite{deMedeirosVarzielas:2019cyj, King:2019vhv}, the double covering of modular groups~\cite{Liu:2019khw}, the $A^{}_{4}$ symmetry from the modular $S^{}_{4}$ symmetry~\cite{Kobayashi:2019mna, Kobayashi:2019xvz}, the modular residual symmetry~\cite{Novichkov:2018yse,Gui-JunDing:2019wap} and the unification of quark and lepton flavors with modular invariance~\cite{Okada:2019uoy}.

In this paper, we investigate the generic MSM with the modular $S^{}_4$ symmetry and explore its implications for lepton mass spectra, flavor mixing and CP violation. The main motivation for such an investigation is to take the advantages of the minimality of the MSM and the predictive power of the modular symmetry. Two salient features of such a theoretical setup should be emphasized. First, those two right-handed neutrinos in the MSM are naturally assigned into the two-dimensional irreducible representation of the $S^{}_4$ group~\cite{Okada:2019lzv,Yang:2011fh}. Without the introduction of the flavon fields, the number of free model parameters will be kept as minimal as possible. Second, we demonstrate that the baryon number asymmetry in the present Universe, which is characterized by the baryon-to-photon density ratio $\eta^{}_{\rm B} \equiv (n^{}_{\rm b} - n^{}_{\overline{\rm b}})/n^{}_\gamma = (6.131 \pm 0.041)\times 10^{-10}$ as observed by the Planck collaboration~\cite{Aghanim:2018eyx}, can be successfully explained via the leptogenesis mechanism~\cite{Fukugita:1986hr}. Different from Ref.~\cite{Asaka:2019vev}, the resonant leptogenesis~\cite{Pilaftsis:1997jf, Pilaftsis:2003gt,GonzalezFelipe:2003fi,Branco:2005ye,Ahn:2006rn} is implemented in the MSM with the modular $S^{}_4$ symmetry such that the seesaw scale can be as low as $\Lambda^{}_{\rm SS} \approx 10^{7}~{\rm GeV}$, evading the gravitino overproduction problem in the gauge-mediated supersymmetry breaking models~\cite{Khlopov:1984pf, Ellis:1984eq, Asaka:2000zh}. As the modular symmetry is intrinsically working in the supersymmetric framework~\cite{Feruglio:2012cw}, the supersymmetry breaking mechanism and the gravitino problem should be properly addressed in order to consistently incorporate the thermal leptogenesis mechanism.

The remaining part of this paper is organized as follows. In Sec.~\ref{sec:modular}, a brief summary of the modular symmetry and the representations of the modular $S^{}_4$ symmetry is given. Two simple but viable scenarios in the MSM with one single modular $S^{}_4$ symmetry are then proposed in Sec.~\ref{sec:models}, where the low-energy phenomenology of lepton mass spectra, flavor mixing pattern and CP violation in these models are also discussed. The renormalization-group running effects on the lepton flavor mixing parameters are examined. In addition, we implement the resonant leptogenesis mechanism in our models to account for the matter-antimatter asymmetry and make a connection between low- and high-energy CP violation in Sec.~\ref{sec:bau}. Finally, we summarize our main conclusions in Sec.~\ref{sec:summary}, and some properties of the modular $S^{}_4$ symmetry group are presented in Appendix~\ref{sec:appA}.

\section{Modular $S^{}_4$ Symmetry} \label{sec:modular}

The basics of modular symmetries have been expounded in previous works~\cite{Feruglio:2012cw}, and the properties of the modular $S^{}_{4}$ symmetry can be found in Refs.~\cite{Novichkov:2018ovf, Novichkov:2019sqv}. In this section, we shall give a brief introduction to the modular symmetry and establish our notations. In the supersymmetric theory with a modular symmetry, the action is given by
\begin{eqnarray}
{\cal S} = \int {\rm d}^4_{}x {\rm d}^2_{} \theta {\rm d}^2 \overline{\theta} {\cal K}(\tau,\overline{\tau}, \chi, \overline{\chi}) + \int {\rm d}^4_{}x {\rm d}^2_{} \theta {\cal W}(\tau, \chi) + \int {\rm d}^4_{}x {\rm d}^2_{} \overline{\theta} \overline{\cal W}(\overline{\tau}, \overline{\chi}) \; ,
\label{eq:action}
\end{eqnarray}
where $x^\mu$ (for $\mu = 0, 1, 2, 3$), $\theta^\alpha$ and $\overline{\theta}^{}_{\dot{\alpha}}$ (for $\alpha, \dot{\alpha} = 1, 2$) are the superspace coordinates, ${\cal K}$ is the K\"ahler potential, and ${\cal W}$ is the superpotential. Additionally, $\tau$ stands for the modulus parameter and the supermultiplet is denoted by $\chi$, for which different supermultiplets will be distinguished by the associated superscript $\chi^{(I)}$. Under the modular transformations,
\begin{eqnarray}
\gamma: \tau \rightarrow \dfrac{a \tau + b}{c \tau + d} \; , \quad
\chi^{(I)}_{} \rightarrow (c \tau +d )^{-k^{}_I} \rho^{(I)}_{} (\gamma) \chi^{(I)} _{} \; ,
\label{eq:actiontransf}
\end{eqnarray}
the action ${\cal S}$ in Eq. (\ref{eq:action}) must be invariant. In Eq.~(\ref{eq:actiontransf}), $\gamma$ is the element of the modular group $\Gamma$ with $a$, $b$, $c$ and $d$ being integers satisfying $ad - bc =1$, $\tau$ is an arbitrary complex number in the upper complex plane, $\rho^{(I)}(\gamma)$ denotes the representation matrix of the modular transformation $\gamma$, and $k^{}_{I}$ is the weight associated with the supermultiplet $\chi^{(I)}$. The modular group $\Gamma$ has two generators $S$ and $T$, corresponding respectively to the basic transformations $\tau \xrightarrow[]{S} -1/\tau$ and $\tau \xrightarrow[]{T} \tau + 1$, and they fulfill the identities $S^{2}_{} = (ST)^{3}_{} = {\bf 1}$ with ${\bf 1}$ being the identity element. The finite modular group of the level $N$ is then defined as $\Gamma^{}_{N} \equiv \Gamma/\Gamma(N)$, where $\Gamma(N)$ is the principal congruence subgroup of $\Gamma$. As we have mentioned in the previous section, $\Gamma^{}_N$ with $N \leq 5$ are isomorphic to the permutation symmetry groups, namely, $\Gamma^{}_2 \simeq S^{}_3$, $\Gamma^{}_3 \simeq A^{}_4$, $\Gamma^{}_4 \simeq S^{}_4$ and $\Gamma^{}_5 \simeq A^{}_5$.

The invariance of ${\cal S}$ under the transformations in Eq.~(\ref{eq:actiontransf}) demands that ${\cal K}(\tau,\overline{\tau}, \chi, \overline{\chi})$ is invariant up to the K\"ahler transformation ${\cal K}(\tau,\overline{\tau}, \chi, \overline{\chi}) \rightarrow {\cal K}(\tau,\overline{\tau}, \chi, \overline{\chi}) + f(\tau, \chi)+f(\overline{\tau}, \overline{\chi})$,\footnote{The most general K\"ahler potential consistent with the modular symmetry may contain additional terms, as recently pointed out in Ref.~\cite{Chen:2019ewa}. However, for a phenomenological purpose, we consider only the simplest form of the K\"ahler potential.} where $f(\tau, \chi)$ itself is invariant under the modular transformation. At the same time, the superpotential ${\cal W}(\tau, \chi)$ is invariant as well and can be expanded in terms of the supermultiplets as follows
\begin{eqnarray}
{\cal W}(\tau, \chi)= \sum_{n}^{}\sum_{\{I^{}_{1},\dots,I^{}_{n}\}}^{} Y^{}_{I^{}_1\dots I^{}_n}(\tau)\chi_{}^{(I^{}_1)}\cdots\chi_{}^{(I^{}_n)} \; ,
\label{eq:surpoten}
\end{eqnarray}
where the coefficients $Y^{}_{I^{}_1\dots I^{}_n}(\tau)$ take the modular forms, transforming under $\Gamma^{}_{N}$ as
\begin{eqnarray}
Y^{}_{I^{}_1\dots I^{}_n}(\tau) \rightarrow (c\tau+d)^{k^{}_Y}_{} \rho^{}_{Y} (\gamma) Y^{}_{I^{}_1 \dots I^{}_n}(\tau) \; ,
\label{eq:Yuktransf}
\end{eqnarray}
where $\rho^{}_Y$ is the representation matrix of $\Gamma^{}_N$ and the even integer $k^{}_{Y}$ is the weight of $Y^{}_{I^{}_1\dots I^{}_n}(\tau)$. In addition, $k^{}_Y$ and $\rho^{}_{Y}$ must satisfy $k^{}_{Y} = k^{}_{I^{}_1} +  \cdots + k^{}_{I^{}_N}$ and $\rho^{}_{Y} \otimes \rho^{}_{I^{}_{1}}  \otimes \cdots \otimes \rho^{}_{I^{}_{N}} \ni {\bf 1}$, respectively.

For the symmetry group $\Gamma^{}_{4} \simeq S^{}_{4}$ of our interest, there are five linearly independent modular forms of the lowest non-trivial weight $k^{}_{Y}=2$, denoted as $Y^{}_i(\tau)$ for $i = 1, 2, \cdots, 5$, which form a doublet ${\bf 2}$ and a triplet ${\bf 3}^{\prime}_{}$ under the modular $S^{}_4$ symmetry transformations~\cite{Penedo:2018nmg}, namely,
\begin{eqnarray}
Y^{}_{\bf 2}(\tau) \equiv \left(\begin{matrix} Y^{}_{1}(\tau) \\ Y^{}_2 (\tau)  \end{matrix}\right) \; , \quad Y^{}_{\bf 3^{\prime}_{}} (\tau) \equiv  \left(\begin{matrix} Y^{}_{3}(\tau) \\ Y^{}_4 (\tau) \\ Y^{}_{5} (\tau) \end{matrix}\right) \; .
\label{eq:S4Y}
\end{eqnarray}
The explicit expressions of $Y^{}_{i}(\tau)$ (for $i=1,2,\dots,5$) are presented in Appendix~\ref{sec:appA}, where one can also find how to derive the modular forms of higher weights from the products of modular forms of the weight $k^{}_Y = 2$.

Finally, we write down the superpotential relevant for lepton masses and flavor mixing in the minimal supersymmetric standard model (MSSM) with the seesaw extension, i.e.,
\begin{eqnarray}
{\cal W} = \sum_i \alpha^{}_i \left[f^{}_{l}(\tau)\widehat{L}^{m}_{} \widehat{H}^{}_{{\rm d} m} \widehat{E}^{\rm C}_{}\right]^{i}_{\bf 1} + \sum_i g^{}_i \left[f^{}_{\rm D}(\tau)\varepsilon^{}_{mn} \widehat{L}^{m}_{} \widehat{H}^{n}_{\rm u} \widehat{N}^{\rm C}\right]^{i}_{\bf 1} + \frac{1}{2}\sum_i \Lambda^{}_i \left[f^{}_{\rm R}(\tau) \widehat{N}^{\rm C}_{} \widehat{N}^{\rm C}_{}\right]^{i}_{\bf 1} \; , \quad
\label{eq:suppoten1}
\end{eqnarray}
where $\widehat{L}$, $\widehat{E}^{\rm C}_{}$ and $\widehat{N}^{\rm C}_{}$ stand for the chiral superfields containing the left-handed lepton doublets, right-handed charged-lepton singlets and right-handed neutrino singlets, respectively. Moreover, $\widehat{H}^{}_{\rm u}$ and $\widehat{H}^{}_{\rm d}$ are the Higgs superfields with the hypercharges $+1/2$ and $-1/2$, and the contraction of the ${\rm SU}(2)^{}_{\rm L}$ doublets has been explicitly shown with the Levi-Civita symbol $\varepsilon^{}_{mn}$ (i.e., $\varepsilon^{}_{12} = - \varepsilon^{}_{21} = 1$ and $\varepsilon^{}_{11} = \varepsilon^{}_{22} = 0$). Note that $f^{}_l(\tau)$, $f^{}_{\rm D}(\tau)$ and $f^{}_{\rm R}(\tau)$ are modular forms, which carry the flavor index and transform nontrivially under the modular flavor symmetry, and they together with the fermion superfields form the flavor singlets, each of which is labelled by the superscript $i$ and the subscript ${\bf 1}$. Therefore, on the right-hand side of Eq.~(\ref{eq:suppoten1}), the summation is taken over all possible flavor singlets $i$ together with the corresponding coefficients $\alpha^{}_i$, $g^{}_i$ and $\Lambda^{}_i$. When the modulus parameter $\tau$ is fixed, the modular symmetry is broken down and the superpotential reads
\begin{eqnarray}
{\cal W} = \lambda^{}_{l} \widehat{L}^{m}_{} \widehat{H}^{}_{{\rm d} m} \widehat{E}^{\rm C}_{} + \lambda^{}_{\rm D} \varepsilon^{}_{mn} \widehat{L}^{m}_{} \widehat{H}^{n}_{\rm u} \widehat{N}^{\rm C} + \frac{1}{2}\lambda^{}_{\rm R} \widehat{N}^{\rm C}_{} \widehat{N}^{\rm C}_{} \; ,
\label{eq:suppoten2}
\end{eqnarray}
where $\lambda^{}_{l}$ and $\lambda^{}_{\rm D}$ turn out to be the charged-lepton and Dirac neutrino Yukawa coupling matrices, respectively, while $\lambda^{}_{\rm R}$ becomes the right-handed neutrino mass matrix. As a result, the flavor structures of $\lambda^{}_l$, $\lambda^{}_{\rm D}$ and $\lambda^{}_{\rm R}$ are dictated by the modular symmetry to a large extent.

After the supersymmetry breaking and the spontaneous breakdown of the ${\rm SU}(2)^{}_{\rm L}\times {\rm U}(1)^{}_{\rm Y}$ gauge symmetry, one can obtain the lepton mass terms from Eq.~(\ref{eq:suppoten2}), whose SM version has been given just below Eq.~(\ref{eq:LagSM}). Converting into our left-right convention for the fermion mass terms, we find the following correspondence between the lepton mass matrices and the Yukawa coupling matrices in the MSSM framework
\begin{eqnarray}
M^{}_l = v^{}_{\rm d} \lambda^{\ast}_l/\sqrt{2} \; , \quad  M^{}_{\rm D} = v^{}_{\rm u} \lambda^{\ast}_{\rm D}/\sqrt{2} \; ,  \quad  M^{}_{\rm R} = \lambda^{\ast}_{\rm R} \; , \label{eq:Mlambda}
\end{eqnarray}
where $v^{}_{\rm d} = v \cos\beta$ and $v^{}_{\rm u} = v\sin\beta$ with $v \approx 246~{\rm GeV}$ are respectively the vev of the neutral scalar component field of $\widehat{H}^{}_{\rm d}$ and that of $\widehat{H}^{}_{\rm u}$, with $\tan\beta \equiv v^{}_{\rm u}/v^{}_{\rm d}$ being their ratio, and ``$\ast$'' denotes the complex conjugation.

\section{Low-energy Phenomenology}\label{sec:models}

After setting up the framework, now we propose two viable scenarios of the MSM with the modular $S^{}_4$ symmetry and explore their implications for the low-energy observables. Some general remarks on the model building are helpful.
\begin{itemize}
\item As we have already stressed, two right-handed neutrino singlets $\{\widehat{N}^{\rm C}_1, \widehat{N}^{\rm C}_2\}$ can naturally be assigned into the unique two-dimensional irreducible representation ${\bf 2}$ of the $S^{}_4$ symmetry group. Such an assignment will always be assumed in the present work.

\item Three lepton doublets $\{\widehat{L}^{}_1, \widehat{L}^{}_2, \widehat{L}^{}_3\}$ are arranged as a triplet ${\bf 3}$ under the $S^{}_4$ symmetry, while three charged-lepton singlets $\{\widehat{E}^{\rm C}_1, \widehat{E}^{\rm C}_2, \widehat{E}^{\rm C}_3\}$ should be assigned into the trivial ${\bf 1}$ or nontrivial ${\bf 1}^\prime$ singlets of $S^{}_4$. Otherwise, it will be difficult to accommodate the strong mass hierarchy of three charged leptons, namely, $m^{}_e \ll m^{}_\mu \ll m^{}_\tau$.

\item The Higgs doublets $\{\widehat{H}^{}_{\rm u}, \widehat{H}^{}_{\rm d}\}$ are naturally assigned into the trivial representation ${\bf 1}$ under the $S^{}_{4}$ symmetry. In this case, their modular weights $k^{}_{\rm u}$ and $k^{}_{\rm d}$ are both vanishing. As a consequence, the remaining part of the MSSM irrelevant for leptonic flavor mixing need not be changed.
\end{itemize}

Given the above assignments, we are then left with the coefficients $f^{}_l(\tau)$, $f^{}_{\rm D}(\tau)$ and $f^{}_{\rm R}(\tau)$ in Eq.~(\ref{eq:suppoten1}), which are modular forms with the modular weights being positive and even integers. Once the weights of supermultiplets are fixed, those of modular forms can be determined from the identity $k^{}_{Y} = k^{}_{I^{}_1} +  \cdots + k^{}_{I^{}_N}$. Since the explicit expressions of $f^{}_l(\tau)$, $f^{}_{\rm D}(\tau)$ and $f^{}_{\rm R}(\tau)$ with specific weights are known, one can figure out the mass matrices of both charged leptons and neutrinos, from which the lepton mass spectra and flavor mixing parameters can be extracted. Due to the freedom in choosing the weights of supermultiplets, we follow the criterion that the number of free model parameters should be as small as possible. More explicitly, we first count the number of low-energy observables: three charged-lepton masses $\{m^{}_e, m^{}_\mu, m^{}_\tau\}$, two independent neutrino mass-squared differences $\{\Delta m^2_{21}, |\Delta m^2_{31}|\}$, three mixing angles $\{\theta^{}_{12}, \theta^{}_{13}, \theta^{}_{23}\}$ and one Dirac CP-violating phase $\delta$. Thus the number of free model parameters should be no more than nine in order to have predictive power for the other parameters, such as the Majorana CP-violating phase. To this end, we restrict the weights of $f^{}_l(\tau)$, $f^{}_{\rm D}(\tau)$ and $f^{}_{\rm R}(\tau)$ into the value ranging from 2 to 6. After a systematic analysis, two classes of distinct models surviving current neutrino oscillation data have been found and will be discussed in detail in the following two subsections.

\subsection{Model A}
\begin{table}[t!]
\centering
\caption{The charge assignments of the chiral superfields and the relevant couplings under the ${\rm SU}(2)^{}_{\rm L}$ gauge symmetry and the modular $S^{}_{4}$ symmetry in {\bf Model A}, with the corresponding modular weights listed in the last row.}
\vspace{0.5cm}
\begin{tabular}{ccccccccc}
\toprule
		& $\widehat{L}$ & $\widehat{E}^{\rm C}_{1}, \widehat{E}^{\rm C}_{2}, \widehat{E}^{\rm C}_{3}$ & $\widehat{N}^{\rm C}_{}$ & $\widehat{H}^{}_{\rm u}$ & $\widehat{H}^{}_{\rm d}$ & $f^{}_{e}(\tau), f^{}_{\mu}(\tau), f^{}_{\tau}(\tau)$ & $f^{}_{\rm D}(\tau), f^\prime_{\rm D}(\tau)$ & $f^{}_{\rm R} (\tau)$\\
\midrule
		${\rm SU}(2)^{}_{\rm L}$ & 2 & 1 & 1 & 2 & 2 & 1 & 1 & 1 \\
		$S^{}_{4}$ & \bf{3} & $\bf{1}^{\prime}_{},\bf{1},\bf{1}$ & \bf{2} & \bf{1} & \bf{1} & $\bf{3^{\prime}_{}},\bf{3},\bf{3}$ & $\bf{3},\bf{3}^{\prime}_{}$ & \bf{2} \\
		$-k^{}_{I}$ & $-5$ & $3,1,-1$ & $-1$ & 0 & 0 & $k^{}_{e,\mu,\tau}=2,4,6$ & $k^{}_{\rm D}=6$ & $k^{}_{\rm R}=2$ \\
\bottomrule
\end{tabular}
\label{Table:CaseA}
\end{table}

In the first model, the charge assignments of the chiral superfields and the couplings under the ${\rm SU}(2)^{}_{\rm L}$ gauge symmetry and the modular $S^{}_{4}$ symmetry have been summarized in Table~\ref{Table:CaseA}, where the corresponding modular weights are listed in the last row. Some explanations for the assignments of all the Yukawa couplings are necessary. The charged-lepton Yukawa couplings $f^{}_{e}(\tau)$, $f^{}_{\mu}(\tau)$, $f^{}_{\tau}(\tau)$ are set to be the triplet ${\bf 3}^\prime$ with a weight of $2$, ${\bf 3}$ with a weight of $4$, and ${\bf 3}$ with a weight of $6$, respectively. On the other hand, two triplets $f^{}_{\rm D}(\tau)$ and $f^\prime_{\rm D}(\tau)$ with the weight of $6$ will be introduced. One of them is set to ${\bf 3}$, while the other is ${\bf 3}^\prime$. However, there are two distinct forms of $\bf 3^{\prime}_{}$ with a weight of $6$, as one can see from Eq.~(\ref{eq:Y6}) in Appendix~\ref{sec:appA}, so both of them should in principle enter into the model. In this case, we could not avoid introducing more than eight free parameters, which makes our model less economical and predictive. Therefore, we will mainly focus on the scenario where only one of these two forms are retained. Numerically we find that $Y^{(6)}_{{\bf 3}^{\prime}_{},2}$ for the ${\bf 3}^{\prime}_{}$ with a weight $6$ in Eq.~(\ref{eq:Y6}) could not lead to realistic flavor mixing and the relatively simple form $Y^{(6)}_{{\bf 3}^{\prime}_{},1}$ is the only choice for $f^{\prime}_{\rm D}(\tau)$ in {\bf Model A}. At the end of this subsection, we shall also consider both $Y^{(6)}_{{\bf 3}^{\prime}_{},1}$ and $Y^{(6)}_{{\bf 3}^{\prime}_{},2}$, and explore their phenomenological implications.

Now it is straightforward to verify that the superpotential is invariant under the modular $S^{}_4$ symmetry and can be decomposed into three parts ${\cal W} = {\cal W}^{}_l + {\cal W}^{}_{\rm D} + {\cal W}^{}_{\rm R}$ with
\begin{eqnarray}
{\cal W}^{}_l &=& \alpha^{}_{1} \left[\left(\widehat{L} \widehat{E}^{\rm C}_{1}\right)^{}_{{\bf 3}^\prime} \left(f^{}_e(\tau)\right)^{}_{{\bf 3}^\prime}\right]^{}_{\bf 1} \widehat{H}^{}_{\rm d} + \alpha^{}_{2} \left[\left(\widehat{L} \widehat{E}^{\rm C}_{2}\right)^{}_{\bf 3} \left(f^{}_{\mu}(\tau)\right)^{}_{\bf 3} \right]^{}_{\bf 1} \widehat{H}^{}_{\rm d} + \alpha^{}_{3} \left[ \left(\widehat{L} \widehat{E}^{\rm C}_{3}\right)^{}_{\bf 3} \left(f^{}_{\tau}(\tau)\right)^{}_{\bf 3} \right]^{}_{\bf 1}  \widehat{H}^{}_{\rm d} \nonumber \; ,
\\
{\cal W}^{}_{\rm D} &=& g^{}_{1}\left[\left(\widehat{L} \widehat{N}^{\rm C}_{}\right)^{}_{\bf 3} \left(f^{}_{\rm D}(\tau) \right)^{}_{\bf 3}\right]^{}_{\bf 1} \widehat{H}^{}_{\rm u} + g^{}_{2}\left[\left( \widehat{L} \widehat{N}^{\rm C}_{}\right)^{}_{{\bf 3}^{\prime}_{}} \left(f^{\prime}_{\rm D}(\tau) \right)^{}_{{\bf 3}^{\prime}_{}}\right]^{}_{\bf 1} \widehat{H}^{}_{\rm u} \nonumber \; ,
\\
{\cal W}^{}_{\rm R} &=& \frac{1}{2}\Lambda \left[\left( \widehat{N}^{\rm C}_{} \widehat{N}^{\rm C}_{}\right)^{}_{\bf 2} \left(f^{}_{\rm R}(\tau)\right)^{}_{\bf 2}\right]^{}_{\bf 1} \; ,
\label{eq:superpotentiaA}
\end{eqnarray}
where the isospin indices of the ${\rm SU}(2)^{}_{\rm L}$ doublets have been suppressed and the representations of the $S^{}_4$ symmetry have been explicitly indicated. By using the product rules of the $S^{}_{4}$ symmetry group collected in Appendix~\ref{sec:appA}, we can obtain the charged-lepton mass matrix
\begin{eqnarray}
M^{}_l = \frac{v^{}_{\rm d}}{\sqrt{2}} \left( \begin{matrix}
\alpha^{}_{1} Y^{}_{3} &&&  -2\alpha^{}_{2} Y^{}_{2}Y^{}_{3}   &&&  \alpha^{}_{3} Y^{}_{1}(Y^{2}_{4}-Y^{2}_{5})   \\ \alpha^{}_{1} Y^{}_{5}
&&& \alpha^{}_{2} (\sqrt{3}Y^{}_{1}Y^{}_{4}+Y^{}_{2}Y^{}_{5})   &&&  -\alpha^{}_{3} Y^{}_{3}(Y^{}_{1}Y^{}_{4}+\sqrt{3}Y^{}_{2}Y^{}_{5}) \\ \alpha^{}_{1} Y^{}_{4}
&&&  \alpha^{}_{2} (\sqrt{3}Y^{}_{1}Y^{}_{5}+Y^{}_{2}Y^{}_{4})   &&& \alpha^{}_{3} Y^{}_{3}(Y^{}_{1}Y^{}_{5}+\sqrt{3}Y^{}_{2}Y^{}_{4})
\end{matrix} \right)^* \; ,
\label{eq:MeA}
\end{eqnarray}
and the Dirac neutrino mass matrix
\begin{eqnarray}
M^{}_{\rm D} = && \frac{v^{}_{\rm u}}{\sqrt{2}}  \left[\rule{0cm}{1.4cm}\right. g^{}_{1}\left(
\begin{matrix}
Y^{}_{1}(Y^{2}_{4}-Y^{2}_{5}) &&& 0 \\ \dfrac{1}{2}Y^{}_{3}(Y^{}_{1}Y^{}_{4}+\sqrt{3}Y^{}_{2}Y^{}_{5})  &&&  \dfrac{\sqrt{3}}{2}Y^{}_{3}(Y^{}_{1}Y^{}_{5}+\sqrt{3}Y^{}_{2}Y^{}_{4}) \\ -\dfrac{1}{2}Y^{}_{3}(Y^{}_{1}Y^{}_{5}+\sqrt{3}Y^{}_{2}Y^{}_{4})
&&& -\dfrac{\sqrt{3}}{2}Y^{}_{3}(Y^{}_{1}Y^{}_{4}+\sqrt{3}Y^{}_{2}Y^{}_{5})
\end{matrix}\right)  \nonumber \\
&& + g^{}_{2} (Y^{2}_{1}+Y^{2}_{2}) \left(
\begin{matrix}
0  &&& -Y^{}_{3} \\ \dfrac{\sqrt{3}}{2}Y^{}_{4} &&&  \dfrac{1}{2}Y^{}_{5}   \\
\dfrac{\sqrt{3}}{2}Y^{}_{5}  &&& \dfrac{1}{2}Y^{}_{4}\end{matrix}\right) \left.\rule{0cm}{1.4cm}\right]^* \; ,
\label{eq:MDA}
\end{eqnarray}
where the complex conjugation ``$*$" on the right-hand side should be noticed as we have explained in Eq.~(\ref{eq:Mlambda}). Without loss of generality, we can choose $\alpha^{}_i$ (for $i = 1, 2, 3$) in Eq.~(\ref{eq:MeA}) to be real and positive. Since the overall phase of any lepton mass matrix is irrelevant for lepton masses and flavor mixing, one can take $g^{}_1$ in Eq.~(\ref{eq:MDA}) to be real and it is convenient to parametrize the other complex parameter as $g^{}_{2}/g^{}_{1} \equiv \widetilde{g} = g {\rm e}^{{\rm i}\phi^{}_{g}}_{}$ with $g = |\widetilde{g}|$ and $\phi^{}_g \equiv \arg(\widetilde{g})$. In addition, the Majorana mass matrix of right-handed neutrinos is given by
\begin{eqnarray}
M^{}_{\rm R} = \Lambda\left(\begin{matrix}
-Y^{}_{1} &&& Y^{}_{2} \\
Y^{}_{2} &&& Y^{}_{1}
\end{matrix}\right)^* \; ,
\label{eq:MRA}
\end{eqnarray}
where $\Lambda$ is real and positive parameter characterizing the absolute scale of heavy Majorana neutrino masses. With the help of the seesaw formula $M^{}_\nu \approx - M^{}_{\rm D} M^{-1}_{\rm R} M^{\rm T}_{\rm D}$, we finally arrive at the effective neutrino mass matrix $M^{}_{\nu}$ and will be able to explore its implications for lepton mass spectra and flavor mixing.

Given the complex parameter $\tau$ (or equivalently two real parameters ${\rm Re}\,\tau$ and ${\rm Im}\,\tau$), one can determine three model parameters $v^{}_{\rm d} \alpha^{}_3/\sqrt{2}$, $\alpha^{}_1/\alpha^{}_3$ and $\alpha^{}_2/\alpha^{}_3$ from the observed charged-lepton masses $m^{}_{e}=0.511~{\rm MeV}$, $m^{}_{\mu}=105.7~{\rm MeV}$ and $m^{}_{\tau} = 1776.86~{\rm MeV}$~\cite{Tanabashi:2018oca}. Since the absolute scale of neutrino masses is fixed by $v^2_{\rm u}g^2_1/(2\Lambda)$, we are left with only two free parameters $g$ and $\phi^{}_g$, which together with ${\rm Re}\,\tau$ and ${\rm Im}\,\tau$ will give rise to three neutrino mixing angles $\{\theta^{}_{12}, \theta^{}_{13}, \theta^{}_{23}\}$, the ratio of two neutrino mass-squared differences $\Delta m^2_{21}/|\Delta m^2_{31}|$, the Dirac CP-violating phase $\delta$ and the Majorana CP-violating phase $\sigma$. As the number of free parameters is two less than that of the observables, the model under consideration should be predictive. Then we proceed to explore the phenomenological implications for lepton mass spectra, flavor mixing and CP violation. To confront our model with the latest neutrino oscillation data, we perform a numerical analysis and demonstrate that the predictions are compatible with the experimental data only in the NO case at the $3\sigma$ level. Our strategy for numerical analysis is outlined below.
\begin{itemize}
\item First of all, the modulus parameter $\tau$ is randomly generated in the fundamental domain ${\cal G}$, which is defined as
    \begin{eqnarray}
	{\cal G} = \left\{ \tau \in \mathbb{C}: \quad {\rm Im}\,\tau > 0, \; | {\rm Re}\,\tau| \leq 0.5, \; |\tau| \geq 1 \right\} \; .
   \label{eq:fundo}
   \end{eqnarray}
This domain can be identified by using the basic properties of the modular forms as clearly explained in Ref.~\cite{Novichkov:2018ovf}. Another interesting feature of the modular forms should be noticed. If the replacement $\tau \to -\tau^{\ast}_{}$ is made in Eq.~(\ref{eq:Y3q}), the modular forms $Y^{}_i(\tau)$ will change to their complex-conjugate counterparts, i.e., $Y^{}_i(-\tau^{\ast_{}}) = Y^{\ast}_{i}(\tau)$. If we further replace $\widetilde{g}$ with $\widetilde{g}^{\ast}$ in $M^{}_{\rm D}$, all the lepton mass matrices become their complex-conjugate counterparts. Under such replacements, the theoretical predictions for all the experimental observables keep unchanged except that the signs of all CP-violating phases should be reversed.
\begin{table}[t]
\begin{center}
\vspace{-0.25cm} \caption{The best-fit values, the
1$\sigma$ and 3$\sigma$ intervals, together with the values of $\sigma^{}_{i}$ being the symmetrized $1\sigma$ uncertainties, for three neutrino mixing angles $\{\theta^{}_{12}, \theta^{}_{13}, \theta^{}_{23}\}$, two mass-squared differences $\{\Delta m^2_{21}, \Delta m^2_{31}~{\rm or}~\Delta m^2_{32}\}$ and the Dirac CP-violating phase $\delta$ from a global-fit analysis of current experimental data~\cite{Esteban:2018azc}.} \vspace{0.5cm}
\begin{tabular}{c|c|c|c|c}
\hline
\hline
Parameter & Best fit & 1$\sigma$ range &  3$\sigma$ range & $\sigma^{}_{i}$ \\
\hline
\multicolumn{5}{c}{Normal neutrino mass ordering
$(m^{}_1 < m^{}_2 < m^{}_3)$} \\ \hline
$\sin^2_{}\theta^{}_{12}$
& $0.310$ & 0.298 --- 0.323 &  0.275 --- 0.350 & 0.0125 \\
$\sin^2_{}\theta^{}_{13}$
& $0.02241$ & 0.02176 --- 0.02307 &  0.02046 --- 0.02440  & 0.000655 \\
$\sin^2_{}\theta^{}_{23}$
& $0.558$  & 0.525 --- 0.578 &  0.427 --- 0.609  & 0.0265 \\
$\delta/^\circ$ &  $222$ & 194 --- 260 &  141 --- 370 & 33 \\
$\Delta m^2_{21}/[10^{-5}~{\rm eV}^2]$ &  $7.39$ & 7.19 --- 7.60 & 6.79 --- 8.01 & 0.205 \\
$\Delta m^2_{31}/[10^{-3}~{\rm eV}^2]$ &  $+2.523$ & +2.493 --- +2.555 & +2.432 --- +2.618 & 0.031 \\\hline
\multicolumn{5}{c}{Inverted neutrino mass ordering
$(m^{}_3 < m^{}_1 < m^{}_2)$} \\ \hline
$\sin^2_{}\theta^{}_{12}$
& $0.310$ & 0.298 --- 0.323 &  0.275 --- 0.350 & 0.0125\\
$\sin^2_{}\theta^{}_{13}$
& $0.02261$ & 0.02197 --- 0.02328 &  0.02066 --- 0.02461 & 0.000655 \\
$\sin^2_{}\theta^{}_{23}$
& $0.563$  & 0.537 --- 0.582 &  0.430 --- 0.612 & 0.0225  \\
$\delta/^\circ$ &  $285$ & 259 --- 309 &  205 --- 354 & 25 \\
$\Delta m^2_{21}/[10^{-5}~{\rm eV}^2]$ &  $7.39$ & 7.19 --- 7.60 & 6.79 --- 8.01 & 0.205 \\
$\Delta m^2_{32}/[10^{-3}~{\rm eV}^2]$ &  $-2.509$ & $-2.539$ --- $-2.477$  & $-2.603$ --- $-2.416$ & $0.031$ \\ \hline\hline
\end{tabular}
\label{table:gfit}
\end{center}
\end{table}
	
Once the values of $\{ {\rm Re}\,\tau, {\rm Im}\,\tau\}$ are randomly chosen, we can extract the parameters $v^{}_{\rm d} \alpha^{}_{3}/\sqrt{2}$, $\alpha^{}_{1}/\alpha^{}_{3}$ and $\alpha^{}_{2}/\alpha^{}_{3}$ from the following identities
	\begin{eqnarray}
	{\rm Tr} \left( M^{}_l M^{\dag}_l \right) &=& m^{2}_{e} + m^{2}_{\mu} + m^{2}_{\tau} \; ,  \label{eq:tr}\\
	{\rm Det}\left( M^{}_l M^{\dag}_l \right) &=& m^{2}_{e} m^{2}_{\mu} m^{2}_{\tau} \; , \label{eq:det}\\
	\dfrac{1}{2}\left[{\rm Tr} \left(M^{}_l M^{\dag}_l\right)\right]^2_{} - \dfrac{1}{2}{\rm Tr}\left[ (M^{}_l M^{\dag}_l)^2_{}\right] &= & m^{2}_{e}m^{2}_{\mu}+m^{2}_{\mu}m^{2}_{\tau}+m^{2}_{\tau}m^{2}_{e} \; \label{eq:tr2}.
	\end{eqnarray}
So far all the parameters in $M^{}_l$ have been determined. It is then easy to diagonalize the charged-lepton mass matrix via $U^{\dag}_l M^{}_l M^{\dag}_l U^{}_l = {\rm Diag} \left\{ m^{2}_{e}, m^{2}_{\mu}, m^{2}_{\tau}\right\}$, from which the unitary matrix $U^{}_l$ can be obtained.

\item Next the values of the other two parameters $g \in [0,10]$ and $\phi^{}_{g} \in [0^{\circ}_{}, 360^{\circ}_{})$ are randomly generated. Therefore, the effective neutrino mass matrix $M^{}_{\nu}$ is determined up to the overall scale parameter $v^2_{\rm u} g^2_1/(2\Lambda)$, which can be fixed by ${\rm Tr} \left[M^{}_\nu M^\dagger_\nu\right] = m^2_2 + m^2_3 = \Delta m^2_{21} + \Delta m^2_{31}$ in the NO case or ${\rm Tr} \left[M^{}_\nu M^\dagger_\nu\right] = m^2_1 + m^2_2 =  2|\Delta m^2_{32}|-\Delta m^2_{21}$ in the IO case. In practice, we also randomly choose two neutrino mass-squared differences in their allowed ranges and then $M^{}_\nu$ is numerically known. After diagonalizing $M^{}_\nu$ via $U^{\dag}_{\nu} M^{}_{\nu} U^{\ast}_{\nu} = {\rm Diag}\left\{m^{}_{1}, m^{}_{2}, m^{}_{3}\right\}$, we get the unitary matrix $U^{}_{\nu}$. Hence, the lepton flavor mixing matrix $U = U^{\dag}_l U^{}_{\nu}$ can be calculated by using $U^{}_l$ and $U^{}_\nu$. In the standard parametrization~\cite{Tanabashi:2018oca}, we have
	\begin{eqnarray}
	U = \left(
	\begin{matrix}
	c^{}_{12}c^{}_{13} && s^{}_{12}c^{}_{13} && s^{}_{13}e^{-{\rm i}\delta}_{} \\
	-s^{}_{12}c^{}_{23}-c^{}_{12}s^{}_{23}s^{}_{13}e^{{\rm i}\delta} && c^{}_{12}c^{}_{23}-s^{}_{12}s^{}_{23}s^{}_{13}e^{{\rm i}\delta}_{} && s^{}_{23}c^{}_{13} \\
	s^{}_{12}s^{}_{23}-c^{}_{12}c^{}_{23}s^{}_{13}e^{{\rm i}\delta} && -c^{}_{12}s^{}_{23}-s^{}_{12}c^{}_{23}s^{}_{13}e^{{\rm i}\delta} &&
	c^{}_{23}c^{}_{13}
	\end{matrix}\right) \cdot \left(
	\begin{matrix}
	1 &&& ~  &&& ~\\
	~ &&& {\rm e}^{{\rm i}\sigma} &&& ~ \\
	~ &&& ~ &&& 1
	\end{matrix}\right) \; ,
	\label{eq:UPMNS}
	\end{eqnarray}
	where $c^{}_{ij} \equiv \cos \theta^{}_{ij}$ and $s^{}_{ij} \equiv \sin \theta^{}_{ij}$ (for $ij =12,13,23$) have been defined, $\delta$ and $\sigma$ are the Dirac and Majorana CP-violating phases, respectively. Note that the lightest neutrino mass in the MSM is vanishing, so we are left with only one Majorana CP-violating phase.
\end{itemize}
\begin{figure}[t!]
\centering		\includegraphics[width=1\textwidth]{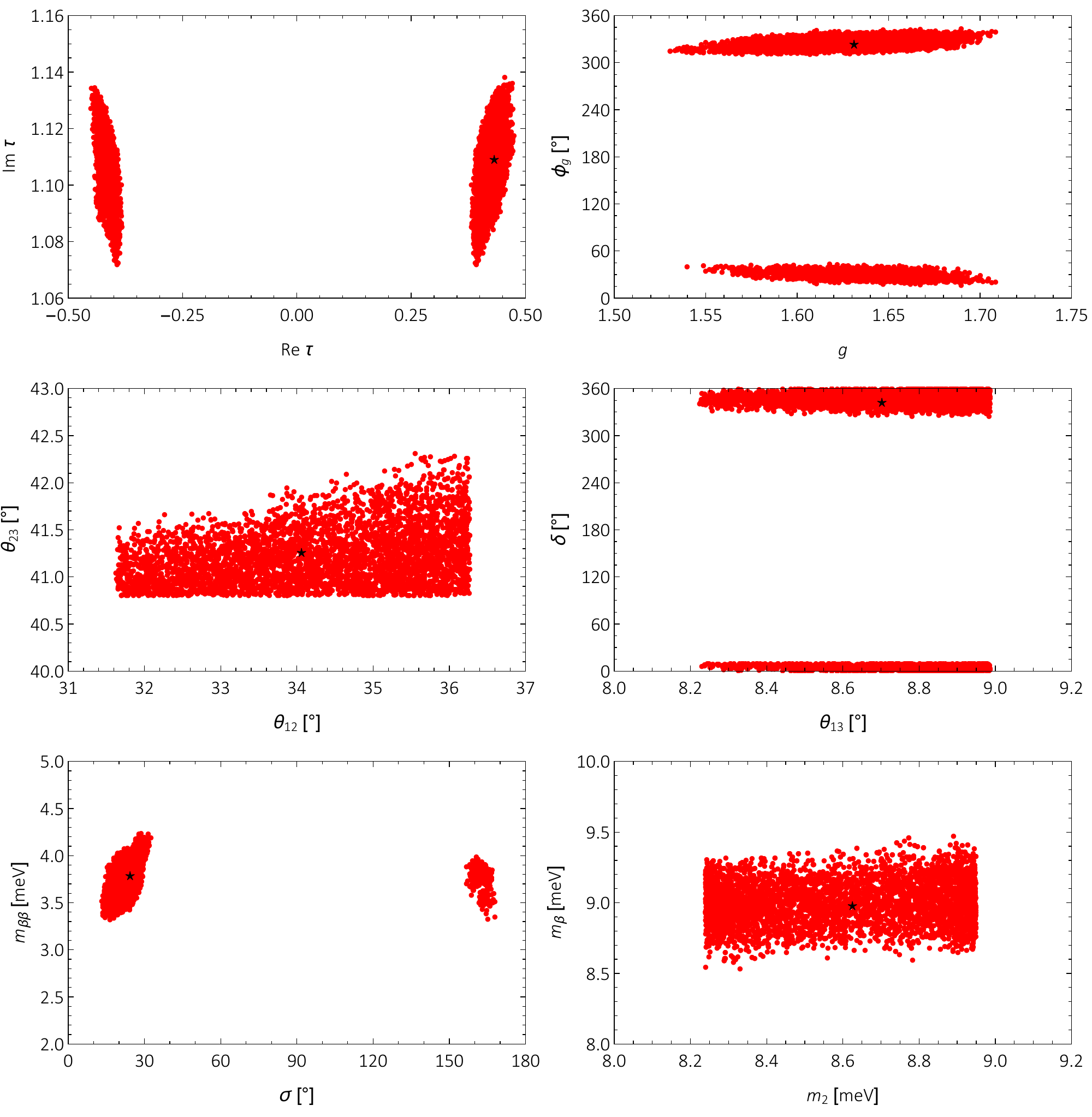}
\vspace{-0.5cm}
\caption{Allowed ranges of the model parameters $\{{\rm Re}\,\tau, {\rm Im}\,\tau\}$ and $\{g, \phi^{}_g\}$ and the constrained ranges of low-energy observables in the NO case in {\bf Model A}, where the $3\sigma$ ranges of neutrino mixing parameters and mass-squared differences from the global-fit analysis of neutrino oscillation data have been input~\cite{Esteban:2018azc}. Notice that the best-fit values from our $\chi^2$-fit analysis are indicated by the black stars.}
\label{fig:CaseA} 
\end{figure}

To find out the allowed parameter space of $\{{\rm Re}\,\tau, {\rm Im}\,\tau\}$ and $\{g, \phi^{}_g\}$, we implement the global-fit results from NuFIT 4.1~\cite{Esteban:2018azc} without including the atmospheric neutrino data from Super-Kamiokande. The best-fit values of three neutrino mixing angles $\{\theta^{}_{12}, \theta^{}_{13}, \theta^{}_{23}\}$, two neutrino mass-squared differences $\{\Delta m^2_{21}, \Delta m^2_{31}\}$ (or $\{\Delta m^2_{21}, \Delta m^2_{32}\}$), the Dirac CP-violating phase $\delta$ and their $1\sigma$ and $3\sigma$ ranges in the NO (or IO) case are summarized in Table~\ref{table:gfit}. Note that although the current constraint on $\delta$ from the global-fit results is rather weak, we still include the information of $\delta$ in the numerical analysis to further restrict the parameter space in our models. The allowed parameter space and the constraint on the mixing parameters and other low-energy observables have been shown in Fig.~\ref{fig:CaseA}. Some comments on the numerical results are in order.
\begin{itemize}
\item To determine the model parameters from neutrino oscillation data and demonstrate how well the model is consistent with observations, we construct the $\chi^2$-function by regarding the best-fit values $q^{\rm bf}_j$ of the oscillation parameters $q^{}_j \in \{\sin^2\theta^{}_{12}, \sin^2\theta^{}_{13}, \sin^2\theta^{}_{23}, \Delta m^2_{21}, \Delta m^2_{31}, \delta\}$ from the global analysis in Ref.~\cite{Esteban:2018azc} as experimental measurements, namely,
\begin{eqnarray}
\chi^2_{}(p^{}_{i}) = \sum^{}_{j}\left(\frac{q^{}_{j}(p^{}_{i})-q^{\rm bf}_{j}}{\sigma^{}_{j}}\right)^2_{} \; ,
\label{eq:chi}
\end{eqnarray}
where $p^{}_{i} \in \{{\rm Re}\,\tau, {\rm Im}\,\tau, g, \phi^{}_g\}$ stand for four free model parameters, and $q^{}_j(p^{}_i)$ denote the model predictions for observables. The uncertainties $\sigma^{}_j$ have been derived by symmetrizing $1\sigma$ uncertainties from the global-fit analysis, which have already been given in Table~\ref{table:gfit}. This setup is for the NO case, while for the IO case the oscillation parameters are instead taken to be $q^{}_j \in \{ \sin^2\theta^{}_{12}, \sin^2\theta^{}_{13}, \sin^2\theta^{}_{23},$ $\Delta m^2_{21}, \Delta m^2_{32}, \delta \}$. Once the model parameters are known, we can get the constraints on the observables $q^{}_j$, as well as the predictions for the Majorana CP-violating phase $\sigma$ and the effective neutrino mass $m^{}_\beta$ for beta decays and $m^{}_{\beta\beta}$ for neutrinoless double-beta decays, which will be discussed later on.
	
\item The allowed parameter space of $\{{\rm Re}\,\tau, {\rm Im}\,\tau\}$ and $\{g, \phi^{}_g\}$ exists only in the NO case at the $3\sigma$ level. As one can observe from the constrained range of $\theta^{}_{23}$ in the second row of the left panel in Fig.~\ref{fig:CaseA}, only relatively small values of $\theta^{}_{23}$ survive. In particular, the maximal value of $\theta^{}_{23}$ is no more than $42.5^{\circ}_{}$, whereas the best-fit value from the global-fit analysis is $48.3^\circ$ in the NO case or $48.6^\circ$ in the IO case.

\item The allowed parameter space of ${\rm Re}\,\tau$ contains two well separated regions $-0.45 < {\rm Re}\,\tau < -0.38$ and $0.38 < {\rm Re}\,\tau < 0.48$. However these two regions are not exactly centered on the axis ${\rm Re}\,\tau=0$ since the constraint on $\delta$ from the global-fit results has been taken into account. Meanwhile, this is also the case for the phase $\phi^{}_{g}$, namely, $17^\circ < \phi^{}_g < 45^\circ$ and $310^\circ < \phi^{}_g < 343^\circ$. Based on the $\chi^2_{}$-fit analysis, we obtain the minimum $\chi^{2}_{\rm min} = 35.4$ in the NO case, which corresponds to the following best-fit values of the model parameters
\begin{eqnarray}
{\rm Re}\,\tau = 0.432\;, \quad {\rm Im}\,\tau = 1.11 \;, \quad g = 1.63 \;, \quad \phi^{}_{g} = 322^{\circ}_{} \;,
\label{eq:bfAN1}
\end{eqnarray}
which together with the charged-lepton masses $m^{}_\alpha$ (for $\alpha = e, \mu, \tau$) lead to $v^{}_{\rm d} \alpha^{}_3 /\sqrt{2} = 1.57~{\rm GeV}$, $\alpha^{}_1/\alpha^{}_3 = 3.10\times 10^{-4}$ and $\alpha^{}_2/\alpha^{}_3 = 3.92\times 10^{-2}$. In addition, the absolute scale of neutrino masses is given by $v^2_{\rm u}g^2_1/(2\Lambda) = 9.89~{\rm meV}$. With these best-fit values of model parameters, we get the neutrino mass spectrum $m^{}_1 = 0$, $m^{}_2 = 8.63~{\rm meV}$ and $m^{}_3 = 50.2~{\rm meV}$, three mixing angles $\theta^{}_{12} = 34.1^{\circ}$, $\theta^{}_{13} = 8.70^{\circ}$ and $\theta^{}_{23} = 41.2^{\circ}$, and two CP-violating phases $\delta = 341^{\circ}$ and $\sigma = 24.4^{\circ}$.

\item Since the neutrino mass spectrum and the mixing parameters are known, we can predict the effective neutrino mass for beta decays, i.e.,
    \begin{eqnarray}
    m^{}_\beta \equiv \sqrt{m^2_1 |U^{}_{e1}|^2 + m^2_2 |U^{}_{e2}|^2 + m^2_3 |U^{}_{e3}|^2} \; .
    \label{eq:mbeta}
    \end{eqnarray}
    For the MSM in the NO case, we have $m^{}_1 = 0$ and thus the effective neutrino mass can be easily calculated. Given the best-fit values $\sin^2 \theta^{}_{12} = 0.313$ and $\sin^2 \theta^{}_{13} = 0.0229$ that we have already obtained, together with $\Delta m^2_{21} = 7.44\times 10^{-5}~{\rm eV}^2$ and $\Delta m^2_{31} = 2.521\times 10^{-3}~{\rm eV}^2$, one arrives at
    \begin{eqnarray}
    m^{}_\beta = \sqrt{\Delta m^2_{21} \cos^2 \theta^{}_{13} \sin^2 \theta^{}_{12} + \Delta m^2_{31} \sin^2 \theta^{}_{13}} = 8.97~{\rm meV} \; .
    \label{eq:mbetaA}
    \end{eqnarray}
    The latest result from the KATRIN experiment, where the electron energy spectrum from tritium beta decays is precisely measured, indicates $m^{}_\beta < 1.1~{\rm eV}$ at the $90\%$ confidence level~\cite{Aker:2019uuj, Aker:2019qfn}. With more data accumulated in KATRIN, the upper bound will be improved to $m^{}_\beta < 0.2~{\rm eV}$. However, taking into account of the $3\sigma$ ranges of neutrino mixing parameters and mass-squared differences, we obtain $8.5~{\rm meV} \lesssim m^{}_\beta \lesssim 9.5~{\rm meV}$ in our model. This is far below the future sensitivity of KATRIN.

\item Furthermore, ordinary neutrinos turn out to be Majorana particles in the seesaw model, indicating that the neutrinoless double-beta decays of some even-even heavy nuclei could take place. The effective neutrino mass relevant for neutrinoless double-beta decays is usually defined as
    \begin{eqnarray}
    m^{}_{\beta\beta} \equiv |m^{}_1 U^2_{e1} + m^{}_2 U^2_{e2} + m^{}_3 U^2_{e3}| \; ,
    \label{eq:mbb}
    \end{eqnarray}
    which can be computed with $m^{}_1 = 0$ in our model. Given the best-fit values of the free model parameters in Eq.~(\ref{eq:bfAN1}), we find that the best-fit value of $m^{}_{\beta\beta}$ in {\bf Model A} is $3.78~{\rm meV}$. The $3\sigma$ range of $m^{}_{\beta\beta}$ is found to be $3.3~{\rm meV} \lesssim m^{}_{\beta\beta} \lesssim 4.3~{\rm meV}$, as can be seen in Fig.~\ref{fig:CaseA}. Hence it is very challenging for the next-generation neutrinoless double-beta decay experiments to reach a high enough sensitivity to observe a positive signal~\cite{Dolinski:2019nrj, Cao:2019hli}.
\end{itemize}

In summary, {\bf Model A} is compatible with neutrino oscillation data only in the NO case at the $3\sigma$ level, but we cannot find any viable parameter space of $\{{\rm Re}\,\tau, {\rm Im}\,\tau\}$ and $\{g, \phi^{}_g\}$ at the $1\sigma$ level. As we shall show later, the successful leptogenesis can be realized in {\bf Model A} only for an extremely-high scale of heavy neutrino masses, namely, the lightest heavy Majorana neutrino mass $M^{}_1 \gtrsim 10^{11}~{\rm GeV}$, requiring a high reheating temperature.

\begin{figure}[t!]
	\centering		\includegraphics[width=1.01\textwidth]{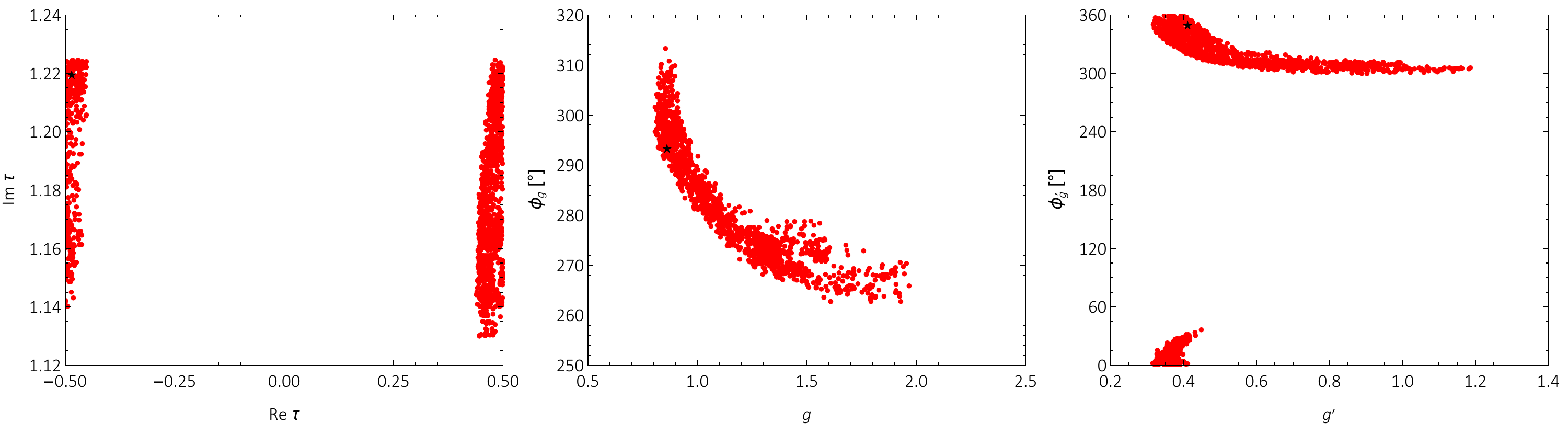}
	\vspace{-0.5cm}
	\caption{Allowed ranges of the model parameters $\{{\rm Re}\,\tau, {\rm Im}\,\tau\}$, $\{g, \phi^{}_g\}$ and $\{g^{\prime}_{}, \phi^{\prime}_g\}$ in the modified version of {\bf Model A} in the NO case , where the $1\sigma$ ranges of neutrino mixing parameters and mass-squared differences from the global-fit analysis of neutrino oscillation data have been input~\cite{Esteban:2018azc}. Note that the best-fit values from our $\chi^2$-fit analysis are indicated by the black stars.}
	\label{fig:CaseAMod1} 
\end{figure}

Before closing this subsection, we briefly discuss the scenario where both $Y^{(6)}_{{\bf 3^{\prime}_{}},1}$ and $Y^{(6)}_{{\bf 3^{\prime}_{}},2}$ are included in the superpotential ${\cal W}^{}_{\rm D}$ in {\bf Model A}. In this scenario, an additional term is added into the expression of ${\cal W}^{}_{\rm D}$, i.e.,
\begin{eqnarray}
{\cal W}^{}_{\rm D} = g^{}_{1}\left[\left(\widehat{L} \widehat{N}^{\rm C}_{}\right)^{}_{\bf 3} \left(f^{}_{\rm D}(\tau) \right)^{}_{\bf 3}\right]^{}_{\bf 1} \widehat{H}^{}_{\rm u} &+& g^{}_{2}\left[\left( \widehat{L} \widehat{N}^{\rm C}_{}\right)^{}_{{\bf 3}^{\prime}_{}} \left(f^{\prime}_{\rm D}(\tau) \right)^{}_{{\bf 3}^{\prime}_{}}\right]^{}_{\bf 1} \widehat{H}^{}_{\rm u} \nonumber \\
&+& g^{}_{3}\left[\left( \widehat{L} \widehat{N}^{\rm C}_{}\right)^{}_{{\bf 3}^{\prime}_{}} \left(f^{\prime\prime}_{\rm D}(\tau) \right)^{}_{{\bf 3}^{\prime}_{}}\right]^{}_{\bf 1} \widehat{H}^{}_{\rm u}\; ,
\end{eqnarray}
where the form of $f^{\prime\prime}_{\rm D}(\tau)$ in the second line is taken to be $Y^{(6)}_{{\bf 3^{\prime}_{}},2}$, as we have mentioned before, and $g^{}_3$ is the extra complex coupling constant. After applying the product rule of the $S^{}_4$ symmetry group, we can get the modified Dirac neutrino mass matrix
\begin{eqnarray}
M^{}_{\rm D} = && \frac{v^{}_{\rm u} g^{}_1}{\sqrt{2}}  \left[\rule{0cm}{1.4cm}\right.\left(
\begin{matrix}
Y^{}_{1}(Y^{2}_{4}-Y^{2}_{5}) && 0 \\ \dfrac{1}{2}Y^{}_{3}(Y^{}_{1}Y^{}_{4}+\sqrt{3}Y^{}_{2}Y^{}_{5})  &&  \dfrac{\sqrt{3}}{2}Y^{}_{3}(Y^{}_{1}Y^{}_{5}+\sqrt{3}Y^{}_{2}Y^{}_{4}) \\ -\dfrac{1}{2}Y^{}_{3}(Y^{}_{1}Y^{}_{5}+\sqrt{3}Y^{}_{2}Y^{}_{4})
&& -\dfrac{\sqrt{3}}{2}Y^{}_{3}(Y^{}_{1}Y^{}_{4}+\sqrt{3}Y^{}_{2}Y^{}_{5})
\end{matrix}\right) + \widetilde{g} (Y^{2}_{1}+Y^{2}_{2}) \left(
\begin{matrix}
0  && -Y^{}_{3} \\ \dfrac{\sqrt{3}}{2}Y^{}_{4} &&  \dfrac{1}{2}Y^{}_{5}   \\
\dfrac{\sqrt{3}}{2}Y^{}_{5}  && \dfrac{1}{2}Y^{}_{4}\end{matrix}\right) \nonumber \\
&&  + \widetilde{g}^{\prime}_{} \left(
\begin{matrix}
0  && -Y^{}_{2}(Y^{2}_5-Y^{2}_4) \\ -\dfrac{\sqrt{3}}{2}Y^{}_3(Y^{}_2Y^{}_5-\sqrt{3}Y^{}_1 Y^{}_4) &&  \dfrac{1}{2}Y^{}_3(Y^{}_2Y^{}_4-\sqrt{3}Y^{}_1Y^{}_5)  \\
\dfrac{\sqrt{3}}{2}Y^{}_3(Y^{}_2Y^{}_4-\sqrt{3}Y^{}_1Y^{}_5)  && -\dfrac{1}{2}Y^{}_3(Y^{}_2Y^{}_5-\sqrt{3}Y^{}_1Y^{}_4) \end{matrix}\right)
\left.\rule{0cm}{1.4cm}\right]^* \; ,
\label{eq:MDAmod}
\end{eqnarray}
where $\widetilde{g}^{\prime}_{} = g^{\prime}_{}{\rm e}^{{\rm i}\phi^{\prime}_{g^{}_{}}} \equiv g^{}_3/g^{}_1$ has been defined in the same way as for $\widetilde{g}$. As one may expect, the inclusion of $Y^{(6)}_{{\bf 3^{\prime}_{}},2}$ brings two extra degrees of freedom into the model, leading to a better fit in the modified model to the experimental data.
\begin{figure}[t!]
	\centering		\includegraphics[width=1.0\textwidth]{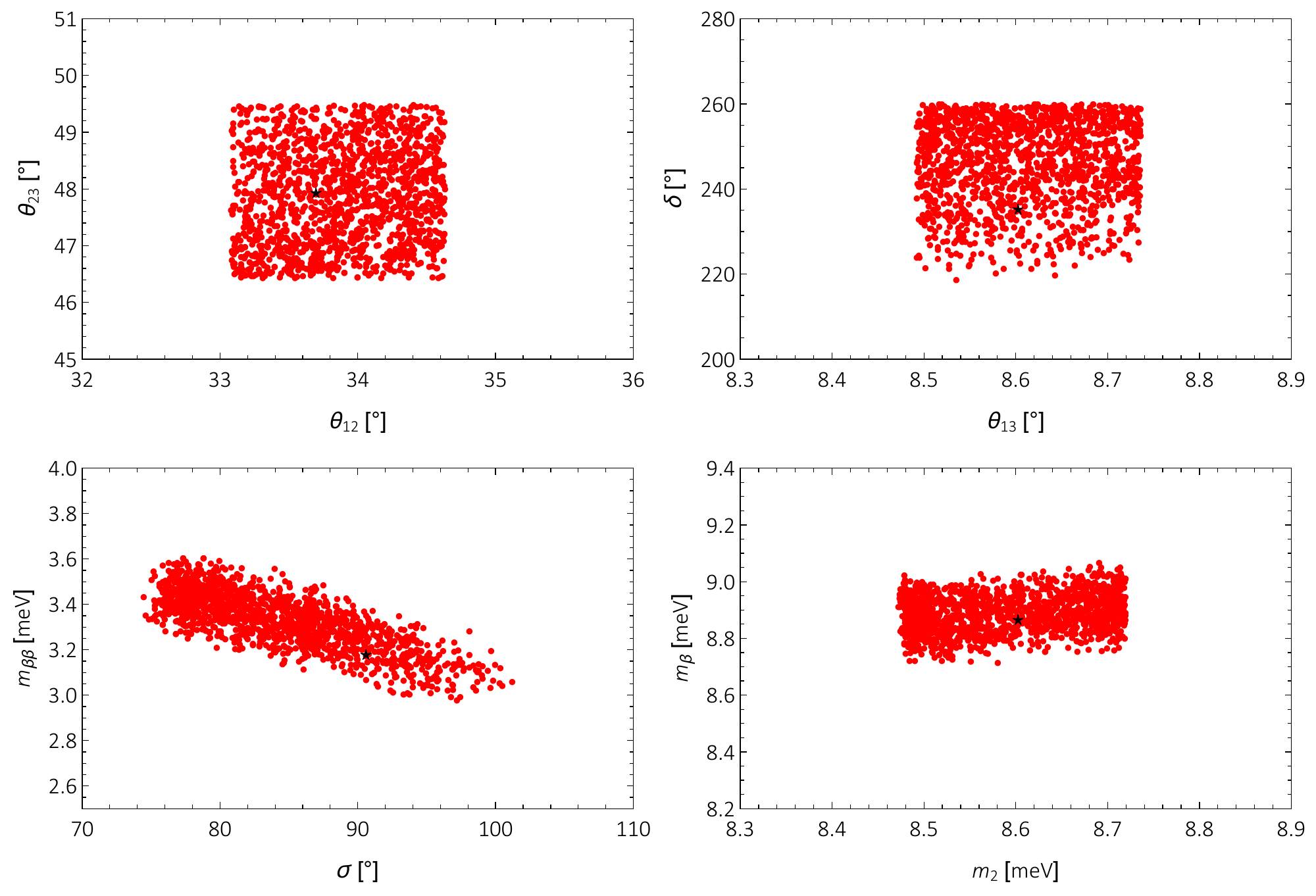}
	\vspace{-0.5cm}
	\caption{The constrained ranges of low-energy observables, including three neutrino mixing angles $\{\theta^{}_{12}, \theta^{}_{13}, \theta^{}_{23}\}$, two CP-violating phases $\{\delta, \sigma\}$, the effective neutrino masses $\{m^{}_\beta, m^{}_{\beta\beta}\}$ and the absolute neutrino mass $m^{}_2$, in the modified version of {\bf Model A} in the NO case, where the $1\sigma$ ranges of neutrino mixing parameters and mass-squared differences from the global-fit analysis of neutrino oscillation data have been input~\cite{Esteban:2018azc}. The best-fit values from our $\chi^2$-fit analysis are indicated by the black stars.}
	\label{fig:CaseAmod2} 
\end{figure}

Following the same strategy as before, we find that the modified model is consistent with the global-fit results even at the $1\sigma$ level in the NO case. The allowed parameter space of $\{{\rm Re}\,\tau, {\rm Im}\,\tau\}$, $\{g, \phi^{}_g\}$ and $\{g^\prime_{}, \phi^{\prime}_g\}$ at the $1\sigma$ level is presented in Fig.~\ref{fig:CaseAMod1}, where we can see that the value of $|{\rm Re}\,\tau|$ is very close to 0.5. The minimum of the $\chi^2_{}$-function is $\chi^2_{\rm min} = 0.335$, corresponding to the following best-fit values of the model parameters
\begin{eqnarray}
{\rm Re}\,\tau = -0.485\;, \quad {\rm Im}\,\tau = 1.22 \;, \quad g = 0.861 \;, \quad \phi^{}_{g} = 293^{\circ}_{} \;, \quad g^{\prime}_{} = 0.411 \;, \quad \phi^{\prime}_{g} =349^{\circ}_{} \;, \quad
\label{eq:bfAN1mod}
\end{eqnarray}
which together with the charged-lepton masses $m^{}_\alpha$ (for $\alpha = e, \mu, \tau$) lead to $v^{}_{\rm d} \alpha^{}_3 /\sqrt{2} = 2.10~{\rm GeV}$, $\alpha^{}_1/\alpha^{}_3 = 2.68\times 10^{-4}$ and $\alpha^{}_2/\alpha^{}_3 = 3.83\times 10^{-2}$. Meanwhile, the absolute scale of neutrino masses is given by $v^2_{\rm u}g^2_1/(2\Lambda) = 35.6~{\rm meV}$. The constrained ranges of low-energy observables are shown in Fig.~\ref{fig:CaseAmod2}. Given the best-fit values of free model parameters in Eq.~(\ref{eq:bfAN1mod}), we can find the corresponding values of three neutrino masses $m^{}_1 = 0$, $m^{}_2 = 8.60~{\rm meV}$ and $m^{}_3 = 50.1~{\rm meV}$, three mixing angles $\theta^{}_{12} = 33.7^{\circ}$, $\theta^{}_{13} = 8.60^{\circ}$ and $\theta^{}_{23} = 47.9^{\circ}$, two CP-violating phases $\delta = 235^{\circ}$ and $\sigma = 90.6^{\circ}$. Additionally, we have the effective neutrino mass $m^{}_{\beta}= 8.86~{\rm meV}$ for beta decays and $m^{}_{\beta\beta} = 3.17~{\rm meV}$ for neutrinoless double-beta decays.

\subsection{Model B}

In {\bf Model B}, the charge assignments of the superfields and the couplings under the ${\rm SU}(2)^{}_{\rm L}$ gauge symmetry and the modular $S^{}_{4}$ symmetry have been given in Table~\ref{Table:CaseB}, where the corresponding modular weights are listed in the last row. As one can see from Table~\ref{Table:CaseB}, the charge assignments are quite similar to those in {\bf Model A}. In particular, all the modular weights are exactly the same. However, these two models differ significantly in the charged-lepton sector. Now that both $\widehat{E}^{\rm C}_{2}$ and $\widehat{E}^{\rm C}_{3}$ are assigned into ${\bf 1}^\prime$ under the $S^{}_4$ symmetry group, the representations of $f^{}_{\mu}(\tau)$ and $f^{}_{\tau}(\tau)$ are accordingly changed to be ${\bf 3}^{\prime}_{}$. Notice that $f^{}_{\tau}(\tau)$ is assigned as the ${\bf 3}^{\prime}_{}$ representation with a weight of 6, so again we are facing with two choices $Y^{(6)}_{{\bf 3}^{\prime}_{},1}$ and $Y^{(6)}_{{\bf 3}^{\prime}_{},2}$ for $f^{}_\tau(\tau)$. Notice also that $Y^{(6)}_{{\bf 3}^{\prime}_{},1} = (Y^{2}_{1}+Y^{2}_{2}) (Y^{}_{3},Y^{}_{4},Y^{}_{5})^{\rm T}_{}$ is linearly proportional to $Y^{(2)}_{{\bf 3}^{\prime}_{}} = (Y^{}_{3},Y^{}_{4},Y^{}_{5})^{\rm T}_{}$. If both $Y^{(2)}_{{\bf 3}^{\prime}_{}}$ for $f^{}_e(\tau)$ and $Y^{(6)}_{{\bf 3}^{\prime}_{},1}$ for $f^{}_\tau(\tau)$ were adopted, the charged-lepton mass matrix $M^{}_l$ would have two different rows that are proportional to each other, reducing the rank of $M^{}_l$ to two and thus leading to $m^{}_e = 0$. Since this is obviously in contradiction with observations, $Y^{(6)}_{{\bf 3}^{\prime}_{},2}$ serves as the unique choice for $f^{}_{\tau}(\tau)$.
\begin{table}[t!]
	\centering
	\caption{The charge assignments of the chiral superfields and the relevant couplings under the ${\rm SU}(2)^{}_{\rm L}$ gauge symmetry and the modular $S^{}_{4}$ symmetry in {\bf Model B}, with the corresponding modular weights listed in the last row.}
	\vspace{0.5cm}
	\begin{tabular}{ccccccccc}
		\toprule
		& $\widehat{L}$ & $\widehat{E}^{\rm C}_{1}, \widehat{E}^{\rm C}_{2}, \widehat{E}^{\rm C}_{3}$ & $\widehat{N}^{\rm C}_{}$ & $\widehat{H}^{}_{\rm u}$ & $\widehat{H}^{}_{\rm d}$ & $f^{}_{e}(\tau), f^{}_{\mu}(\tau), f^{}_{\tau}(\tau)$ & $f^{}_{\rm D}(\tau), f^\prime_{\rm D}(\tau)$ & $f^{}_{\rm R} (\tau)$\\
		\midrule
		${\rm SU}(2)^{}_{\rm L}$ & 2 & 1 & 1 & 2 & 2 & 1 & 1 & 1 \\
		$S^{}_{4}$ & \bf{3} & $\bf{1}^{\prime}_{},\bf{1}^{\prime}_{},\bf{1}^{\prime}_{}$ & \bf{2} & \bf{1} & \bf{1} & $\bf{3^{\prime}_{}},\bf{3^{\prime}_{}},\bf{3^{\prime}_{}}$ & $\bf{3},\bf{3}^{\prime}_{}$ & \bf{2} \\
		$-k^{}_{I}$ & $-5$ & $3,1,-1$ & $-1$ & 0 & 0 & $k^{}_{e,\mu,\tau}=2,4,6$ & $k^{}_{\rm D}=6$ & $k^{}_{\rm R}=2$ \\
		\bottomrule
	\end{tabular}
	\label{Table:CaseB}
\end{table}

In a similar way as for {\bf Model A}, we can write down the modular invariant superpotential in the lepton sector and derive the charged-lepton mass matrix
\begin{eqnarray}
M^{}_l &=& \frac{v^{}_{\rm d}}{\sqrt{2}} \left( \begin{matrix}
\alpha^{}_{1} Y^{}_{3} &&&  2\alpha^{}_{2} Y^{}_{1}Y^{}_{3}   &&&   \alpha^{}_{3} Y^{}_{2}(Y^{2}_{5}-Y^{2}_{4})   \\ \alpha^{}_{1} Y^{}_{5}
&&& \alpha^{}_{2} (\sqrt{3}Y^{}_{2}Y^{}_{4}-Y^{}_{1}Y^{}_{5})   &&&  \alpha^{}_{3} Y^{}_{3}(Y^{}_{2}Y^{}_{4}-\sqrt{3}Y^{}_{1}Y^{}_{5})  \\ \alpha^{}_{1} Y^{}_{4}
&&& \alpha^{}_{2} (\sqrt{3}Y^{}_{2}Y^{}_{5}-Y^{}_{1}Y^{}_{4})   &&& -\alpha^{}_{3} Y^{}_{3}(Y^{}_{2}Y^{}_{5}-\sqrt{3}Y^{}_{1}Y^{}_{4})
\end{matrix} \right)^* \; ,
\label{eq:MeB}
\end{eqnarray}
and the Dirac neutrino mass matrix
\begin{eqnarray}
M^{}_{\rm D} =
&& \frac{v^{}_{\rm u} g^{}_{1}}{\sqrt{2}} \left[\rule{0cm}{1.4cm}\right. \left(
\begin{matrix}
Y^{}_{1}(Y^{2}_{4}-Y^{2}_{5}) &&& 0 \\ \dfrac{1}{2}Y^{}_{3}(Y^{}_{1}Y^{}_{4}+\sqrt{3}Y^{}_{2}Y^{}_{5})  &&&  \dfrac{\sqrt{3}}{2}Y^{}_{3}(Y^{}_{1}Y^{}_{5}+\sqrt{3}Y^{}_{2}Y^{}_{4}) \\ -\dfrac{1}{2}Y^{}_{3}(Y^{}_{1}Y^{}_{5}+\sqrt{3}Y^{}_{2}Y^{}_{4})
&&& -\dfrac{\sqrt{3}}{2}Y^{}_{3}(Y^{}_{1}Y^{}_{4}+\sqrt{3}Y^{}_{2}Y^{}_{5})
\end{matrix}\right) \nonumber \\
&& + \widetilde{g} (Y^{2}_{1}+Y^{2}_{2}) \left(
\begin{matrix}
0  &&& -Y^{}_{3} \\ \dfrac{\sqrt{3}}{2}Y^{}_{4} &&&  \dfrac{1}{2}Y^{}_{5}   \\
\dfrac{\sqrt{3}}{2}Y^{}_{5}  &&& \dfrac{1}{2}Y^{}_{4}\end{matrix}\right) \left.\rule{0cm}{1.4cm}\right]^* \; ,
\label{eq:MDB}
\end{eqnarray}
where the real coefficient $g^{}_1$ has been factorized out from the square brackets on the right-hand side of Eq.~(\ref{eq:MDB}) and the complex parameter $\widetilde{g} \equiv g^{}_2/g^{}_1 = g e^{{\rm i}\phi^{}_g}$ has been introduced as before. In addition, the Majorana mass matrix of right-handed neutrinos is given by
\begin{eqnarray}
M^{}_{\rm R}&=&\Lambda\left(\begin{matrix}
-Y^{}_{1} &&& Y^{}_{2} \\
Y^{}_{2} &&& Y^{}_{1}
\end{matrix}\right)^* \; .
\label{eq:MRB}
\end{eqnarray}

\begin{figure}[t!]
\centering \includegraphics[width=1\textwidth]{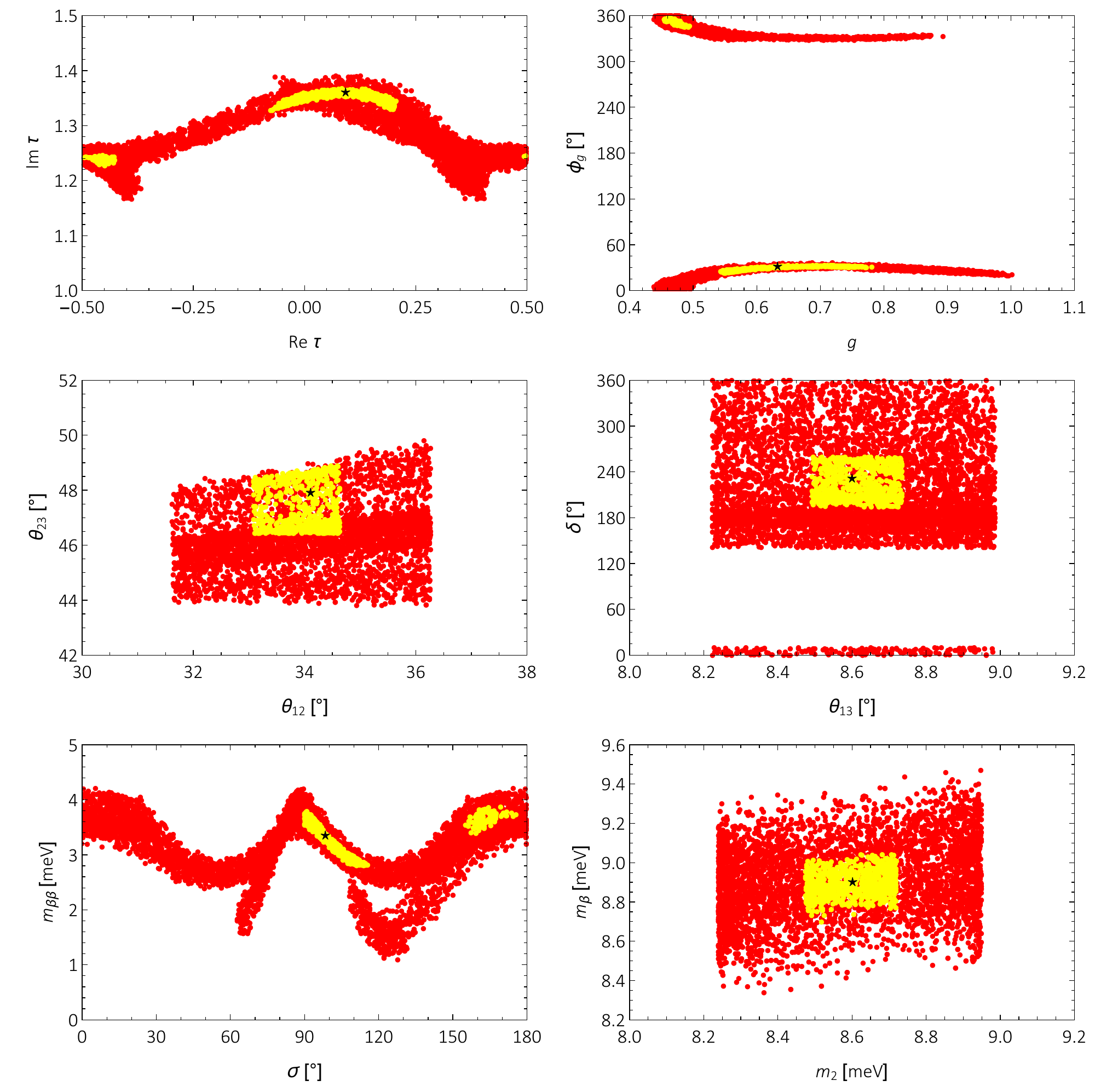}
\vspace{-0.5cm}
\caption{Allowed ranges of the model parameters $\{{\rm Re}\,\tau, {\rm Im}\,\tau\}$ and $\{g, \phi^{}_g\}$ and the constrained ranges of low-energy observables in the \underline{NO} case in {\bf Model B}, where the $1\sigma$ (yellow dots) and $3\sigma$ ranges (red dots) of neutrino mixing parameters and mass-squared differences from the global-fit analysis of neutrino oscillation data have been input~\cite{Esteban:2018azc}. The best-fit values from our $\chi^2$-fit analysis are indicated by the black stars.}
\label{fig:CaseB_NO} 
\end{figure}

\begin{figure}[t!]
\centering \includegraphics[width=1\textwidth]{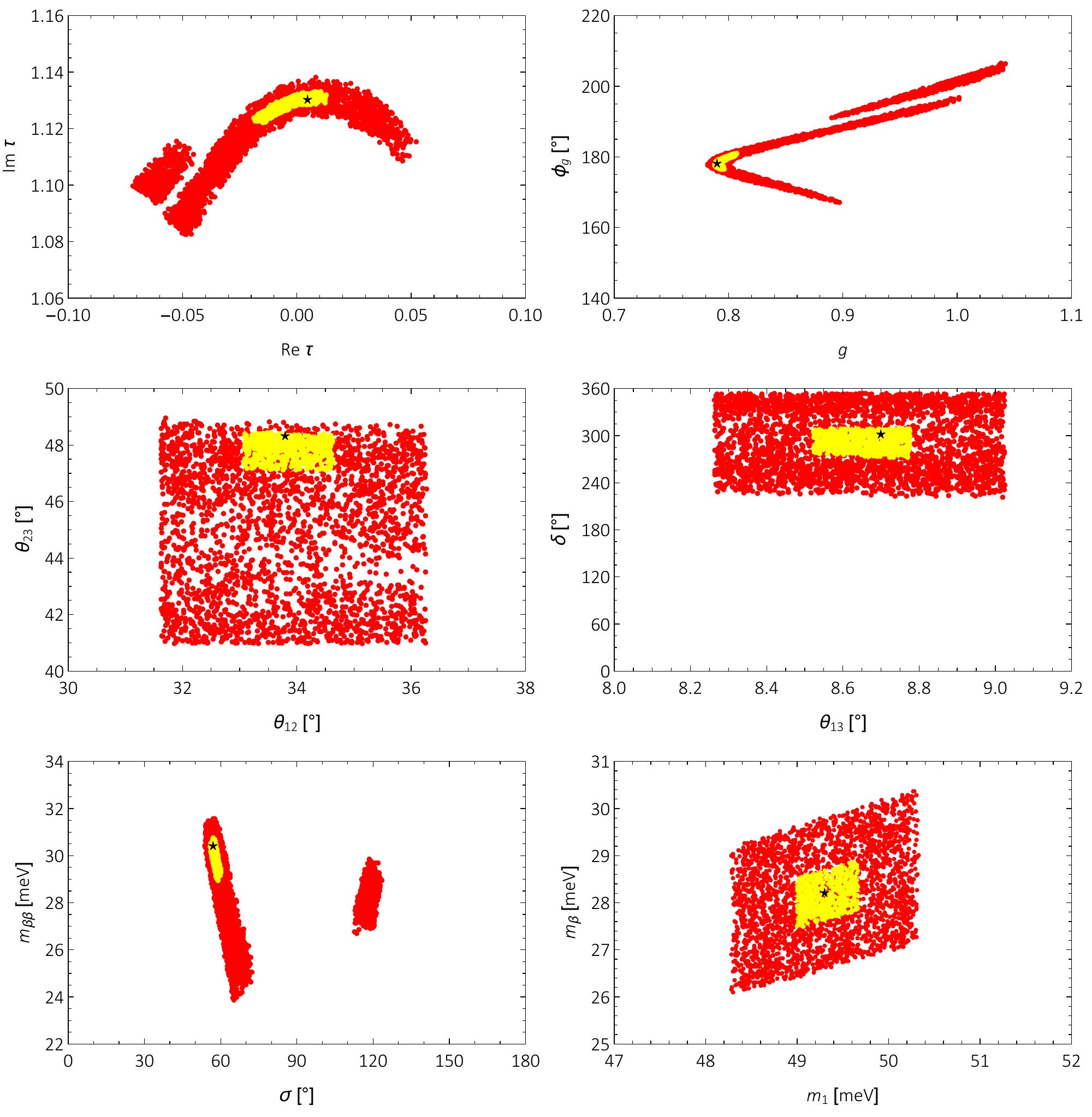}
\vspace{-0.8cm}
\caption{Allowed ranges of the model parameters $\{{\rm Re}\,\tau, {\rm Im}\,\tau\}$ and $\{g, \phi^{}_g\}$ and the constrained ranges of low-energy observables in the \underline{IO} case in {\bf Model B}, where the $1\sigma$ (yellow dots) and $3\sigma$ ranges (red dots) of neutrino mixing parameters and mass-squared differences from the global-fit analysis of neutrino oscillation data have been input~\cite{Esteban:2018azc}. The best-fit values from our $\chi^2$-fit analysis are indicated by the black stars.}
\label{fig:CaseB_IO} 
\end{figure}

The strategy to explore the allowed parameter space and the low-energy phenomenology in {\bf Model B} follows that in {\bf Model A}, so we shall not repeat it here. It turns out that both NO and IO cases are allowed in {\bf Model B} even at the $1\sigma$ level. The allowed ranges of the model parameters in the NO and IO cases are shown in Fig.~\ref{fig:CaseB_NO} and Fig.~\ref{fig:CaseB_IO}, respectively. Some helpful comments on the numerical results are in order.

\begin{itemize}
\item The numerical results in the NO case are shown in Fig.~\ref{fig:CaseB_NO}. As one can observe from two plots in the first row, the whole range $[-0.5, 0.5]$ of ${\rm Re}\,\tau$ is allowed at the $3\sigma$ level while the phase $\phi^{}_g$ is restricted into the narrow region  $[0^\circ, 40^\circ]$ or $[320^\circ, 360^\circ]$. However, at the $1\sigma$ level, two disconnected regions $-0.42<{\rm Re}\,\tau<-0.08$ and $0.20<{\rm Re}\,\tau<0.49$ are forbidden because either the mixing angle $\theta^{}_{23}$ or the Dirac CP-violating phase $\delta$ can not reach their respective $1\sigma$ ranges in these regions. We find that the minimum $\chi^{2}_{\rm min} = 0.30$ is obtained in the NO case with the following best-fit values of the model parameters
\begin{eqnarray}
{\rm Re}\,\tau = 0.0934\;, \quad {\rm Im}\,\tau = 1.36 \;, \quad g = 0.633 \;, \quad \phi^{}_{g} = 31.0^{\circ}_{} \;,
\label{eq:bfBN1}
\end{eqnarray}
which together with the charged-lepton masses $m^{}_\alpha$ (for $\alpha = e, \mu, \tau$) give rise to $v^{}_{\rm d} \alpha^{}_3 /\sqrt{2} = 2.06~{\rm GeV}$, $\alpha^{}_1/\alpha^{}_3 = 1.00\times 10^{-3}$ and $\alpha^{}_2/\alpha^{}_3 = 2.68\times 10^{-2}$. In addition, the absolute scale of neutrino masses is given by $v^2_{\rm u}g^2_1/(2\Lambda) = 75.3~{\rm meV}$. With these best-fit values of model parameters, we get the neutrino mass spectrum $m^{}_1 = 0$, $m^{}_2 = 8.60~{\rm meV}$ and $m^{}_3 = 50.2~{\rm meV}$, three mixing angles $\theta^{}_{12} = 34.1^{\circ}$, $\theta^{}_{13} = 8.60^{\circ}$ and $\theta^{}_{23} = 47.9^{\circ}$, and two CP-violating phases $\delta = 231^{\circ}$ and $\sigma = 98.6^{\circ}$. The effective neutrino masses $m^{}_{\beta} = 8.90~{\rm meV}$ and $m^{}_{\beta \beta} = 3.34~{\rm meV}$ are quite small, as usually expected in the NO case.

\item The numerical results in the IO case are shown in Fig. \ref{fig:CaseB_IO}. Unlike the NO case, the allowed range of ${\rm Re}\,\tau$ is highly constrained and only the narrow region $-0.02 < {\rm Re}\,\tau < 0.01$ survives at the $1\sigma$ level. As one can observe from the top-left panel, the modulus parameter $\tau$ is very close to a pure imaginary number, i.e., $\tau \approx {\rm i}$. Moreover, the phase $\phi^{}_g$ is restricted into the region around $180^\circ$ at the $1\sigma$ level. These two parameters essentially lead to very constrained regions of the CP-violating phases, namely, $\delta \approx 300^\circ$ (or $60^\circ$) and $\sigma \approx 60^\circ$ (or $120^\circ$). Performing the $\chi^2$-fit analysis in the IO case, we find that the minimum $\chi^{2}_{\rm min}=2.22$ is achieved at
    \begin{eqnarray}
    {\rm Re}\,\tau = 4.85\times 10^{-3} \;, \quad {\rm Im}\,\tau = 1.13 \;, \quad g = 0.790 \;, \quad \phi^{}_{g} = 178^{\circ}_{} \; ,
   \label{eq:bfBI1}
   \end{eqnarray}
which together with the charged-lepton masses $m^{}_\alpha$ (for $\alpha = e, \mu, \tau$) determine the other parameters $v^{}_{\rm d} \alpha^{}_3 / \sqrt{2} = 0.121~{\rm GeV}$, $\alpha^{}_1/\alpha^{}_3 = 13.4$ and $\alpha^{}_2/\alpha^{}_3 = 2.38\times 10^{-3}$. The absolute neutrino mass scale is given by $v^2_{\rm u}g^{2}_{1}/(2\Lambda) = 60.8\;{\rm meV}$, and the neutrino mass spectrum is found to be $m^{}_3 = 0$, $m^{}_1 = 49.3~{\rm meV}$ and $m^{}_{2} = 50.1~{\rm meV}$. In addition, three mixing angles are $\theta^{}_{12} = 33.9^{\circ}$, $\theta^{}_{13} = 8.70^{\circ}$ and $\theta^{}_{23} = 48.3^{\circ}$, while two CP-violating phases are $\delta = 301^{\circ}$ and $\sigma = 57.0^{\circ}$. The effective neutrino masses turn out to be $m^{}_{\beta} = 28.2~{\rm meV}$ and $m^{}_{\beta \beta} = 30.4~{\rm meV}$, which will hopefully be tested in the next-generation beta decay and neutrinoless double-beta decay experiments.

\item In Appendix~\ref{sec:appA}, Eq.~(\ref{eq:Y3q}) reveals that the modular forms $Y^{}_{i}(\tau)$ (for $i=1,2,\dots,5$) depend on $\tau$ via $q={\rm e}^{2\pi{\rm i}\tau}$. Since all the expansion coefficients are real, the complex phases of $Y^{}_{i}(\tau)$ will be determined only by ${\rm Re}\,\tau$. Therefore there are totally two sources of CP violation in the lepton sector, namely, ${\rm Re}\,\tau$ and $\phi^{}_{g}$. From Eq.~(\ref{eq:bfBI1}) we can see the best-fit value of ${\rm Re}\,\tau$ is quite small in the IO case while $\phi^{}_{g}$ is very close to $180^{\circ}_{}$. If ${\rm Re}\,\tau=0$ and $\phi^{}_g=180^{\circ}_{}$ exactly hold, {\bf Model B} should posses the so-called generalized CP symmetry~\cite{Novichkov:2019sqv}, which gives trivially $\delta=0$ or $180^\circ_{}$. Then one may wonder how they can lead to relatively large values of two CP-violating phases $\delta$ and $\sigma$, as shown in Fig.~\ref{fig:CaseB_IO}. In fact, the contributions to $\delta$ and $\sigma$ from $M^{}_{l}$ and $M^{}_{\rm R}$ are negligible since the imaginary parts of $M^{}_{l}$ and $M^{}_{\rm R}$ are suppressed by ${\rm Re}\,\tau$, as we can observe from Eqs.~(\ref{eq:MeB}) and (\ref{eq:MRB}). For this reason, CP violation should be governed by nontrivial complex phases of $M^{}_{\rm D}$. Although both ${\rm Re}\,\tau$ and ${\rm Im}\,\widetilde{g} = g \sin \phi^{}_{g}$ are small, $\phi^{}_{g} \approx 180^{\circ}_{}$ implies that those two matrices in the square brackets on the right-hand side of Eq.~(\ref{eq:MDB}) have the opposite signs, resulting in a comparable real and imaginary part of $M^{}_{\rm D}$. This is why relatively large values of CP-violating phases $\delta$ and $\sigma$ come out eventually, which is also consistent with the conclusion drawn in Ref.~\cite{Novichkov:2019sqv}.
\end{itemize}

In summary, {\bf Model B} is perfectly consistent with current neutrino oscillation data in both NO and IO cases. A generic feature of this model is that the best-fit value of ${\rm Re}\,\tau$ is very small (i.e., ${\rm Re}\,\tau = 0.0934$ in the NO case and ${\rm Re}\,\tau = 4.85\times 10^{-3}$ in the IO case). Moreover, one can observe that $\phi^{}_g \approx 180^\circ$ holds in the IO case, indicating that those two matrices in the square brackets on the right-hand side of Eq.~(\ref{eq:MDB}) contribute destructively to the Dirac neutrino mass matrix $M^{}_{\rm D}$ but lead to significant CP-violating phases. Since the complex phases in the lepton mass matrices are completely controlled by two complex parameters $\tau$ and $\tilde{g}$, we may be able to establish a direct connection between the CP violation at low energies and that at the high-energy seesaw scale, as will be shown in the next section.

\subsection{Renormalization-group Running Effects}

In this subsection, let us study radiative corrections to the flavor mixing parameters in our models via the renormalization-group (RG) equations. Such a study is mainly motivated by the fact that both the modular symmetry and the seesaw mechanism may work at a very high energy scale $\Lambda^{}_{\rm SS}$, whereas the flavor mixing parameters are measured at the electroweak scale characterized by the mass of $Z$ gauge boson $m^{}_Z \approx 91.2~{\rm GeV}$. As is well known, the RG running effects could be significant, particularly for the nearly-degenerate neutrino mass spectrum and a large value of $\tan \beta$~\cite{Ohlsson:2013xva}. Actually the RG running effects have already been considered in the models with a modular symmetry in Ref.~\cite{Criado:2018thu}, while those in the MSM without modular symmetries can be found in Ref.~\cite{Mei:2003gn}. To be specific, we carry out a quantitative analysis of running effects on the flavor mixing parameters for {\bf Model B} in both NO and IO cases, and explain how large the RG running effects can be in this model for $\tan\beta = 10$. For illustration, the value $\tan\beta = 10$ has simply been chosen in accordance with our discussions about the matter-antimatter asymmetry in the next section.

In the flavor basis where the charged-lepton Yukawa coupling matrix $\widetilde{Y}^{}_l \equiv {\rm Diag}\{y^{}_e, y^{}_\mu, y^{}_\tau\}$ with $y^{}_\alpha = \sqrt{2} m^{}_\alpha/v^{}_{\rm d}$ (for $\alpha = e, \mu, \tau$) is diagonal, the Dirac neutrino Yukawa coupling matrix turns out to be $\widetilde{Y}^{}_\nu = U^\dagger_l Y^{}_\nu$, where the diagonalization of the original charged-lepton Yukawa coupling matrix $Y^{}_l$ via $U^\dagger_l Y^{}_l Y^\dagger_l U^{}_l = \widetilde{Y}^2_l$ should be noticed. Below the seesaw scale $\Lambda^{}_{\rm SS}$, heavy right-handed neutrinos will be integrated out, and thus the effective neutrino mass parameter is given by ${\cal M} \equiv U^{\dag}_{l}Y^{}_{\nu}M^{-1}_{\rm R}Y^{\rm T}_{\nu}U^\ast_l$. Therefore, the radiative corrections to lepton flavor mixing parameters are governed by the one-loop RG equations of $\widetilde{Y}^{}_l$ and ${\cal M}$~\cite{Machacek:1983fi, Arason:1991ic, Castano:1993ri, Chankowski:1993tx, Babu:1993qv, Antusch:2001ck}, namely,
\begin{eqnarray}
16\pi^2_{}\frac{{\rm d} \widetilde{Y}^{}_l }{{\rm d} \mathfrak{t}} &=&  \left[\alpha^{}_{l} + 3 \left(\widetilde{Y}^{}_l \widetilde{Y}^{\dag}_{l}\right)\right]\widetilde{Y}^{}_l \; , \label{eq:MRGE1} \\
16\pi^2_{}\frac{{\rm d} {\cal M}}{{\rm d} \mathfrak{t}} &=& \alpha^{}_{\nu}{\cal M} + \left[ \left(\widetilde{Y}^{}_l \widetilde{Y}^{\dag}_{l}\right){\cal M}+{\cal M} \left(\widetilde{Y}^{}_l \widetilde{Y}^{\dag}_{l}\right)^{\rm T}_{}\right] \; ,
\label{eq:MRGE}
\end{eqnarray}
where $\mathfrak{t} \equiv \ln (\mu/\Lambda^{}_{\rm SS})$ with $\mu$ being the renormalization scale, $\alpha^{}_{l} = - 1.8 \mathfrak{g}^2_1 - 3 \mathfrak{g}^2_2 + 3 {\rm Tr}(Y^{}_{\rm d}Y^{\dag}_{\rm d}) + {\rm Tr}(\widetilde{Y}^{}_l \widetilde{Y}^{\dag}_l)$ and $\alpha^{}_{\nu}= -1.2 \mathfrak{g}^2_1 - 6 \mathfrak{g}^2_2 + 6 {\rm Tr}(Y^{}_{\rm u}Y^{\dag}_{\rm u})$. Notice that $\mathfrak{g}^{}_{2}$ and $\mathfrak{g}^{}_{1}$ represent respectively the ${\rm SU}(2)^{}_{\rm L}$ and ${\rm U}(1)^{}_{\rm Y}$ gauge couplings, $Y^{}_{\rm u}$ and $Y^{}_{\rm d}$ denote the up- and down-type quark Yukawa coupling matrices. From Eq.~(\ref{eq:MRGE1}), one can easily observe that $\widetilde{Y}^{}_l$ remains to be diagonal during the RG running. As a result, the running of ${\cal M}$ is responsible for the corrections to neutrino masses and lepton flavor mixing~\cite{Mei:2003gn}.

Now that only the diagonal elements on both sides of Eq.~(\ref{eq:MRGE1}) survive in the chosen flavor basis, the corresponding three differential equations can be solved. Then the solutions to the RG equations in Eq.~(\ref{eq:MRGE}) can be found out with the help of the following two evolution functions~\cite{Mei:2003gn}
\begin{eqnarray}
I^{}_\nu & = & {\rm exp}\left[-\frac{1}{16\pi^2_{}}\int^{\ln (\Lambda^{}_{\rm SS}/m^{}_Z)}_{0}\alpha^{}_{\nu}(\mathfrak{t}) {\rm d}\mathfrak{t}\right] \; , \nonumber \\
I^{}_\alpha & = & {\rm exp}\left[-\frac{1}{16\pi^2_{}}\int^{\ln (\Lambda^{}_{\rm SS}/m^{}_Z)}_{0} y^2_\alpha (\mathfrak{t}) {\rm d}\mathfrak{t}\right] \; ,
\label{eq:evofun}
\end{eqnarray}
with $y^{}_\alpha$ (for $\alpha = e,\mu,\tau$) being the charged-lepton Yukawa couplings. It is straightforward to derive the approximate solution to Eq.~(\ref{eq:MRGE}) as follows
\begin{eqnarray}
{\cal M}(m^{}_Z) = I^{}_\nu \left(
\begin{matrix}
I^{}_e &&& 0 &&& 0 \\ 0 &&& I^{}_\mu &&& 0 \\ 0 &&& 0 &&& I^{}_\tau
\end{matrix}
\right) {\cal M}(\Lambda^{}_{\rm SS}) \left(
\begin{matrix}
I^{}_e &&& 0 &&& 0 \\ 0 &&& I^{}_\mu &&& 0 \\ 0 &&& 0 &&& I^{}_\tau
\end{matrix}
\right) \; ,
\label{eq:solRGE}
\end{eqnarray}
where $I^{}_\nu$ affects only the absolute scale of light neutrino masses while $I^{}_\alpha$ (for $\alpha = e, \mu, \tau$) can modify both neutrino masses and flavor mixing parameters. Since the electron and muon Yukawa couplings are extremely small, $I^{}_e \approx I^{}_\mu \approx 1$ holds as an excellent approximation and $I^{}_\tau$ contributes dominantly to the running effects of ${\cal M}$. By using Eq.~(\ref{eq:solRGE}), one can easily obtain ${\cal M}(m^{}_Z)$ at the electroweak scale, given the input ${\cal M}(\Lambda^{}_{\rm SS})$ at the seesaw scale.

\begin{figure}[t!]
	\centering
	\begin{overpic}[scale=0.85]{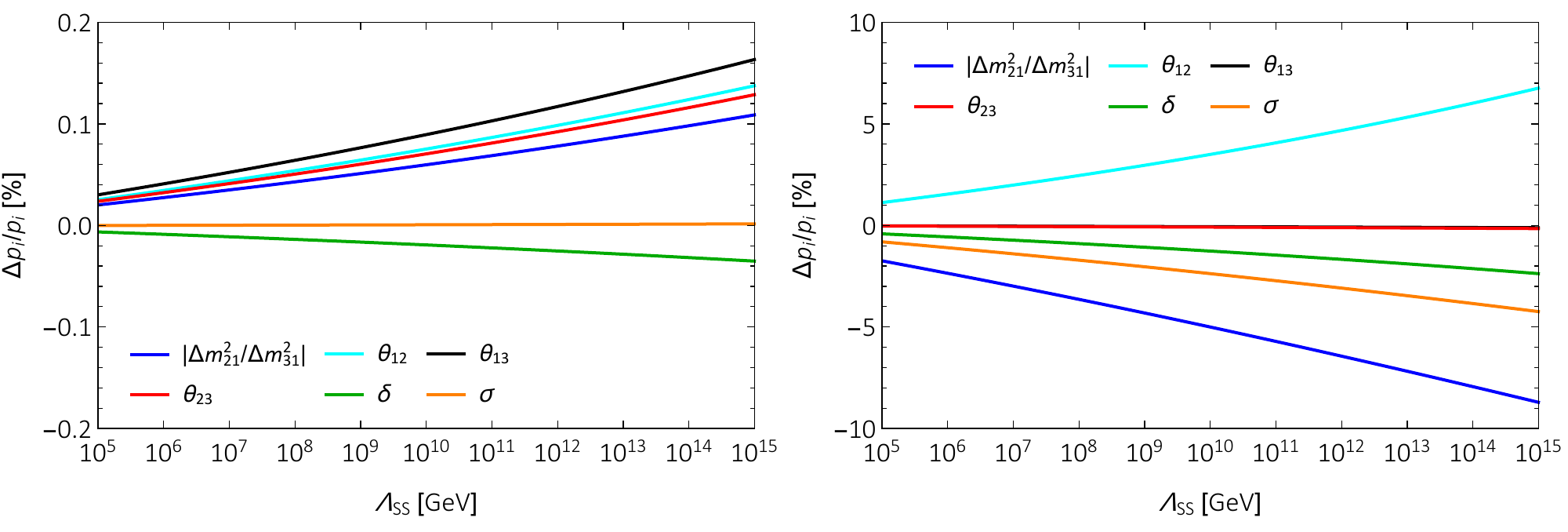}	
		\put(9,29){\underline{NO}}	
		\put(93,29){\underline{IO}}	
	\end{overpic}
	\caption{Illustration for the relative changes $\Delta p^{}_i/p^{}_i \equiv [p^{}_i(m^{}_Z)-p^{}_{i}(\Lambda^{}_{\rm SS})]/p^{}_i(\Lambda^{}_{\rm SS})$ of lepton flavor mixing parameters $p^{}_i=\{|\Delta m^2_{21}/\Delta m^2_{31}|, \theta^{}_{12}, \theta^{}_{13}, \theta^{}_{23}, \delta, \sigma\}$, running from the high-energy scale $\Lambda^{}_{\rm SS}$ to the electroweak scale $m^{}_Z$ in {\bf Model B}. The results in the \underline{NO} and \underline{IO} case are shown in the left and right panel, respectively, where the high-energy scale $\Lambda^{}_{\rm SS}$ varies from $10^5_{}~{\rm GeV}$ to $10^{15}_{}~{\rm GeV}$ and $\tan\beta=10$ is fixed.}
	\label{fig:RGE}
\end{figure}

In order to illustrate how the radiative corrections affect the flavor mixing parameters, we have numerically solved the RG equations in {\bf Model B}. The basic strategy is to take the best-fit values of model parameters obtained in the previous subsection as the initial conditions at the seesaw scale $\Lambda^{}_{\rm SS}$, and then calculate the corresponding values of low-energy observables at $m^{}_Z$ via the RG running. The final numerical results are shown in Fig.~\ref{fig:RGE}, where we have introduced the relative changes $\Delta p^{}_i/p^{}_i \equiv [p^{}_i(m^{}_Z)-p^{}_{i}(\Lambda^{}_{\rm SS})]/p^{}_i(\Lambda^{}_{\rm SS})$ of the flavor mixing parameters $p^{}_i=\{|\Delta m^2_{21}/\Delta m^2_{31}|, \theta^{}_{12}, \theta^{}_{13}, \theta^{}_{23}, \delta, \sigma\}$ with $\Lambda^{}_{\rm SS}$ varying from $10^5_{}~{\rm GeV}$ to $10^{15}_{}~{\rm GeV}$ and $\tan\beta=10$. From the left panel of Fig.~\ref{fig:RGE} for the NO case, one can see that the RG running effects of all the flavor mixing parameters are negligible. Even for the extremely-high seesaw scale $\Lambda^{}_{\rm SS} = 10^{15}~{\rm GeV}$, the relative changes of all parameters at the electroweak scale are no more than $0.2\%$. In the IO case, relatively large corrections to the flavor mixing parameters are obtained, especially for $|\Delta m^2_{21}/\Delta m^2_{31}|$ and $\theta^{}_{12}$, as shown in the right panel of Fig.~\ref{fig:RGE}. For instance, the relative changes for $|\Delta m^2_{21}/\Delta m^2_{31}|$ and $\theta^{}_{12}$ can be larger than $5\%$ when $\Lambda^{}_{\rm SS}$ is at $10^{15}_{}~{\rm GeV}$. However, for a relatively low seesaw scale (e.g., $\Lambda^{}_{\rm SS} \sim 10^{7}~{\rm GeV}$), the running effects are also small in the IO case. For comparison, we list in Table~\ref{table:RGE} the best-fit values of lepton flavor mixing parameters at the seesaw scale $\Lambda^{}_{\rm SS} = 10^{7}~{\rm GeV}$,
together with the corresponding values at the electroweak scale $m^{}_Z = 91.2~{\rm GeV}$, in both NO and IO cases for {\bf Model B} with $\tan\beta = 10$. One can immediately recognize that the RG running effects in the IO case are less than $3\%$. It is worthwhile to mention that we have not attempted to thoroughly investigate the RG running effects by varying the $\tan\beta$ value. A systematic study of the RG running together with the supersymmetric threshold effects is interesting~\cite{Criado:2018thu} and will be left for future works.

\begin{table}[t]
	\begin{center}
		\vspace{-0.25cm} \caption{Comparison between the best-fit values of $\{|\Delta m^2_{21}/\Delta m^2_{31}|, \theta^{}_{12}, \theta^{}_{13}, \theta^{}_{23}, \delta, \sigma\}$ at the seesaw scale $\Lambda^{}_{\rm SS} = 10^7_{}~{\rm GeV}$ and their corresponding values at the electroweak scale $m^{}_Z = 91.2~{\rm GeV}$ after taking into account the RG running in both the NO and IO cases for {\bf Model B}, where $\tan\beta=10$ is assumed.} \vspace{0.5cm}
		\begin{tabular}{c|c|c|c|c|c|c|c}
			\hline
			\hline
			\multicolumn{2}{c|}{} & $|\Delta m^2_{21}/\Delta m^2_{31}|$ & $\theta^{}_{12}/^\circ$ & $\theta^{}_{13}/^\circ$ & $\theta^{}_{23}/^\circ$ & $\delta/^\circ$ & $\sigma/^\circ$ \\
			\hline
			\multirow{2}*{NO}
			& $\Lambda^{}_{\rm SS}$ & 0.0293 & 34.1 & 8.60 & 47.9 & 231 & 98.6\\
			~ & $m^{}_Z$ & 0.0293 & 34.1 & 8.61 & 47.9 & 231 & 98.6 \\
			\hline
			\multirow{2}*{IO}
			& $\Lambda^{}_{\rm SS}$ & 0.0306 & 33.9 & 8.70 & 48.3 & 301 & 57.0\\
			~ & $m^{}_Z$& 0.0297 & 34.6 & 8.70 & 48.2 & 299 & 56.2\\
			\hline
			\hline
		\end{tabular}
		\label{table:RGE}
	\end{center}
\end{table}

\section{Matter-antimatter Asymmetry}\label{sec:bau}

As is well known, another salient feature of the canonical seesaw model with heavy right-handed neutrinos is offering a natural and attractive framework to explain the observed matter-antimatter asymmetry in our Universe via thermal leptogenesis~\cite{Fukugita:1986hr, Luty:1992un}. The basic idea is that the CP-violating and out-of-equilibrium decays of heavy Majorana neutrinos in the early Universe generate the lepton number asymmetry, which will be subsequently converted into the baryon number asymmetry by the baryon- and lepton-number-violating sphaleron processes~\cite{Klinkhamer:1984di, Kuzmin:1985mm, Arnold:1987mh}. In the conventional scenario of thermal leptogenesis in the MSM, a strong mass hierarchy for two heavy Majorana neutrinos $N^{}_1$ and $N^{}_2$ is assumed, i.e., $M^{}_{1} \ll M^{}_{2}$, such that the lepton number asymmetry produced in the decays of the heavier Majorana neutrino $N^{}_2$ will be washed out by the lepton-number-violating processes mediated by the lighter one $N^{}_1$. Therefore, it is the lightest heavy Majorana neutrino $N^{}_1$ that is responsible for the matter-antimatter asymmetry and a successful leptogenesis requires a very large mass of $N^{}_1$, e.g., $M^{}_1 \gtrsim (10^{11}\cdots 10^{13})~{\rm GeV}$~\cite{Buchmuller:2004nz, Davidson:2008bu}.

The viable scenario of thermal leptogenesis with such high-scale masses of heavy Majorana neutrinos needs the reheating temperature in the post-inflation era to be high enough $T^{}_{\rm rh} \gtrsim M^{}_1$ so as to thermalize the heavy Majorana neutrino $N^{}_1$. Nevertheless, such a high reheating temperature may cause the overproduction problem of gravitinos in the gauge-mediated supersymmetric models~\cite{Khlopov:1984pf, Ellis:1984eq, Asaka:2000zh}. To avoid this problem in our models with the modular $S^{}_4$ symmetry, we can lower the seesaw scale down to $\Lambda^{}_{\rm SS} \sim 10^{7}~{\rm GeV}$ by implementing the resonant leptogenesis~\cite{Pilaftsis:1997jf, Pilaftsis:2003gt,GonzalezFelipe:2003fi,Branco:2005ye,Ahn:2006rn}, for which a nearly-degenerate mass spectrum $M^{}_2 \approx M^{}_1$ for two heavy Majorana neutrinos is necessary. In the following discussions, we shall demonstrate that the resonant leptogenesis works well only in {\bf Model B} but not in {\bf Model A}.

\begin{figure}[t!]
	\centering \includegraphics[width=0.48\textwidth]{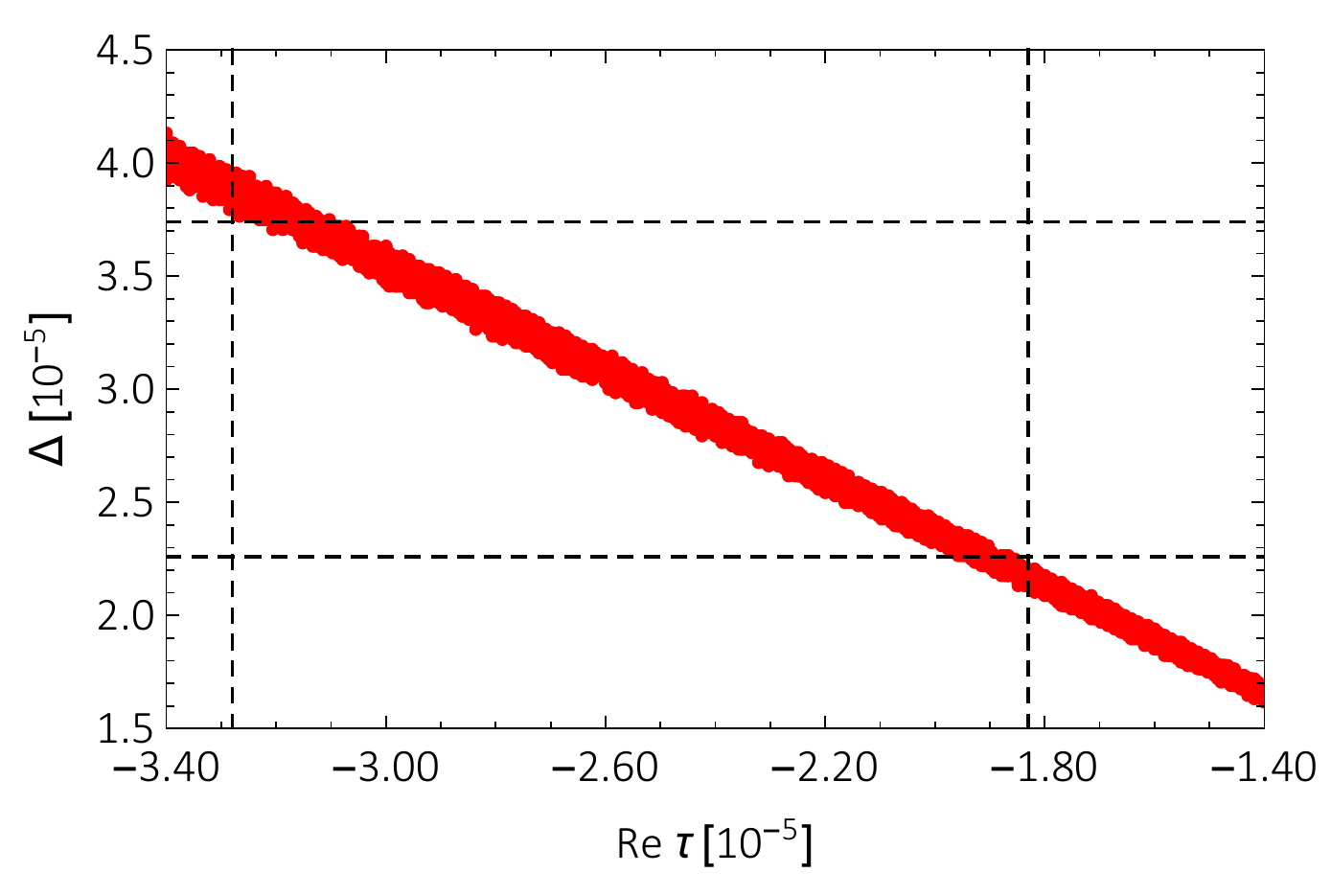}	\includegraphics[width=0.48\textwidth]{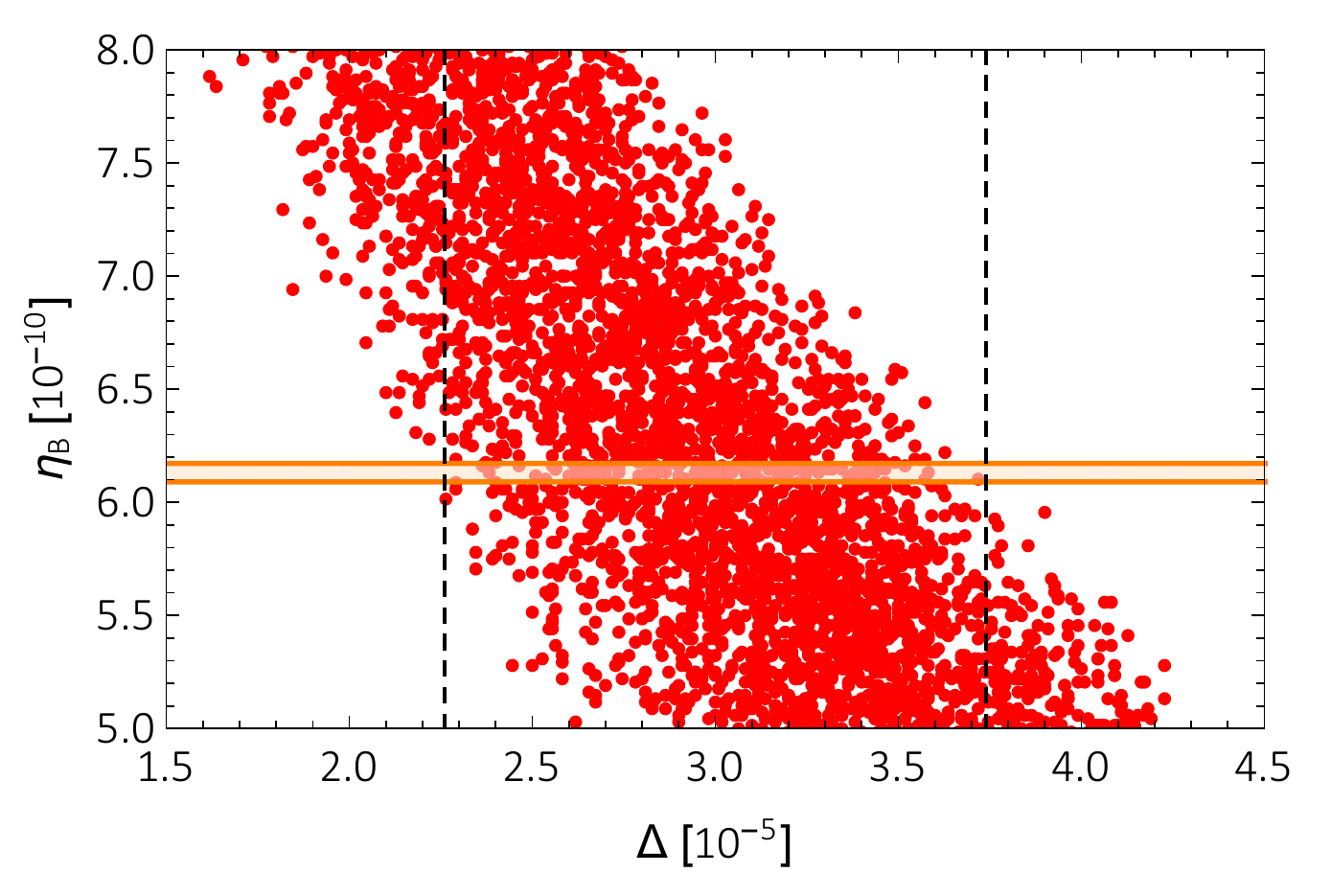}
	\vspace{-0.2cm}
	\caption{In the left panel, the correlation between the degeneracy parameter $\Delta \equiv (M^{}_2 - M^{}_1)/M^{}_1$ and the real part ${\rm Re}\,\tau$ of the modulus parameter has been shown for \underline{NO} in {\bf Model B}, while the model predictions from resonant leptogenesis for the baryon-to-photon ratio $\eta^{}_{\rm B}$ in the present Universe are given in the right panel. The shaded region in the right panel denotes the $1\sigma$ range of $\eta^{}_{\rm B} \in (6.008\cdots 6.254)\times 10^{-10}$ and two vertical dashed lines indicate the required range of $\Delta$ for successful leptogenesis, while the corresponding range of ${\rm Re}\,\tau$ is also indicated in the left panel. For illustration, $M^{}_1 = 10^{7}~{\rm GeV}$ and $\tan\beta = 10$ have been assumed in numerical calculations.}
	\label{fig:Lepgen1} 
\end{figure}

\begin{figure}[t!]
	\centering	\includegraphics[width=0.48\textwidth]{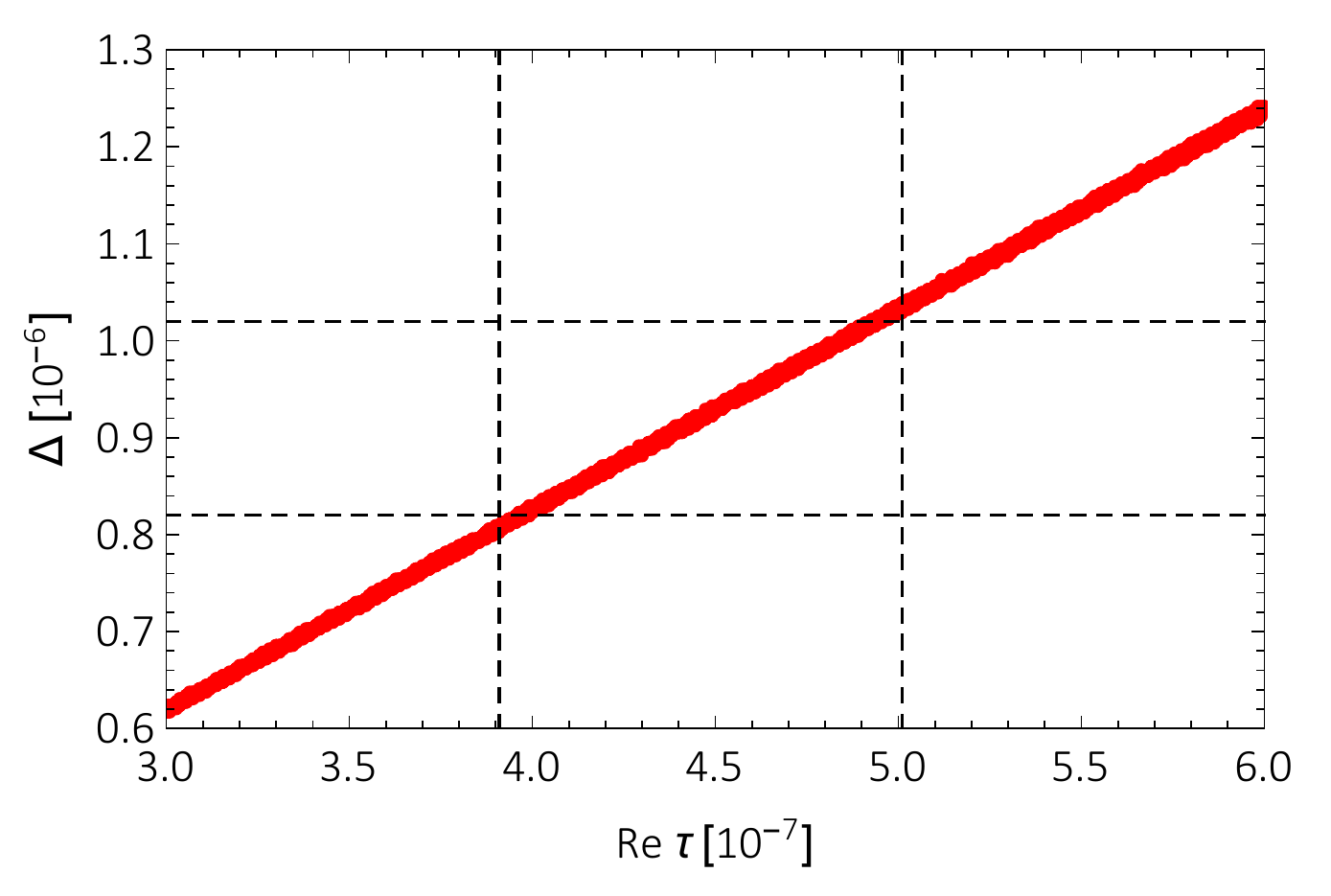}	\includegraphics[width=0.48\textwidth]{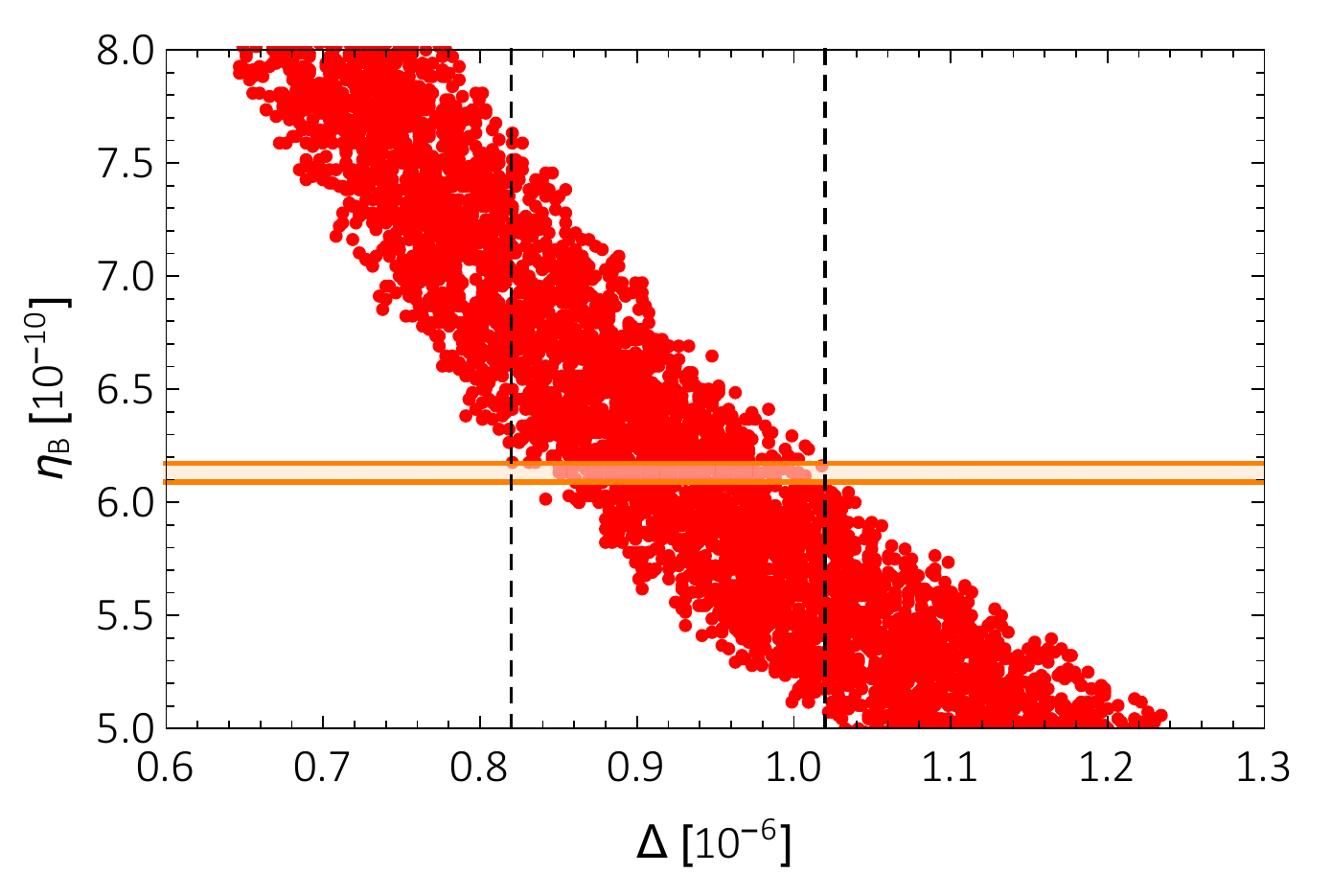}
	\vspace{-0.2cm}
	\caption{In the left panel, the correlation between the degeneracy parameter $\Delta \equiv (M^{}_2 - M^{}_1)/M^{}_1$ and the real part ${\rm Re}\,\tau$ of the modulus parameter has been shown for \underline{IO} in {\bf Model B}, while the model predictions from resonant leptogenesis for the baryon-to-photon ratio $\eta^{}_{\rm B}$ in the present Universe are given in the right panel. The shaded region in the right panel denotes the $1\sigma$ range of $\eta^{}_{\rm B} \in (6.008\cdots 6.254)\times 10^{-10}$ and two vertical dashed lines indicate the required range of $\Delta$ for successful leptogenesis, while the corresponding range of ${\rm Re}\,\tau$ is also indicated in the left panel. For illustration, $M^{}_1 = 10^{7}~{\rm GeV}$ and $\tan\beta = 10$ have been assumed in numerical calculations.}
	\label{fig:Lepgen2} 
\end{figure}

First, we have to figure out the mass spectrum of two heavy Majorana neutrinos. Given the mass matrix $M^{}_{\rm R}$ of neutrinos in Eq.~(\ref{eq:MRB}) for {\bf Model B}, which is exactly the same as that in Eq.~(\ref{eq:MRA}) for {\bf Model A}, one can immediately diagonalize it via $U^{\dag}_{\rm R}M^{}_{\rm R}U^{\ast}_{\rm R} = {\rm Diag}\{M^{}_1, M^{}_2\}$, where $M^{}_i$ denotes the mass eigenvalues of the heavy Majorana neutrinos $N^{}_i$ (for $i = 1, 2$) and the unitary matrix $U^{}_{\rm R}$ is given by
\begin{eqnarray}
U^{}_{\rm R}=\frac{1}{\sqrt{2}} \left(\begin{matrix}
\rm{i} &&& \rm{i} \\
1 &&& -1
\end{matrix}\right)
\left(\begin{matrix}
{\rm e}^{{\rm i}(\psi-\varphi)/2} &&& 0 \\
0 &&& 1
\end{matrix}\right) \; ,
\label{eq:UN}
\end{eqnarray}
with $\psi \equiv \arg[Y^{\ast}_{1}(\tau) - {\rm i}Y^{\ast}_{2}(\tau)]$ and $\varphi \equiv \arg[Y^{\ast}_{1}(\tau) + {\rm i}Y^{\ast}_{2}(\tau)]$. The masses of two heavy Majorana neutrinos are found to be $M^{}_{1,2} = \Lambda\sqrt{|Y^{}_{1}(\tau)|^{2}_{} + |Y^{}_{2}(\tau)|^{2}_{} \mp 2 {\rm Im}[Y^{\ast}_{1}(\tau) Y^{}_{2}(\tau)]}$, which depend solely on the modulus parameter $\tau$. Notice that we always assume $M^{}_{1}$ to be the lighter one of two mass eigenvalues, so ${\rm Im}[Y^{\ast}_{1}(\tau) Y^{}_{2}(\tau)]>0$ should be satisfied. For ${\rm Im}[Y^{\ast}_{1}(\tau) Y^{}_{2}(\tau)]<0$, two mass eigenvalues have to be changed as $M^{}_{1,2} =\Lambda \sqrt{|Y^{}_{1}(\tau)|^{2}_{} + |Y^{}_{2}(\tau)|^{2}_{} \pm 2 {\rm Im}[Y^{\ast}_{1}(\tau) Y^{}_{2}(\tau)]}$ and thus the diagonalizing unitary matrix in Eq.~(\ref{eq:UN}) should be replaced by
\begin{eqnarray}
U^{}_{\rm R}=\frac{1}{\sqrt{2}} \left(\begin{matrix}
\rm{i} &&& \rm{i} \\
1 &&& -1
\end{matrix}\right)
\left(\begin{matrix}
{\rm e}^{{\rm i}(\psi-\varphi)/2} &&& 0 \\
0 &&& 1
\end{matrix}\right)
\left(\begin{matrix}
0 &&& 1 \\
1 &&& 0
\end{matrix}\right) \; .
\label{eq:UN1}
\end{eqnarray}
As we have already seen from the previous section, ${\rm Re}\,\tau$ can be restricted to very small values by neutrino oscillation data. In the limit of ${\rm Re}\,\tau \rightarrow 0$ and ${\rm Im}\,\tau>0$, the modular forms $Y^{}_1(\tau)$ and $Y^{}_2(\tau)$ in Eq.~ (\ref{eq:Y3q}) well approximate to
\begin{eqnarray}
Y^{}_{1}(\tau) \approx -\frac{3\pi}{8} \; , \quad
Y^{}_{2}(\tau) \approx  3\sqrt{3}\pi t^{2}_{}(1 + {\rm i}\pi {\rm Re}\,\tau) \; , \label{eq:Y12a}
\end{eqnarray}
where $t \equiv {\rm{e}^{-(\pi {\rm Im}\,\tau)/2}}$ is defined and only the terms up to ${\cal O}(t^{2})$ are retained. Taking the best-fit value of ${\rm Im}\,\tau = 1.36$ (or $1.13$) in the NO (or IO) case, we get $t = 0.12$ (or $0.17$) and thus it is safe to neglect higher-order terms of ${\cal O}(t^4)$. With the help of Eq.~(\ref{eq:Y12a}), we can obtain
\begin{eqnarray}
\tan\left(\frac{\psi-\varphi}{2}\right) \approx 8\sqrt{3}t^{2}_{} \; , \quad
\Delta \equiv \frac{M^{}_{2} - M^{}_{1}}{M^{}_{1}} \approx 16\sqrt{3}\pi t^{2}_{} |{\rm Re}\,\tau |\; , \label{eq:eigenappro}
\end{eqnarray}
where it is interesting to notice that the mass degeneracy parameter $\Delta$ is proportional to $|{\rm Re}\,\tau|$ in the limit of ${\rm Re}\,\tau \rightarrow 0$. Given $t = 0.12$ (or $0.17$) in the NO (or IO) case, we have $\Delta \sim {\cal O}(|{\rm Re}\,\tau|)$. The correlation between the mass degeneracy parameter $\Delta$ and ${\rm Re}\,\tau$ in the NO and IO cases of {\bf Model B} is shown in the left panel of Fig.~\ref{fig:Lepgen1} and that of Fig.~\ref{fig:Lepgen2}, respectively. As given by Eq.~(\ref{eq:eigenappro}), $\Delta \sim {\cal O}(|{\rm Re}\,\tau|)$ is expected for small values of $|{\rm Re}\,\tau|$. Therefore, the correlation between $\Delta$ and $|{\rm Re}\,\tau|$ is perfectly linear in both NO and IO cases, and it is possible to obtain $\Delta \sim 10^{-5}$ (or $\Delta \sim 10^{-6}$) for $|{\rm Re}\,\tau| \sim 10^{-5}$ (or $|{\rm Re}\,\tau| \sim 10^{-7}$) in the NO (or IO) case. We conclude that a strong mass degeneracy of heavy Majorana neutrinos is allowed in {\bf Model B} and this model is also well consistent with neutrino oscillation data at low energies. However, the values of $|{\rm Re}\,\tau|$ in {\bf Model A} are restricted by neutrino oscillation data to the region $0.38<|{\rm Re}\,\tau| < 0.48$, for which the approximate expression of $\Delta$ in Eq.~(\ref{eq:eigenappro}) is no longer valid. Numerically, we find $1.25<\Delta<1.64$ for $0.38<|{\rm Re}\,\tau| < 0.48$, implying that the resonant leptogenesis cannot work well in {\bf Model A}. Although for a relatively large value of $\Delta$ in {\bf Model A} the successful leptogenesis is still possible, the lightest heavy Majorana neutrino mass will be required to be greater than $10^{11}_{}~{\rm GeV}$.

Next, we calculate the lepton number asymmetries of different lepton flavors from the CP-violating decays of heavy Majorana neutrinos. Due to the interference between the tree- and one-loop-level amplitudes, the asymmetries between the rates of heavy Majorana neutrino decays $N^{}_{i} \to \ell^{}_\alpha + H$ and those of the CP-conjugate decays $N^{}_i \to \overline{\ell^{}_\alpha} + \overline{H}$ can be found~\cite{Davidson:2008bu}
\begin{eqnarray}
\epsilon^{}_{i \alpha} \equiv \frac{\Gamma(N^{}_{i} \rightarrow \ell^{}_{\alpha} + H) - \Gamma(N^{}_{i} \rightarrow \overline{\ell^{}_{\alpha}} + \overline{H})}{\displaystyle \sum^{}_{\alpha} \left[\Gamma(N^{}_{i} \rightarrow \ell^{}_{\alpha} + H) + \Gamma(N^{}_{i} \rightarrow \overline{\ell^{}_{\alpha}} + \overline{H})\right]} \; ,
\label{eq:lepdef}
\end{eqnarray}
where $\Gamma(N^{}_{i} \rightarrow \ell^{}_{\alpha} + H)$ and $\Gamma(N^{}_{i} \rightarrow \overline{\ell^{}_{\alpha}} + \overline{H})$  denote the decay rates of $N^{}_{i}$ into leptons and anti-leptons for $\alpha = e, \mu, \tau$. In the MSSM, there are three extra sources of CP asymmetries $\epsilon^{}_{\widetilde{i}\alpha}$, $\epsilon^{}_{i\widetilde{\alpha}}$ and $\epsilon^{}_{\widetilde{i}\widetilde{\alpha}}$, arising from the decays of sneutrinos into leptons and Higgsinos, neutrinos into sleptons and Higgsinos, and sneutrinos into sleptons and Higgs, respectively~\cite{Antusch:2006cw}. All these four types of CP asymmetries are identical and can be expressed as~\cite{Covi:1996wh, Plumacher:1997ru, Zhang:2015tea, Bambhaniya:2016rbb, Dev:2017wwc, Borah:2017qdu, Asaka:2018hyk, Brdar:2019iem, Brivio:2019hrj}
\begin{eqnarray}
\epsilon^{}_{i \alpha} = \frac{1}{8\pi  \left(\widetilde{Y}^{\dag}_{\nu}\widetilde{Y}^{}_{\nu}\right)^{}_{ii}}\; \sum^{}_{k \neq i}{\rm Im}  \left\{ \left(\widetilde{Y}^{}_{\nu}\right)^{*}_{\alpha i} \left(\widetilde{Y}^{}_{\nu}\right)^{}_{\alpha k}   \left[\left(\widetilde{Y}^{\dag}_{\nu}\widetilde{Y}^{}_{\nu}\right)^{}_{i k} f(x^{}_{ki}) + \left(\widetilde{Y}^{\dag}_{\nu}\widetilde{Y}^{}_{\nu}\right)^{\ast}_{i k} g(x^{}_{ki})\right] \right\} \; ,
\label{eq:lepexp}
\end{eqnarray}
where $x^{}_{ki} \equiv M^{2}_{k}/M^{2}_{i}$ (for $k,i=1,2$ but $k \neq i$) and $\widetilde{Y}^{}_{\nu} \equiv U^{\dag}_{l}Y^{}_{\nu}U^{\ast}_{\rm R}$ with $Y^{}_{\nu} \equiv \sqrt{2}M^{}_{\rm D}/v^{}_{\rm u}$ being the Dirac neutrino Yukawa coupling matrix. The effective coupling matrix $\widetilde{Y}^{}_\nu$ has been obtained by transforming into the basis where both the charged-lepton Yukawa coupling matrix and the heavy neutrino mass matrix are diagonal. In Eq.~(\ref{eq:lepexp}), two relevant loop functions have been defined as
\begin{eqnarray}
f(x^{}_{ki}) &=& \sqrt{x^{}_{ki}}\left[ \frac{2(1-x^{}_{ki})}{(1-x^{}_{ki})^2_{}+r^2_{ki}}\right] \; , \label{eq:loopfun} \\
g(x^{}_{ki}) &=& \frac{2(1-x^{}_{ki})}{(1-x^{}_{ki})^2_{}+r^2_{ki}} \; ,\label{eq:loopfun1}
\end{eqnarray}
where $r^{}_{ki} \equiv \Gamma^{}_{k}/M^{}_{i}$ with $\Gamma^{}_{k} =(\widetilde{Y}^{\dag}_{\nu}\widetilde{Y}^{}_{\nu})^{}_{kk} M^{}_{k}/(8\pi)$ being the tree-level total decay width of $N^{}_{k}$ and it serves as the regulator to remove any singularity in the limit of exact mass degeneracy $M^{2}_{k}=M^{2}_{i}$ or equivalently $x^{}_{ki}=1$~\cite{Pilaftsis:1997jf, Pilaftsis:2003gt}. The contribution from vertex corrections has already been neglected in Eq.~(\ref{eq:loopfun}). In view of $\Delta \sim |{\rm Re}\,\tau| \ll 1$, we have $x^{}_{21} \equiv M^2_2/M^2_1 = (1 + \Delta)^2 \approx 1+32\sqrt{3}\pi t^{2}_{}|{\rm Re}\,\tau|$, where the second identity in Eq.~(\ref{eq:eigenappro}) has been used. Furthermore, if the condition $\Delta \gg r^{}_{ki}$ or equivalently $(1 - x^{}_{ik})^2 \gg r^2_{ki}$ is satisfied, the loop functions in Eqs.~(\ref{eq:loopfun}) and (\ref{eq:loopfun1}) can be approximately written as
\begin{eqnarray}
f(x^{}_{12}) \approx -f(x^{}_{21}) \approx g(x^{}_{12}) \approx -g(x^{}_{21}) \approx \dfrac{1}{16\sqrt{3}\pi t^{2}_{}|{\rm Re}\,\tau|} \; ,
\label{eq:loopfappro1}
\end{eqnarray}
where the resonant enhancement of the CP asymmetries from the one-loop self-energy corrections can be realized by the high mass degeneracy of decaying heavy Majorana neutrinos.

It has been emphasized in Refs.~\cite{Dev:2014laa,Dev:2014wsa} that, apart from the resonant mixing of two nearly-degenerate heavy Majorana neutrinos, the oscillations between different flavors of heavy neutrinos offer a physically-distinct source of CP asymmetries. In the strong washout regime where the initial lepton number asymmetries will be essentially destroyed, both mixing and oscillation sources contribute additively and thus the total CP asymmetries $\epsilon^{\rm tot}_{i\alpha}$ read
\begin{eqnarray}
\epsilon^{\rm tot}_{i\alpha} = \epsilon^{}_{i\alpha}+ \epsilon^{\rm osc}_{i\alpha} \; ,
\label{eq:lepexp1}
\end{eqnarray}
where $\epsilon^{\rm osc}_{i\alpha}$ denote the CP asymmetries stemming from heavy neutrino oscillations. The expressions of these asymmetries take the same form as $\epsilon^{}_{i \alpha}$ in Eq.~(\ref{eq:lepexp}) but with the different regulators in the loop functions, which are defined as below~\cite{Dev:2014laa}
\begin{eqnarray}
r^{\rm osc}_{ki} \equiv \frac{M^{}_{i}\Gamma^{}_i+M^{}_k\Gamma^{}_k}{M^{2}_i}\sqrt{\frac{\det [{\rm Re} (\widetilde{Y}^{\dag}_\nu \widetilde{Y}^{}_\nu)]} {(\widetilde{Y}^{\dag}_{\nu} \widetilde{Y}^{}_\nu)^{}_{ii} (\widetilde{Y}^{\dag}_{\nu} \widetilde{Y}^{}_\nu)^{}_{kk}}} \; .
\label{eq:regosc}
\end{eqnarray}

Finally, we proceed to estimate the baryon number asymmetry via resonant leptogenesis. Since we are interested in the scenario of thermal leptogenesis working at the energy scale of $\Lambda^{}_{\rm SS} \sim M^{}_i \approx 10^{7}~{\rm GeV}$, the Yukawa interactions of muon and tau charged leptons turn out to be in thermal equilibrium and the lepton asymmetries of different lepton flavors from $N^{}_i$ decays should be distinguished~\cite{Barbieri:1999ma, Endoh:2003mz, Abada:2006fw, Nardi:2006fx, Abada:2006ea}. The generation of the lepton number asymmetry in each flavor and its subsequent depletion by the inverse decays and the lepton-number-violating scattering processes should be studied by solving the complete set of Boltzmann equations~\cite{Luty:1992un, Giudice:2003jh, Davidson:2008bu}. As an order-of-magnitude estimation, the baryon number asymmetry in the present Universe, characterized by the baryon-to-photon ratio, is given by~\cite{Giudice:2003jh}
\begin{eqnarray}
\eta^{}_{\rm B} \approx -1.04 \times 10^{-2}_{} \sum^{}_{i}\sum^{}_{\alpha}\epsilon^{\rm tot}_{i \alpha} \kappa^{}_{i \alpha} \; ,
\label{eq:bayasy}
\end{eqnarray}
where $\kappa^{}_{i \alpha}$ (for $i = 1, 2$ and $\alpha = e, \mu, \tau$) are the efficiency factors taking account of the washout effects on the lepton number asymmetries. In the scenario of resonant leptogenesis, both heavy Majorana neutrinos will contribute significantly to the generation and washout of lepton number asymmetries. Hence we have $\kappa^{}_{1\alpha} \approx \kappa^{}_{2\alpha} \approx \kappa(K^{}_\alpha)$, and $K^{}_{\alpha} \equiv K^{}_{1\alpha}+K^{}_{2\alpha}$ with $K^{}_{i\alpha} \equiv P^{}_{i\alpha}K^{}_{i}$ are the flavor-dependent decay parameters, where $P^{}_{i\alpha} = |(\widetilde{Y}^{}_{\nu})^{}_{\alpha i}|^2_{}/(\widetilde{Y}^{\dag}_{\nu} \widetilde{Y}^{}_{\nu})^{}_{ii}$ and $K^{}_{i}= \Gamma^{}_{i}/H(M^{}_{i})$ have been defined. The Hubble parameter $H(T) = 1.66\sqrt{g^{\ast}_{}(T)}T^{2}_{}/M^{}_{\rm pl}$ is evaluated at the temperature $T = M^{}_i$, where $M^{}_{\rm pl}=1.2 \times 10^{19}_{} \; {\rm GeV}$ is the Planck mass and $g^{\ast}_{}(T)$ is the number of relativistic degrees of freedom at the temperature $T$. Taking into account the particle content of the MSSM, one gets $g^{\ast}_{}(T) = 228.75$ in the radiation-dominated epoch.

In the parameter space of our interest, where ${\rm Re}\,\tau \in (-3.4 \cdots -1.8)\times 10^{-5}_{}$ or ${\rm Re}\,\tau \in (3.9 \cdots 5.0)\times 10^{-7}_{}$ in the NO or IO case in {\bf Model B} is perfectly allowed by neutrino oscillation data, we can calculate the corresponding ranges of the decay parameters $K^{}_{\alpha}$, i.e.,
\begin{eqnarray}
\underline{\rm NO}&:& \; 1.80 \lesssim K^{}_{e} \lesssim 2.02 \; , \quad 26.5 \lesssim K^{}_{\mu} \lesssim 28.5 \; , \quad 12.8 \lesssim K^{}_{\tau} \lesssim 13.8 \; ; \label{eq:approxK1} \\
\underline{\rm IO}&:& \; 18.7 \lesssim K^{}_{e} \lesssim 19.1 \; , \quad 34.9 \lesssim K^{}_{\mu} \lesssim 35.6 \; , \quad 36.5 \lesssim K^{}_{\tau} \lesssim 37.0 \; . \label{eq:approxK2}
\end{eqnarray}
One can see from Eqs.~(\ref{eq:approxK1}) and (\ref{eq:approxK2}) that except for $K^{}_{e}$ in the NO case, all the decay parameters are much larger than 1. Therefore this points to the strong washout regime. Furthermore, if the heavy Majorana neutrinos are in thermal equilibrium at $T \gg M^{}_{1}$, we have~\cite{Blanchet:2006dq, Blanchet:2006be}
\begin{eqnarray}
\eta^{}_{\rm B} \approx -1.04 \times 10^{-2}_{} \sum^{}_{\alpha}(\epsilon^{\rm tot}_{1 \alpha}+\epsilon^{\rm tot}_{2 \alpha}) \kappa(K^{}_{\alpha}) \; ,
\label{eq:bayappro}
\end{eqnarray}
with
\begin{equation}
\kappa(K^{}_{\alpha}) \approx \frac{1}{K^{}_{\alpha} z^{}_{\rm B}(K^{}_{\alpha})}\left[ 1 - \exp \left(- \frac{K^{}_{\alpha} z^{}_{\rm B}(K^{}_{\alpha})}{2}\right)\right] \; ,
\label{eq:kappa}
\end{equation}
where $z^{}_{\rm B}(K^{}_{\alpha}) = 2 + 4 K^{0.13}_{\alpha} \exp (-2.5/K^{}_\alpha)$.  Note that the right-hand side of Eq.~(\ref{eq:kappa}) is less by a factor of one half when compared to the formula of  $\kappa(K^{}_{\alpha})$ for the SM shown in Ref.~\cite{Blanchet:2006be}, since the associated inverse-decay reactions in the supersymmetric version of MSM will double the washout rates, reducing the asymmetry by a factor of two ~\cite{Davidson:2008bu}.

With the help of Eq.~(\ref{eq:bayappro}), we are ready to compute the baryon number asymmetry $\eta^{}_{\rm B}$ for both NO and IO cases in {\bf Model B}. For illustration, we fix the mass of $N^{}_{1}$ at $M^{}_{1} = 10^{7}_{}~{\rm GeV}$ and $\tan\beta = 10$. As we have explained before, the high mass degeneracy of heavy Majorana neutrinos is required for resonant leptogenesis to work, and thus only small values of $|{\rm Re}\,\tau|$ are viable. For this reason, we first randomly choose the values of $\{{\rm Re}\,\tau, {\rm Im}\,\tau, g, \phi^{}_{g}\}$ within their $1\sigma$ ranges allowed by neutrino oscillation data and further impose the restriction $0<|{\rm Re} \, \tau| <10^{-4} _{}$. Once the values of $\{{\rm Re}\,\tau, {\rm Im}\,\tau, g, \phi^{}_{g}\}$ are given, the coupling matrix $\widetilde{Y}^{}_{\nu}$ and the heavy Majorana neutrino masses can be immediately determined. Then, we calculate the CP asymmetries $\epsilon^{}_ {i\alpha}$ by using Eq.~(\ref{eq:lepexp}) and the baryon-to-photon ratio $\eta^{}_{\rm B}$ in Eq.~(\ref{eq:bayappro}). Finally, we compare the prediction for $\eta^{}_{\rm B}$ with the observed value
\begin{eqnarray}
\eta^{}_{\rm B} = (6.131 \pm 0.041)\times 10^{-10}_{}\; ,
\label{eq:etaobs}
\end{eqnarray}
from the latest Planck observations~\cite{Aghanim:2018eyx}.

\begin{figure}[t!]
	\centering
	\begin{overpic}[scale=0.85]{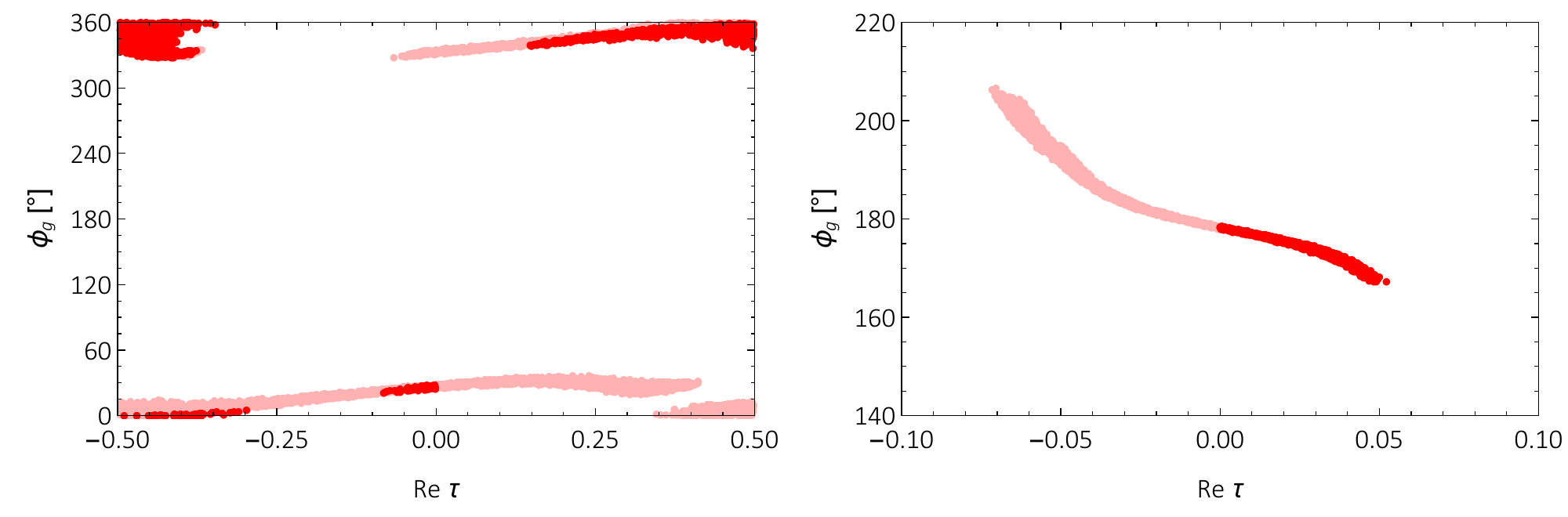}	
		\put(43,27){\underline{NO}}	
		\put(94,27){\underline{IO}}	
	\end{overpic}
\vspace{-0.2cm}
	\caption{Illustration for the $3\sigma$ allowed regions of ${\rm Re}\,\tau$ and $\phi^{}_g$, which determine the sign of the baryon-to-photon ratio $\eta^{}_{\rm B}$ and that of the Jarlskog invariant ${\cal J}$. In the left panel, the dark red dots correspond to $\eta^{}_{\rm B} > 0$ and ${\cal J} < 0$, while the light red ones to either $\eta^{}_{\rm B} < 0$ or ${\cal J} > 0$, in the \underline{NO} case of {\bf Model B}. The results in the \underline{IO} case are shown in the right panel.}
	\label{fig:Rephig}
\end{figure}

In the right panels of Fig.~\ref{fig:Lepgen1} and Fig.~\ref{fig:Lepgen2}, we show our numerical results of $\eta^{}_{\rm B}$ against the mass splitting $\Delta$ of two heavy Majorana neutrinos in the NO and IO cases of {\bf Model B}, respectively. The red dots represent the theoretical predictions for $\eta^{}_{\rm B}$, while the shaded band denotes the $1\sigma$ range of the currently observed value of $\eta^{}_{\rm B}$ in Eq.~(\ref{eq:etaobs}). In the chosen region of ${\rm Re}\,\tau$, the required range of the mass degeneracy parameter $\Delta$ for successful resonant leptogenesis is bounded by two vertical dashed lines. From Fig.~\ref{fig:Lepgen1} and Fig.~\ref{fig:Lepgen2}, we can observe that the observed baryon number asymmetry can be well accommodated in {\bf Model B}, if the mass degeneracy parameter $\Delta$ is of ${\cal O}(10^{-5})$ in the NO case or ${\cal O}(10^{-6})$ in the IO case. More specifically, $\Delta$ should be lying in the range of $(2.26 \cdots 3.74)\times10^{-5}_{}$ corresponding to $-3.28\times 10^{-5}_{}<{\rm Re}\,\tau<-1.83\times 10^{-5}_{}$ in the NO case, whereas $0.82\times 10^{-6}_{}<\Delta<1.02\times 10^{-6}_{}$ corresponding to $3.91\times 10^{-7}_{}<{\rm Re}\,\tau<5.01\times 10^{-7}_{}$ in the IO case. Therefore, we conclude that {\bf Model B} can successfully account for the baryon number asymmetry in our Universe in both NO and IO cases via resonant leptogenesis at a relatively-low energy scale $\Lambda^{}_{\rm SS} \sim 10^{7}_{}~{\rm GeV}$, thus avoiding the potential problem of the gravitino overproduction. Although we have fixed the masses of heavy Majorana neutrinos at $10^7~{\rm GeV}$, it is straightforward to see that an even lower mass scale can be realized by further reducing the mass degeneracy parameter $\Delta$ or equivalently ${\rm Re}\,\tau$.

A final remark on the correlation between low- and high-energy CP violation is helpful. Since CP violation in our models is governed by two parameters ${\rm Re}\,\tau$ and $\phi^{}_g$, they determine both the sign of $\eta^{}_{\rm B}$ and that of the Jarlskog invariant ${\cal J} \equiv \sin \theta^{}_{12} \cos \theta^{}_{12} \sin \theta^{}_{23} \cos \theta^{}_{23} \sin \theta^{}_{13} \cos^2 \theta^{}_{13} \sin \delta$ for leptonic CP violation in neutrino oscillations~\cite{Branco:2011zb}. In Fig.~\ref{fig:Rephig}, the $3\sigma$ ranges of ${\rm Re}\,\tau$ and $\phi^{}_{g}$ allowed by neutrino oscillation data have been presented for NO (left panel) and IO (right panel), but only the dark red dots lead to $\eta^{}_{\rm B}>0$ and ${\cal J} <0$ in {\bf Model B}. It is evident that we need ${\rm Re}\, \tau \rightarrow 0^{-}_{}$ in the NO case (or ${\rm Re}\, \tau \rightarrow 0^{+}_{}$ in the IO case) to guarantee $\eta^{}_{\rm B}>0$ as well as ${\cal J} <0$, the latter of which is preferred by the global-fit analysis of neutrino oscillation experiments in Table~\ref{table:gfit}. Moreover, the observed baryon number asymmetry also constrains the parameter space, particularly that of ${\rm Re}\,\tau$, as indicated in Fig.~\ref{fig:Lepgen1} and Fig.~\ref{fig:Lepgen2}, so we can explore its implications for the Dirac CP-violating phase $\delta$. Numerically, we find a very tight constraint, namely, $208^{\circ}_{}<\delta<212^{\circ}_{}$ in the NO case and $293^{\circ}_{}<\delta<299^{\circ}_{}$ in the IO case, which will be readily tested in the future long-baseline accelerator neutrino oscillation experiments~\cite{Branco:2011zb}. As we have explained in the last section, if the RG running effects are taken into account, the allowed parameter space will be changed at the percent level in the IO case, but essentially unchanged in the NO case.

\section{Summary}\label{sec:summary}

The finite modular symmetry provides us with an attractive and novel way to understand lepton flavor mixing, and has recently attracted a lot of attention. In this paper, we introduce the modular $S^{}_4$ symmetry to the supersymmetric version of MSM, where only two heavy right-handed neutrino singlets are added into the SM. Within this framework, two right-handed neutrino singlets are assigned into the unique two-dimensional representation of the $\Gamma^{}_{4} \simeq S^{}_{4}$ group. Two classes of viable models are constructed to account for lepton mass spectra, flavor mixing, and the matter-antimatter asymmetry.

First, to keep our models as economical as possible, we start with the lowest non-trivial weight $k^{}_{Y} = 2$ of all the relevant modular forms, namely, $f^{}_{e}(\tau)$, $f^{}_{\mu}(\tau)$ and $f^{}_{\tau}(\tau)$ for charged-lepton Yukawa couplings, $f^{}_{\rm D}(\tau)$ and $f^\prime_{\rm D}(\tau)$ for the Dirac neutrino Yukawa couplings, and $f^{}_{\rm R}(\tau)$ for the right-handed neutrino mass matrix. It turns out the minimal set of the weights of these modular forms allowed by current neutrino oscillation data is $(k^{}_{e}, k^{}_{\mu}, k^{}_{\tau}, k^{}_{\rm D}, k^{}_{\rm R}) = (2,4,6,6,2)$. There are two distinct forms of the triplet ${\bf 3}^{\prime}_{}$ with a weight of 6, i.e., $Y^{(6)}_{{\bf 3},1}$ and $Y^{(6)}_{{\bf 3},2}$. If only one of them is included in the superpotential, we have found only two viable models, denoted as {\bf Model A} and {\bf Model B}, for which the low-energy phenomenology has been studied in detail in Sec.~\ref{sec:models}. While {\bf Model A} is consistent with the global-fit results of neutrino oscillation data only at the $3\sigma$ level in the NO case, {\bf Model B} works excellently in both NO and IO cases even at the $1\sigma$ level. We also briefly discuss the scenario where both $Y^{(6)}_{{\bf 3},1}$ and $Y^{(6)}_{{\bf 3},2}$ are added in {\bf Model A} and conclude that the modified model can fit the experimental data even at the $1\sigma$ level. The allowed parameter space of the model parameters, namely, the modulus parameter $\tau = {\rm Re}\,\tau + {\rm i}{\rm Im}\,\tau$ and the coupling constant $\widetilde{g} = g e^{{\rm i}\phi^{}_g}$ has been obtained. Moreover, the constrained regions of three neutrino mixing angles $\{\theta^{}_{12}, \theta^{}_{13}, \theta^{}_{23}\}$ and two CP-violating phases $\{\delta, \sigma\}$, as well as the predictions for the effective neutrino masses $m^{}_{\beta}$ in beta decays and $m^{}_{\beta\beta}$ in neutrinoless double-beta decays, are given. The precision measurements of these parameters will be able to verify or rule out these economical models. Furthermore, we examine the RG running effects of lepton flavor mixing parameters in {\bf Model B}, and find negligible corrections in the NO case but relative changes at the percent level in the IO case.

Second, as the modular symmetry is intrinsically embedded in the supersymmetric theories, we demonstrate that the resonant leptogenesis can be implemented to successfully explain the observed baryon number asymmetry in our Universe, while avoiding the potential problem of the gravitino overproduction usually encountered in the gauge-mediated supersymmetric models. We find that the mass splitting of two heavy Majorana neutrinos is solely governed by the parameter ${\rm Re}\,\tau$. As an excellent approximation, the mass degeneracy parameter $\Delta \equiv (M^{}_2 - M^{}_1)/M^{}_1$ is proportional to $|{\rm Re}\,\tau|$ in the limit of $|{\rm Re}\,\tau| \ll 1$. The successful resonant leptogenesis at a relatively-low seesaw scale can only be realized in {\bf Model B}. The reason is simply that small values of ${\rm Re}\,\tau$ in {\bf Model A} are excluded by neutrino oscillation data. For illustration, we choose $M^{}_1 = 10^7~{\rm GeV}$ and $\tan\beta = 10$, and show that the baryon number asymmetries $\eta^{}_{\rm B} = (6.131 \pm 0.041)\times 10^{-10}_{}$ can be well accommodated in {\bf Model B} for $-3.28\times 10^{-5}_{}<{\rm Re}\,\tau<-1.83\times 10^{-5}_{}$ in the NO case or $3.91\times 10^{-7}_{}<{\rm Re}\,\tau<5.01\times 10^{-7}_{}$ in the IO case. Meanwhile, the observed value of $\eta^{}_{\rm B}$ implies a very narrow range of the Dirac CP-violating phase, namely, $208^{\circ}_{}<\delta<212^{\circ}_{}$ in the NO case and $293^{\circ}_{}<\delta<299^{\circ}_{}$ in the IO case. This is ready for test in the future long-baseline accelerator neutrino oscillation experiments.

Further investigations of the resonant leptogenesis in our models by numerically solving the complete set of Boltzmann equations will be interesting. In this connection, it is worth stressing that successful thermal leptogenesis is one of the salient features of canonical seesaw models and must be incorporated in the models with modular symmetries. As we have already seen, the baryon number asymmetry as an observable offers an independent constraint on the model parameters. Future precision data from neutrino oscillations, beta decays, neutrinoless double-beta decays and cosmological observations will hopefully be able to further narrow down the parameter space and test the theoretical predictions from modular symmetries.

\section*{Acknowledgements}

The authors would like to thank Prof. Gui-Jun Ding, Prof. Stephen F. King, Dr. Newton Nath and Di Zhang for helpful discussions. This work was supported in part by the National Natural Science Foundation of China under grant No.~11775232 and No.~11835013, and by the CAS Center for Excellence in Particle Physics.

\newpage

\appendix

\section{The $\Gamma^{}_{4} \simeq S^{}_{4}$ Symmetry Group}\label{sec:appA}

The permutation symmetry group $S^{}_{4}$ has twenty-four elements and five irreducible representations, which are denoted as ${\bf 1}$, ${\bf 1^{\prime}_{}}$, ${\bf 2}$, ${\bf 3}$ and ${\bf 3^{\prime}_{}}$. In the present work, we choose the same basis for the representation matrices of two generators $S$ and $T$ as in Ref.~\cite{Novichkov:2019sqv}, namely,
\begin{eqnarray}
\begin{array}{cclcl}
  {\bf 1} &: ~\quad & \rho(S) = + 1 \; , & ~\quad & \rho(T) = + 1 \; , \\
  {\bf 1^{\prime_{}}} &: ~\quad & \rho(S)=-1 \; , & ~\quad & \rho(T)=-1 \; , \\
  {\bf 2} &: ~\quad & \displaystyle \rho(S)=\frac{1}{2}\left(\begin{matrix}
-1 &&& \sqrt{3} \\ \sqrt{3} &&& 1
\end{matrix}\right) \; , &~\quad & \rho(T)=\left(\begin{matrix}
1 &&& 0 \\ 0 &&& -1
\end{matrix}\right) \; , \\
  {\bf 3} &: ~\quad & \displaystyle \rho(S)=-\frac{1}{2}\left(\begin{matrix}
0 &&& \sqrt{2} &&& \sqrt{2} \\
\sqrt{2} &&& -1 &&& 1 \\
\sqrt{2} &&& 1 &&& -1
\end{matrix}\right) \; , &~\quad & \rho(T)=-\left(\begin{matrix}
1 &&& 0 &&& 0 \\
0 &&& \rm{i} &&& 0 \\
0 &&& 0 &&& \rm{-i}
\end{matrix}\right) \; , \\
  {\bf 3}^\prime &: ~\quad & \displaystyle \rho(S)=+\frac{1}{2}\left(\begin{matrix}
0 &&& \sqrt{2} &&& \sqrt{2} \\
\sqrt{2} &&& -1 &&& 1 \\
\sqrt{2} &&& 1 &&& -1
\end{matrix}\right) \; , &~\quad & \rho(T) = + \left(\begin{matrix}
1 &&& 0 &&& 0 \\
0 &&& \rm{i} &&& 0 \\
0 &&& 0 &&& \rm{-i}
\end{matrix}\right) \; .
\end{array}
\end{eqnarray}
In this basis, we can explicitly write down the decomposition rules of the Kronecker products of any two $S^{}_{4}$ multiplets.
\begin{itemize}
	\item For the Kronecker products of the singlet ${\bf 1}$ or ${\bf 1^{\prime}_{}}$ and another $S^{}_{4}$ multiplet ${\bf r} = \{{\bf 1}, {\bf 1^{\prime}_{}}, {\bf 2}, {\bf 3}, {\bf 3}^{\prime}_{}\}$:
	\begin{align}
	{\bf 1}\otimes {\bf r} &= {\bf r} \; ,  \label{eq:1r}\\
	{\bf 1^{\prime}_{}} \otimes {\bf 1^{\prime}_{}} &= {\bf 1} \; ,  \label{eq:1p1p}\\
	(\zeta)^{}_{\bf 1^{\prime}_{}} \otimes
	\left(\begin{matrix}
	\xi^{}_{1} \\ \xi^{}_{2}
	\end{matrix}\right)^{}_{\bf 2}
	&= \left(\begin{matrix}
	\zeta\xi^{}_{2} \\ -\zeta\xi^{}_{1}
	\end{matrix}\right)^{}_{\bf 2}  \; , \label{eq:1p2}\\
	(\zeta)^{}_{\bf 1^{\prime}_{}} \otimes
	\left( \begin{matrix}
	\xi^{}_{1} \\ \xi^{}_{2} \\ \xi^{}_{3}
	\end{matrix}\right)^{}_{\bf 3} &=
	\left( \begin{matrix}
	\zeta\xi^{}_{1} \\ \zeta\xi^{}_{2} \\ \zeta\xi^{}_{3}
	\end{matrix}\right)^{}_{\bf 3^{\prime}_{}} \; ,  \label{eq:1p3}\\
	(\zeta)^{}_{\bf 1^{\prime}_{}} \otimes
	\left( \begin{matrix}
	\xi^{}_{1} \\ \xi^{}_{2} \\ \xi^{}_{3}
	\end{matrix}\right)^{}_{\bf 3^{\prime}_{}} &=
	\left( \begin{matrix}
	\zeta\xi^{}_{1} \\ \zeta\xi^{}_{2} \\ \zeta\xi^{}_{3}
	\end{matrix}\right)^{}_{\bf 3} \; ; \label{eq:1p3p}
	\end{align}
	\item For the Kronecker products of the doublet {\bf 2} and another $S^{}_{4}$ multiplet:
	\begin{align}
	\left(\begin{matrix}
	\zeta^{}_{1} \\ \zeta^{}_{2}
	\end{matrix}\right)^{}_{\bf 2} \otimes
	\left(\begin{matrix}
	\xi^{}_{1} \\ \xi^{}_{2}
	\end{matrix}\right)^{}_{\bf 2} &= (\zeta^{}_{1}\xi^{}_{1}+\zeta^{}_{2}\xi^{}_{2})^{}_{\bf 1} \oplus (\zeta^{}_{1}\xi^{}_{2}-\zeta^{}_{2}\xi^{}_{1})^{}_{\bf 1^{\prime}_{}} \oplus \left(\begin{matrix}
	\zeta^{}_{2}\xi^{}_{2}-\zeta^{}_{1}\xi^{}_{1} \\
	\zeta^{}_{1}\xi^{}_{2}+\zeta^{}_{2}\xi^{}_{1}
	\end{matrix}\right)^{}_{\bf 2} \; , \label{eq:22} \\
	\left(\begin{matrix}
	\zeta^{}_{1} \\ \zeta^{}_{2}
	\end{matrix}\right)^{}_{\bf 2} \otimes
	\left(\begin{matrix}
	\xi^{}_{1} \\ \xi^{}_{2} \\ \xi^{}_{3}
	\end{matrix}\right)^{}_{\bf 3} &=
	\left(\begin{matrix}
	\zeta^{}_{1}\xi^{}_{1} \\ (\sqrt{3}/2)\zeta^{}_{2}\xi^{}_{3}-(1/2)\zeta^{}_{1}\xi^{}_{2} \\ (\sqrt{3}/2)\zeta^{}_{2}\xi^{}_{2}-(1/2)\zeta^{}_{1}\xi^{}_{3}
	\end{matrix}\right)^{}_{\bf 3} \oplus	\left(\begin{matrix}
	-\zeta^{}_{2}\xi^{}_{1} \\ (\sqrt{3}/2)\zeta^{}_{1}\xi^{}_{3}+(1/2)\zeta^{}_{2}\xi^{}_{2} \\ (\sqrt{3}/2)\zeta^{}_{1}\xi^{}_{2}+(1/2)\zeta^{}_{2}\xi^{}_{3}
	\end{matrix}\right)^{}_{\bf 3^{\prime}_{}} \; , \label{eq:23} \\
	\left(\begin{matrix}
	\zeta^{}_{1} \\ \zeta^{}_{2}
	\end{matrix}\right)^{}_{\bf 2} \otimes
	\left(\begin{matrix}
	\xi^{}_{1} \\ \xi^{}_{2} \\ \xi^{}_{3}
	\end{matrix}\right)^{}_{\bf 3^{\prime}_{}} &=
	\left(\begin{matrix}
	-\zeta^{}_{2}\xi^{}_{1} \\ (\sqrt{3}/2)\zeta^{}_{1}\xi^{}_{3}+(1/2)\zeta^{}_{2}\xi^{}_{2} \\ (\sqrt{3}/2)\zeta^{}_{1}\xi^{}_{2}+(1/2)\zeta^{}_{2}\xi^{}_{3}
	\end{matrix}\right)^{}_{\bf 3} \oplus	\left(\begin{matrix}
	\zeta^{}_{1}\xi^{}_{1} \\ (\sqrt{3}/2)\zeta^{}_{2}\xi^{}_{3}-(1/2)\zeta^{}_{1}\xi^{}_{2} \\ (\sqrt{3}/2)\zeta^{}_{2}\xi^{}_{2}-(1/2)\zeta^{}_{1}\xi^{}_{3}
	\end{matrix}\right)^{}_{\bf 3^{\prime}_{}} \; ; \label{eq:23p}
	\end{align}
	\item For the Kronecker products of the triplet ${\bf 3}$ or ${\bf 3^{\prime}_{}}$ and another $S^{}_{4}$ triplet:
	\begin{align}
	\left(\begin{matrix}
	\zeta^{}_{1} \\ \zeta^{}_{2} \\ \zeta^{}_{3}
	\end{matrix}\right)^{}_{\bf 3} \otimes
	\left(\begin{matrix}
	\xi^{}_{1} \\ \xi^{}_{2} \\ \xi^{}_{3}
	\end{matrix}\right)^{}_{\bf 3} = &\left(\begin{matrix}
	\zeta^{}_{1} \\ \zeta^{}_{2} \\ \zeta^{}_{3}
	\end{matrix}\right)^{}_{\bf 3^{\prime}_{}} \otimes
	\left(\begin{matrix}
	\xi^{}_{1} \\ \xi^{}_{2} \\ \xi^{}_{3}
	\end{matrix}\right)^{}_{\bf 3^{\prime}_{}} \nonumber \\
	=&~(\zeta^{}_{1}\xi^{}_{1}+\zeta^{}_{2}\xi^{}_{3}+\zeta^{}_{3}\xi^{}_{2})^{}_{\bf 1} \oplus \left( \begin{matrix}
	\zeta^{}_{1}\xi^{}_{1}-(1/2)(\zeta^{}_{2}\xi^{}_{3}+\zeta^{}_{3}\xi^{}_{2}) \\ (\sqrt{3}/2)(\zeta^{}_{2}\xi^{}_{2}+\zeta^{}_{3}\xi^{}_{3})
	\end{matrix}\right)^{}_{\bf 2} \nonumber \\
	& \oplus \left(\begin{matrix}
	\zeta^{}_{3}\xi^{}_{3}-\zeta^{}_{2}\xi^{}_{2} \\
	\zeta^{}_{1}\xi^{}_{3}+\zeta^{}_{3}\xi^{}_{1} \\
	-\zeta^{}_{1}\xi^{}_{2}-\zeta^{}_{2}\xi^{}_{1}
	\end{matrix}\right)^{}_{\bf 3} \oplus \left(\begin{matrix}
	\zeta^{}_{3}\xi^{}_{2}-\zeta^{}_{2}\xi^{}_{3} \\ \zeta^{}_{2}\xi^{}_{1}-\zeta^{}_{1}\xi^{}_{2} \\
	\zeta^{}_{1}\xi^{}_{3}-\zeta^{}_{3}\xi^{}_{1}
	\end{matrix}\right)^{}_{\bf 3^{\prime}_{}} \; ,  \label{eq:33} \\
	\left(\begin{matrix}
	\zeta^{}_{1} \\ \zeta^{}_{2} \\ \zeta^{}_{3}
	\end{matrix}\right)^{}_{\bf 3} \otimes
	\left(\begin{matrix}
	\xi^{}_{1} \\ \xi^{}_{2} \\ \xi^{}_{3}
	\end{matrix}\right)^{}_{\bf 3^{\prime}_{}}
	=&~(\zeta^{}_{1}\xi^{}_{1}+\zeta^{}_{2}\xi^{}_{3}+\zeta^{}_{3}\xi^{}_{2})^{}_{\bf 1^{\prime}_{}} \oplus \left( \begin{matrix}
	(\sqrt{3}/2)(\zeta^{}_{2}\xi^{}_{2}+\zeta^{}_{3}\xi^{}_{3})\\ -\zeta^{}_{1}\xi^{}_{1}+(1/2)(\zeta^{}_{2}\xi^{}_{3}+\zeta^{}_{3}\xi^{}_{2})
	\end{matrix}\right)^{}_{\bf 2} \nonumber \\
	& \oplus \left(\begin{matrix}
	\zeta^{}_{3}\xi^{}_{2}-\zeta^{}_{2}\xi^{}_{3} \\ \zeta^{}_{2}\xi^{}_{1}-\zeta^{}_{1}\xi^{}_{2} \\
	\zeta^{}_{1}\xi^{}_{3}-\zeta^{}_{3}\xi^{}_{1}
	\end{matrix}\right)^{}_{\bf 3} \oplus \left(\begin{matrix}
	\zeta^{}_{3}\xi^{}_{3}-\zeta^{}_{2}\xi^{}_{2} \\
	\zeta^{}_{1}\xi^{}_{3}+\zeta^{}_{3}\xi^{}_{1} \\
	-\zeta^{}_{1}\xi^{}_{2}-\zeta^{}_{2}\xi^{}_{1}
	\end{matrix}\right)^{}_{\bf 3^{\prime}_{}} \; . \label{eq:33p}
	\end{align}
\end{itemize}
With the above decomposition rules and the assignments of relevant fields and modular forms, one can easily find out the Lagrangian invariant  under the modular $S^{}_4$ symmetry group.

As has been mentioned in Sec.~\ref{sec:modular}, there exist five linearly-independent modular forms of the lowest non-trivial weight $k^{}_{Y}=2$, denoted as $Y^{}_i(\tau)$ for $i = 1, 2, \cdots, 5$. They transform as a doublet ${\bf 2}$ and a triplet ${\bf 3}^{\prime}_{}$ under the $S^{}_4$ symmetry, namely~\cite{Penedo:2018nmg},
\begin{eqnarray}
Y^{}_{\bf 2}(\tau) \equiv \left(\begin{matrix} Y^{}_{1}(\tau) \\ Y^{}_2 (\tau)  \end{matrix}\right) \; , \quad Y^{}_{\bf 3^{\prime}_{}} (\tau) \equiv  \left(\begin{matrix} Y^{}_{3}(\tau) \\ Y^{}_4 (\tau) \\ Y^{}_{5} (\tau) \end{matrix}\right) \; .
\label{eq:S4Y1}
\end{eqnarray}
The expressions of modular forms can be derived with the help of the Dedekind $\eta$ function~\cite{Novichkov:2019sqv}
\begin{eqnarray}
\eta(\tau) \equiv q^{1/24}_{} \prod_{n=1}^{\infty}(1-q^{n}_{}) \; ,
\label{eq:eta}
\end{eqnarray}
with $q \equiv e^{2\pi {\rm i} \tau}_{}$, and its derivative~\cite{Penedo:2018nmg}
\begin{equation}
\begin{split}
Y(a^{}_{1}, \dots ,a^{}_{6}|\tau) \equiv
& \frac{\rm d}{\rm{d}\tau}\bigg[ a^{}_{1} \log \eta \left(\tau+\frac{1}{2} \right) + a^{}_{2} \log \eta \left(4\tau\right) + a^{}_{3} \log \eta \left(\frac{\tau}{4} \right) \\
& + a^{}_{4} \log \eta \left(\frac{\tau + 1}{4} \right) + a^{}_{5} \log \eta \left(\frac{\tau + 2}{4} \right) + a^{}_{6} \log \eta \left(\frac{\tau + 3 }{4} \right) \bigg] \; ,
\end{split}
\label{eq:geneY}
\end{equation}
with the coefficients $a^{}_i$ (for $i = 1, 2, \cdots, 6$) fulfilling $a^{}_{1}+ \cdots +a^{}_{6} = 0$. More explicitly, we have~\cite{Novichkov:2019sqv}
\begin{eqnarray}
Y^{}_{1}(\tau)  & \equiv & {\rm i} Y(1,1,-1/2,-1/2,-1/2,-1/2|\tau) \; , \nonumber \\
Y^{}_{2}(\tau)  & \equiv & {\rm i} Y(0,0,\sqrt{3}/2,-\sqrt{3}/2,\sqrt{3}/2,-\sqrt{3}/2|\tau) \; , \nonumber \\
Y^{}_{3}(\tau)  & \equiv & {\rm i} Y(1,-1,0,0,0,0|\tau) \; , \\
Y^{}_{4}(\tau)  & \equiv & {\rm i} Y(0,0,-1/\sqrt{2},{\rm i}/\sqrt{2},1/\sqrt{2},-{\rm i}/\sqrt{2}|\tau) \; , \nonumber \\
Y^{}_{5}(\tau)  & \equiv & {\rm i} Y(0,0,-1/\sqrt{2},-\rm{i}/\sqrt{2},1/\sqrt{2},\rm{i}/\sqrt{2}|\tau) \; , \nonumber
\label{eq:Yexp}
\end{eqnarray}
which can be expanded as the Fourier series~\cite{Novichkov:2019sqv}, i.e.,
\begin{eqnarray}
Y^{}_{1}(\tau) &=& -3\pi\left(1/8 + 3q + 3q^{2}_{} + 12q^{3}_{} + 3q^{4}_{} + 18q^{5}_{} + 12q^{6}_{} + 24q^{7}_{} + 3q^{8}_{} + 39q^{9}_{} + \cdots\right) \; , \nonumber  \\
Y^{}_{2}(\tau) &=&  3\sqrt{3}\pi q^{1/2}_{}(1 + 4q + 6q^{2}_{} + 8q^{3}_{} + 13q^{4}_{} + 12q^{5}_{} + 14q^{6}_{} + 24q^{7}_{} + 18q^{8}_{} + \cdots) \; , \nonumber  \\
Y^{}_{3}(\tau) &=& \pi\left(1/4 - 2q + 6q^{2}_{} - 8q^{3}_{} + 6q^{4}_{} - 12q^{5}_{} + 24q^{6}_{} - 16q^{7}_{} + 6q^{8}_{} + 26q^{9}_{} + \cdots\right) \; , \label{eq:Y3q} \\
Y^{}_{4}(\tau) &=& -\sqrt{2}\pi q^{1/4}_{}(1 + 6q + 13q^{2}_{} + 14q^{3}_{} + 18q^{4}_{} + 32q^{5}_{} + 31q^{6}_{} + 30q^{7}_{} + 48q^{8}_{} + \cdots) \; , \nonumber  \\
Y^{}_{5}(\tau) &=& -4\sqrt{2}\pi q^{3/4}_{}(1 + 2q + 3q^{2}_{} + 6q^{3}_{} + 5q^{4}_{} + 6q^{5}_{} + 10q^{6}_{} + 8q^{7}_{} + 12q^{8}_{}+\cdots) \; . \nonumber
\end{eqnarray}
Based on the modular forms $Y^{}_{i}(\tau)$ for $ i =1,2, \dots ,5$ of weight $k^{}_Y = 2$, one can construct the modular forms of higher weights, such as $k^{}_Y = 4$ and $k^{}_Y = 6$. For $k^{}_Y = 4$, there are totally nine independent modular forms, which transform as {\bf 1}, {\bf 2}, {\bf 3} and ${\bf 3}^{\prime}_{}$ under the $S^{}_{4}$ symmetry, namely~\cite{Novichkov:2019sqv},
\begin{equation}
\begin{split}
Y^{(4)}_{\bf 1} = Y^{2}_{1}+Y^{2}_{2} \; &, \quad Y^{(4)}_{\bf 2} = \left( \begin{matrix} Y^2_2-Y^2_1 \\ 2Y^{}_1 Y^{}_2 \end{matrix} \right) \; , \\
Y^{(4)}_{\bf 3} =\left( \begin{matrix} -2Y^{}_{2}Y^{}_{3} \\ \sqrt{3}Y^{}_{1}Y^{}_{5}+Y^{}_{2}Y^{}_{4} \\ \sqrt{3}Y^{}_{1}Y^{}_{4}+Y^{}_{2}Y^{}_{5} \end{matrix} \right) \; &, \quad Y^{(4)}_{\bf 3^{\prime}_{}} =\left( \begin{matrix} 2Y^{}_{1}Y^{}_{3} \\ \sqrt{3}Y^{}_{2}Y^{}_{5} - Y^{}_{1}Y^{}_{4} \\ \sqrt{3}Y^{}_{2}Y^{}_{4} - Y^{}_{1}Y^{}_{5} \end{matrix} \right) \; ,
\end{split}
\label{eq:Y4}
\end{equation}
where the argument $\tau$ of all the modular forms is suppressed. For $k^{}_Y = 6$, we have thirteen independent modular forms, whose assignments under the $S^{}_4$ symmetry are as follows~\cite{Novichkov:2019sqv}
\begin{equation}
\begin{split}
Y^{(6)}_{\bf 1} = Y^{}_{1}(3Y^{2}_{2}-Y^{2}_{1}) \; &, \quad Y^{(6)}_{\bf 1^{\prime}_{}} = Y^{}_{2}(3Y^{2}_{1}-Y^{2}_{2}) \; ,\\
Y^{(6)}_{\bf 2} = (Y^{2}_{1}+Y^{2}_{2})\left( \begin{matrix} Y^{}_{1} \\ Y^{}_{2} \end{matrix} \right) \;  &, \quad Y^{(6)}_{\bf 3} =\left( \begin{matrix} Y^{}_{1}(Y^{2}_{4}-Y^{2}_{5}) \\ Y^{}_{3}(Y^{}_{1}Y^{}_{5}+\sqrt{3}Y^{}_{2}Y^{}_{4}) \\ -Y^{}_{3}(Y^{}_{1}Y^{}_{4}+\sqrt{3}Y^{}_{2}Y^{}_{5})\end{matrix} \right) \; ,\\
Y^{(6)}_{{\bf 3^{\prime}_{}},1} =(Y^{2}_{1}+Y^{2}_{2})\left( \begin{matrix} Y^{}_{3} \\ Y^{}_{4} \\ Y^{}_{5} \end{matrix} \right) \; &, \quad Y^{(6)}_{{\bf 3^{\prime}_{}},2} =\left( \begin{matrix} Y^{}_{2}(Y^{2}_{5}-Y^{2}_{4}) \\ -Y^{}_{3}(Y^{}_{2}Y^{}_{5}-\sqrt{3}Y^{}_{1}Y^{}_{4}) \\ Y^{}_{3}(Y^{}_{2}Y^{}_{4}-\sqrt{3}Y^{}_{1}Y^{}_{5}) \end{matrix} \right) \; .
\end{split}
\label{eq:Y6}
\end{equation}


\newpage

\begin{thebibliography}{99}
\bibitem{Xing:2011zza}
  Z.~z.~Xing and S.~Zhou,
  ``Neutrinos in particle physics, astronomy and cosmology,''
  Springer-Verlag, Berlin Heidelberg (2011).

\bibitem{Tanabashi:2018oca}
M.~Tanabashi {\it et al.} [Particle Data Group],
``Review of Particle Physics,''
Phys.\ Rev.\ D {\bf 98} (2018) no.3,  030001.

	\bibitem{Minkowski:1977sc}
	P.~Minkowski,
	``$\mu \to e\gamma$ at a Rate of One Out of $10^{9}$ Muon Decays?,''
	Phys.\ Lett.\  {\bf 67B} (1977) 421.
	

	\bibitem{Yanagida:1979}
	T. Yanagida, in \emph{Proc. Workshop on the Baryon Number
	of the Universe and Unified Theories}, edited by
	O. Sawada and A. Sugamoto (1979), p. 95.
	
	\bibitem{GellMan1979}
	M. Gell-Mann, P. Ramond, and R. Slansky, in \emph{Supergravity}
	, edited by P. van Nieuwenhuizen and D. Freedman
	(1979), p. 315.
	
	\bibitem{Mohapatra:1979ia}
	R.~N.~Mohapatra and G.~Senjanovic,
	``Neutrino Mass and Spontaneous Parity Nonconservation,''
	Phys.\ Rev.\ Lett.\  {\bf 44} (1980) 912.

\bibitem{Xing:2019vks}
  Z.~z.~Xing,
  ``Flavor structures of charged fermions and massive neutrinos,''
  arXiv:1909.09610.
	
	\bibitem{King:1999mb}
	S.~F.~King,
	``Large mixing angle MSW and atmospheric neutrinos from single right-handed neutrino dominance and U(1) family symmetry,''
	Nucl.\ Phys.\ B {\bf 576} (2000) 85
	[hep-ph/9912492].
	
	\bibitem{Frampton:2002qc}
	P.~H.~Frampton, S.~L.~Glashow and T.~Yanagida,
	``Cosmological sign of neutrino CP violation,''
	Phys.\ Lett.\ B {\bf 548}, (2002) 119
	[hep-ph/0208157].
	
	\bibitem{Guo:2006qa}
	W.~l.~Guo, Z.~z.~Xing and S.~Zhou,
	``Neutrino Masses, Lepton Flavor Mixing and Leptogenesis in the Minimal Seesaw Model,''
	Int.\ J.\ Mod.\ Phys.\ E {\bf 16} (2007) 1
	[hep-ph/0612033].
	
	\bibitem{Esteban:2018azc}
	I.~Esteban, M.~C.~Gonzalez-Garcia, A.~Hernandez-Cabezudo, M.~Maltoni and T.~Schwetz,
	``Global analysis of three-flavour neutrino oscillations: synergies and tensions in the determination of $\theta_{23}, \delta_{CP}$, and the mass ordering,''
	JHEP {\bf 1901} (2019) 106
	[arXiv:1811.05487].

\bibitem{Xing:2006ms}
  Z.~z.~Xing and S.~Zhou,
  ``Tri-bimaximal Neutrino Mixing and Flavor-dependent Resonant Leptogenesis,''
  Phys.\ Lett.\ B {\bf 653}, 278 (2007)
  [hep-ph/0607302].

	\bibitem{Zhang:2009ac}
	H.~Zhang and S.~Zhou,
	``The Minimal Seesaw Model at the TeV Scale,''
	Phys.\ Lett.\ B {\bf 685} (2010) 297
	[arXiv:0912.2661].
	
	
	\bibitem{Altarelli:2010gt}
	G.~Altarelli and F.~Feruglio,
	``Discrete Flavor Symmetries and Models of Neutrino Mixing,''
	Rev.\ Mod.\ Phys.\  {\bf 82} (2010) 2701
	[arXiv:1002.0211].
	
	\bibitem{Ishimori:2010au}
	H.~Ishimori, T.~Kobayashi, H.~Ohki, Y.~Shimizu, H.~Okada and M.~Tanimoto,
	``Non-Abelian Discrete Symmetries in Particle Physics,''
	Prog.\ Theor.\ Phys.\ Suppl.\  {\bf 183} (2010) 1
	[arXiv:1003.3552].
	
	\bibitem{King:2013eh}
	S.~F.~King and C.~Luhn,
	``Neutrino Mass and Mixing with Discrete Symmetry,''
	Rept.\ Prog.\ Phys.\  {\bf 76} (2013) 056201
	[arXiv:1301.1340].

\bibitem{King:2014nza}
S.~F.~King, A.~Merle, S.~Morisi, Y.~Shimizu and M.~Tanimoto,
``Neutrino Mass and Mixing: from Theory to Experiment,''
New J.\ Phys.\  {\bf 16} (2014) 045018
[arXiv:1402.4271].

\bibitem{King:2017guk}
S.~F.~King,
``Unified Models of Neutrinos, Flavour and CP Violation,''
Prog.\ Part.\ Nucl.\ Phys.\  {\bf 94} (2017) 217
[arXiv:1701.04413].

\bibitem{Lam:2008rs}
C.~S.~Lam,
``Determining Horizontal Symmetry from Neutrino Mixing,''
Phys.\ Rev.\ Lett.\  {\bf 101} (2008) 121602
[arXiv:0804.2622].

\bibitem{Lam:2008sh}
C.~S.~Lam,
``The Unique Horizontal Symmetry of Leptons,''
Phys.\ Rev.\ D {\bf 78} (2008) 073015
[arXiv:0809.1185].

\bibitem{Ge:2011ih}
S.~F.~Ge, D.~A.~Dicus and W.~W.~Repko,
``$Z_2$ Symmetry Prediction for the Leptonic Dirac CP Phase,''
Phys.\ Lett.\ B {\bf 702} (2011) 220
[arXiv:1104.0602].

\bibitem{Ge:2011qn}
S.~F.~Ge, D.~A.~Dicus and W.~W.~Repko,
``Residual Symmetries for Neutrino Mixing with a Large $\theta_{13}$ and Nearly Maximal $\delta_D$,''
Phys.\ Rev.\ Lett.\  {\bf 108} (2012) 041801
[arXiv:1108.0964].

	\bibitem{Hernandez:2012ra}
	D.~Hernandez and A.~Y.~Smirnov,
	``Lepton mixing and discrete symmetries,''
	Phys.\ Rev.\ D {\bf 86} (2012) 053014
	[arXiv:1204.0445].

	\bibitem{Feruglio:2012cw}
	F.~Feruglio, C.~Hagedorn and R.~Ziegler,
	``Lepton Mixing Parameters from Discrete and CP Symmetries,''
	JHEP {\bf 1307} (2013) 027
	[arXiv:1211.5560].
	
	
	\bibitem{Feruglio:2017spp}
	F.~Feruglio,
	``Are neutrino masses modular forms?,''
	arXiv:1706.08749.
	
	\bibitem{Kobayashi:2018vbk}
	T.~Kobayashi, K.~Tanaka and T.~H.~Tatsuishi,
	``Neutrino mixing from finite modular groups,''
	Phys.\ Rev.\ D {\bf 98} (2018) no.1,  016004
	[arXiv:1803.10391].
	
	\bibitem{Kobayashi:2018wkl}
	T.~Kobayashi, Y.~Shimizu, K.~Takagi, M.~Tanimoto, T.~H.~Tatsuishi and H.~Uchida,
	``Finite modular subgroups for fermion mass matrices and baryon/lepton number violation,''
	Phys.\ Lett.\ B {\bf 794} (2019) 114
	[arXiv:1812.11072].
	
	\bibitem{Kobayashi:2019rzp}
	T.~Kobayashi, Y.~Shimizu, K.~Takagi, M.~Tanimoto and T.~H.~Tatsuishi,
	``Modular $S_3$ invariant flavor model in SU(5) GUT,''
	arXiv:1906.10341.

\bibitem{Okada:2019xqk}
  H.~Okada and Y.~Orikasa,
  ``Modular $S_3$ symmetric radiative seesaw model,''
  Phys.\ Rev.\ D {\bf 100}, no. 11, 115037 (2019)
  [arXiv:1907.04716].

	\bibitem{Kobayashi:2018scp}
	T.~Kobayashi, N.~Omoto, Y.~Shimizu, K.~Takagi, M.~Tanimoto and T.~H.~Tatsuishi,
	JHEP {\bf 1811} (2018) 196
	[arXiv:1808.03012].
	
	\bibitem{Criado:2018thu}
	J.~C.~Criado and F.~Feruglio,
	``Modular Invariance Faces Precision Neutrino Data,''
	SciPost Phys.\  {\bf 5} (2018) no.5,  042
	[arXiv:1807.01125].
	
\bibitem{deAnda:2018ecu}
  F.~J.~de Anda, S.~F.~King and E.~Perdomo,
  ``$SU(5)$ grand unified theory with $A_4$ modular symmetry,''
  Phys.\ Rev.\ D {\bf 101}, no. 1, 015028 (2020)
  [arXiv:1812.05620].
	
	\bibitem{Okada:2018yrn}
	H.~Okada and M.~Tanimoto,
	``CP violation of quarks in $A_4$ modular invariance,''
	Phys.\ Lett.\ B {\bf 791} (2019) 54
	[arXiv:1812.09677].
	
	\bibitem{Nomura:2019jxj}
	T.~Nomura and H.~Okada,
	``A modular $A_4$ symmetric model of dark matter and neutrino,''
	Phys.\ Lett.\ B {\bf 797} (2019) 134799
	[arXiv:1904.03937].
	
	\bibitem{Nomura:2019yft}
	T.~Nomura and H.~Okada,
	``A two loop induced neutrino mass model with modular $A_4$ symmetry,''
	arXiv:1906.03927.
	
	\bibitem{Ding:2019zxk}
	G.~J.~Ding, S.~F.~King and X.~G.~Liu,
	``Modular A$_{4}$ symmetry models of neutrinos and charged leptons,''
	JHEP {\bf 1909} (2019) 074
	[arXiv:1907.11714].
	
\bibitem{Nomura:2019lnr}
  T.~Nomura, H.~Okada and O.~Popov,
  ``A modular $A_4$ symmetric scotogenic model,''
  Phys.\ Lett.\ B {\bf 803}, 135294 (2020)
  [arXiv:1908.07457].
	
	\bibitem{Okada:2019mjf}
	H.~Okada and Y.~Orikasa,
	``A radiative seesaw model in modular $A_4$ symmetry,''
	arXiv:1907.13520.
	
\bibitem{Asaka:2019vev}
  T.~Asaka, Y.~Heo, T.~H.~Tatsuishi and T.~Yoshida,
  ``Modular $A_4$ invariance and leptogenesis,''
  JHEP {\bf 2001}, 144 (2020)
  [arXiv:1909.06520].

\bibitem{Zhang:2019ngf}
  D.~Zhang,
  ``A modular $A_4$ symmetry realization of two-zero textures of the Majorana neutrino mass matrix,''
  Nucl.\ Phys.\ B {\bf 952}, 114935 (2020)
  [arXiv:1910.07869].
	\bibitem{Penedo:2018nmg}
	J.~T.~Penedo and S.~T.~Petcov,
	``Lepton Masses and Mixing from Modular $S_4$ Symmetry,''
	Nucl.\ Phys.\ B {\bf 939} (2019) 292
	[arXiv:1806.11040].
	
	\bibitem{Novichkov:2018ovf}
	P.~P.~Novichkov, J.~T.~Penedo, S.~T.~Petcov and A.~V.~Titov,
	``Modular S$_{4}$ models of lepton masses and mixing,''
	JHEP {\bf 1904} (2019) 005
	[arXiv:1811.04933].
	
	\bibitem{Okada:2019lzv}
	H.~Okada and Y.~Orikasa,
	``Neutrino mass model with a modular $S_4$ symmetry,''
	arXiv:1908.08409.
	
	\bibitem{Novichkov:2018nkm}
	P.~P.~Novichkov, J.~T.~Penedo, S.~T.~Petcov and A.~V.~Titov,
	``Modular A$_{5}$ symmetry for flavour model building,''
	JHEP {\bf 1904} (2019) 174
	[arXiv:1812.02158].
	
\bibitem{Ding:2019xna}
  G.~J.~Ding, S.~F.~King and X.~G.~Liu,
  ``Neutrino mass and mixing with $A_5$ modular symmetry,''
  Phys.\ Rev.\ D {\bf 100}, no. 11, 115005 (2019)
  [arXiv:1903.12588].
  	
\bibitem{Criado:2019tzk}
  J.~C.~Criado, F.~Feruglio and S.~J.~D.~King,
  ``Modular Invariant Models of Lepton Masses at Levels 4 and 5,''
  JHEP {\bf 2002}, 001 (2020)
  [arXiv:1908.11867].
  	
	\bibitem{Novichkov:2019sqv}
	P.~P.~Novichkov, J.~T.~Penedo, S.~T.~Petcov and A.~V.~Titov,
	``Generalised CP Symmetry in Modular-Invariant Models of Flavour,''
	JHEP {\bf 1907} (2019) 165
	[arXiv:1905.11970].
	
	\bibitem{deMedeirosVarzielas:2019cyj}
	I.~De Medeiros Varzielas, S.~F.~King and Y.~L.~Zhou,
	``Multiple modular symmetries as the origin of flavour,''
	arXiv:1906.02208.
	
\bibitem{King:2019vhv}
  S.~F.~King and Y.~L.~Zhou,
  ``Trimaximal TM$_1$ mixing with two modular $S_4$ groups,''
  Phys.\ Rev.\ D {\bf 101}, no. 1, 015001 (2020)
  [arXiv:1908.02770].
  	
	\bibitem{Liu:2019khw}
	X.~G.~Liu and G.~J.~Ding,
	``Neutrino Masses and Mixing from Double Covering of Finite Modular Groups,''
	JHEP {\bf 1908} (2019) 134
	[arXiv:1907.01488].
	
\bibitem{Kobayashi:2019mna}
  T.~Kobayashi, Y.~Shimizu, K.~Takagi, M.~Tanimoto and T.~H.~Tatsuishi,
  ``New $A_4$ lepton flavor model from $S_4$ modular symmetry,''
  JHEP {\bf 2002}, 097 (2020)
  [arXiv:1907.09141].
  	
\bibitem{Kobayashi:2019xvz}
  T.~Kobayashi, Y.~Shimizu, K.~Takagi, M.~Tanimoto and T.~H.~Tatsuishi,
  ``$A_4$ lepton flavor model and modulus stabilization from $S_4$ modular symmetry,''
  Phys.\ Rev.\ D {\bf 100}, no. 11, 115045 (2019)
  Erratum: [Phys.\ Rev.\ D {\bf 101}, no. 3, 039904 (2020)]
  [arXiv:1909.05139].
  	
	\bibitem{Novichkov:2018yse}
	P.~P.~Novichkov, S.~T.~Petcov and M.~Tanimoto,
	``Trimaximal Neutrino Mixing from Modular A4 Invariance with Residual Symmetries,''
	Phys.\ Lett.\ B {\bf 793} (2019) 247
	[arXiv:1812.11289].
	
\bibitem{Gui-JunDing:2019wap}
  G.~J.~Ding, S.~F.~King, X.~G.~Liu and J.~N.~Lu,
  ``Modular S$_{4}$ and A$_{4}$ symmetries and their fixed points: new predictive examples of lepton mixing,''
  JHEP {\bf 1912}, 030 (2019)
  [arXiv:1910.03460].

	\bibitem{Okada:2019uoy}
	H.~Okada and M.~Tanimoto,
	``Towards unification of quark and lepton flavors in $A_4$ modular invariance,''
	arXiv:1905.13421.
	
	\bibitem{Yang:2011fh}
	R.~Z.~Yang and H.~Zhang,
	``Minimal seesaw model with $S_4$ flavor symmetry,''
	Phys.\ Lett.\ B {\bf 700} (2011) 316
	[arXiv:1104.0380].

	\bibitem{Aghanim:2018eyx}
	N.~Aghanim {\it et al.} [Planck Collaboration],
	``Planck 2018 results. VI. Cosmological parameters,''
	arXiv:1807.06209.
	
	\bibitem{Fukugita:1986hr}
	M.~Fukugita and T.~Yanagida,
	``Baryogenesis Without Grand Unification,''
	Phys.\ Lett.\ B {\bf 174} (1986) 45.

	\bibitem{Pilaftsis:1997jf}
	A.~Pilaftsis,
	``CP violation and baryogenesis due to heavy Majorana neutrinos,''
	Phys.\ Rev.\ D {\bf 56} (1997) 5431
	[hep-ph/9707235].
	
	\bibitem{Pilaftsis:2003gt}
	A.~Pilaftsis and T.~E.~J.~Underwood,
	``Resonant leptogenesis,''
	Nucl.\ Phys.\ B {\bf 692} (2004) 303
	[hep-ph/0309342].
	
	\bibitem{GonzalezFelipe:2003fi}
	R.~Gonzalez Felipe, F.~R.~Joaquim and B.~M.~Nobre,
	``Radiatively induced leptogenesis in a minimal seesaw model,''
	Phys.\ Rev.\ D {\bf 70} (2004) 085009
	[hep-ph/0311029].
	
	\bibitem{Branco:2005ye}
	G.~C.~Branco, R.~Gonzalez Felipe, F.~R.~Joaquim and B.~M.~Nobre,
	``Enlarging the window for radiative leptogenesis,''
	Phys.\ Lett.\ B {\bf 633} (2006) 336
	[hep-ph/0507092].
	
	\bibitem{Ahn:2006rn}
	Y.~H.~Ahn, C.~S.~Kim, S.~K.~Kang and J.~Lee,
	``mu- tau Symmetry and Radiatively Generated Leptogenesis,''
	Phys.\ Rev.\ D {\bf 75} (2007) 013012
	[hep-ph/0610007].
	
%
%
%
%
	
	\bibitem{Khlopov:1984pf}
	M.~Y.~Khlopov and A.~D.~Linde,
	``Is It Easy to Save the Gravitino?,''
	Phys.\ Lett.\  {\bf 138B} (1984) 265.
	
	\bibitem{Ellis:1984eq}
	J.~R.~Ellis, J.~E.~Kim and D.~V.~Nanopoulos,
	``Cosmological Gravitino Regeneration and Decay,''
	Phys.\ Lett.\  {\bf 145B} (1984) 181.

\bibitem{Asaka:2000zh}
T.~Asaka, K.~Hamaguchi and K.~Suzuki,
``Cosmological gravitino problem in gauge mediated supersymmetry breaking models,''
Phys.\ Lett.\ B {\bf 490} (2000) 136
[hep-ph/0005136].
	
\bibitem{Chen:2019ewa}
M.~C.~Chen, S.~Ramos-S\'anchez and M.~Ratz,
``A note on the predictions of models with modular flavor symmetries,''
Phys. Lett. B \textbf{801}, 135153 (2020)
[arXiv:1909.06910].
  
\bibitem{Aker:2019uuj}
  M.~Aker {\it et al.} [KATRIN Collaboration],
  ``Improved Upper Limit on the Neutrino Mass from a Direct Kinematic Method by KATRIN,''
  Phys.\ Rev.\ Lett.\  {\bf 123}, no. 22, 221802 (2019)
  [arXiv:1909.06048].

\bibitem{Aker:2019qfn}
  M.~Aker {\it et al.} [KATRIN Collaboration],
  ``First operation of the KATRIN experiment with tritium,''
  arXiv:1909.06069.

\bibitem{Dolinski:2019nrj}
  M.~J.~Dolinski, A.~W.~P.~Poon and W.~Rodejohann,
  ``Neutrinoless Double-Beta Decay: Status and Prospects,''
  Ann.\ Rev.\ Nucl.\ Part.\ Sci.\  {\bf 69}, 219 (2019)
  [arXiv:1902.04097].
  
\bibitem{Cao:2019hli}
  J.~Cao, G.~Y.~Huang, Y.~F.~Li, Y.~Wang, L.~J.~Wen, Z.~Z.~Xing, Z.~H.~Zhao and S.~Zhou,
  ``Towards the meV limit of the effective neutrino mass in neutrinoless double-beta decays,''
  Chin.\ Phys.\ C {\bf 44}, no. 3, 031001 (2020)
  [arXiv:1908.08355].
  
  \bibitem{Ohlsson:2013xva}
  T.~Ohlsson and S.~Zhou,
  ``Renormalization group running of neutrino parameters,''
  Nature Commun.\  {\bf 5} (2014) 5153
  [arXiv:1311.3846].

  \bibitem{Mei:2003gn}
  J.~w.~Mei and Z.~z.~Xing,
  ``Radiative corrections to neutrino mixing and CP violation in the minimal seesaw model with leptogenesis,''
  Phys.\ Rev.\ D {\bf 69} (2004) 073003
  [hep-ph/0312167].

  \bibitem{Machacek:1983fi}
  M.~E.~Machacek and M.~T.~Vaughn,
  ``Two Loop Renormalization Group Equations in a General Quantum Field Theory. 2. Yukawa Couplings,''
  Nucl.\ Phys.\ B {\bf 236} (1984) 221.

  \bibitem{Arason:1991ic}
  H.~Arason, D.~J.~Castano, B.~Keszthelyi, S.~Mikaelian, E.~J.~Piard, P.~Ramond and B.~D.~Wright,
  ``Renormalization group study of the standard model and its extensions. 1. The Standard model,''
  Phys.\ Rev.\ D {\bf 46} (1992) 3945.

  \bibitem{Castano:1993ri}
  D.~J.~Castano, E.~J.~Piard and P.~Ramond,
  ``Renormalization group study of the Standard Model and its extensions. 2. The Minimal supersymmetric Standard Model,''
  Phys.\ Rev.\ D {\bf 49} (1994) 4882
  [hep-ph/9308335].

  \bibitem{Chankowski:1993tx}
  P.~H.~Chankowski and Z.~Pluciennik,
  ``Renormalization group equations for seesaw neutrino masses,''
  Phys.\ Lett.\ B {\bf 316} (1993) 312
  [hep-ph/9306333].

  \bibitem{Babu:1993qv}
  K.~S.~Babu, C.~N.~Leung and J.~T.~Pantaleone,
  ``Renormalization of the neutrino mass operator,''
  Phys.\ Lett.\ B {\bf 319} (1993) 191
  [hep-ph/9309223].

  \bibitem{Antusch:2001ck}
  S.~Antusch, M.~Drees, J.~Kersten, M.~Lindner and M.~Ratz,
  ``Neutrino mass operator renormalization revisited,''
  Phys.\ Lett.\ B {\bf 519} (2001) 238
  [hep-ph/0108005].

	\bibitem{Luty:1992un}
	M.~A.~Luty,
	``Baryogenesis via leptogenesis,''
	Phys.\ Rev.\ D {\bf 45} (1992) 455.

	\bibitem{Klinkhamer:1984di}
	F.~R.~Klinkhamer and N.~S.~Manton,
	``A Saddle Point Solution in the Weinberg-Salam Theory,''
	Phys.\ Rev.\ D {\bf 30} (1984) 2212.

\bibitem{Kuzmin:1985mm}
V.~A.~Kuzmin, V.~A.~Rubakov and M.~E.~Shaposhnikov,
``On the Anomalous Electroweak Baryon Number Nonconservation in the Early Universe,''
Phys.\ Lett.\  {\bf 155B} (1985) 36.
	
	\bibitem{Arnold:1987mh}
	P.~B.~Arnold and L.~D.~McLerran,
	``Sphalerons, Small Fluctuations and Baryon Number Violation in Electroweak Theory,''
	Phys.\ Rev.\ D {\bf 36} (1987) 581.
	
	\bibitem{Buchmuller:2004nz}
	W.~Buchmuller, P.~Di Bari and M.~Plumacher,
	``Leptogenesis for pedestrians,''
	Annals Phys.\  {\bf 315} (2005) 305
	[hep-ph/0401240].
	
	\bibitem{Davidson:2008bu}
	S.~Davidson, E.~Nardi and Y.~Nir,
	``Leptogenesis,''
	Phys.\ Rept.\  {\bf 466} (2008) 105
	[arXiv:0802.2962].
		
	\bibitem{Antusch:2006cw}
	S.~Antusch, S.~F.~King and A.~Riotto,
	``Flavour-Dependent Leptogenesis with Sequential Dominance,''
	JCAP {\bf 0611} (2006) 011
	[hep-ph/0609038].
	
	\bibitem{Covi:1996wh}
	L.~Covi, E.~Roulet and F.~Vissani,
	``CP violating decays in leptogenesis scenarios,''
	Phys.\ Lett.\ B {\bf 384} (1996) 169
	[hep-ph/9605319].
	
	\bibitem{Plumacher:1997ru}
	M.~Plumacher,
	``Baryon asymmetry, neutrino mixing and supersymmetric SO(10) unification,''
	Nucl.\ Phys.\ B {\bf 530} (1998) 207
	[hep-ph/9704231].
	
	\bibitem{Zhang:2015tea}
	J.~Zhang and S.~Zhou,
	``A Further Study of the Frampton-Glashow-Yanagida Model for Neutrino Masses, Flavor Mixing and Baryon Number Asymmetry,''
	JHEP {\bf 1509} (2015) 065
	[arXiv:1505.04858].
	
	\bibitem{Bambhaniya:2016rbb}
	G.~Bambhaniya, P.~S.~Bhupal Dev, S.~Goswami, S.~Khan and W.~Rodejohann,
	``Naturalness, Vacuum Stability and Leptogenesis in the Minimal Seesaw Model,''
	Phys.\ Rev.\ D {\bf 95} (2017) no.9,  095016
	[arXiv:1611.03827].

\bibitem{Dev:2017wwc}
B.~Dev, M.~Garny, J.~Klaric, P.~Millington and D.~Teresi,
``Resonant enhancement in leptogenesis,''
Int.\ J.\ Mod.\ Phys.\ A {\bf 33} (2018) 1842003
[arXiv:1711.02863].

  \bibitem{Borah:2017qdu}
  D.~Borah, M.~K.~Das and A.~Mukherjee,
  ``Common origin of nonzero $\theta_{13}$ and baryon asymmetry of the Universe in a TeV scale seesaw model with $A_4$ flavor symmetry,''
  Phys.\ Rev.\ D {\bf 97} (2018) no.11,  115009
  [arXiv:1711.02445].

  \bibitem{Asaka:2018hyk}
  T.~Asaka and T.~Yoshida,
  ``Resonant leptogenesis at TeV-scale and neutrinoless double beta decay,''
  JHEP {\bf 1909} (2019) 089
  [arXiv:1812.11323].

\bibitem{Brdar:2019iem}
  V.~Brdar, A.~J.~Helmboldt, S.~Iwamoto and K.~Schmitz,
  ``Type-I Seesaw as the Common Origin of Neutrino Mass, Baryon Asymmetry, and the Electroweak Scale,''
  Phys.\ Rev.\ D {\bf 100}, 075029 (2019)
  [arXiv:1905.12634].
  
  \bibitem{Brivio:2019hrj}
  I.~Brivio, K.~Moffat, S.~Pascoli, S.~T.~Petcov and J.~Turner,
  ``Leptogenesis in the Neutrino Option,''
  JHEP {\bf 1910} (2019) 059
  [arXiv:1905.12642].

  \bibitem{Dev:2014laa}
  P.~S.~Bhupal Dev, P.~Millington, A.~Pilaftsis and D.~Teresi,
  Nucl.\ Phys.\ B {\bf 886} (2014) 569
  [arXiv:1404.1003].

  \bibitem{Dev:2014wsa}
  P.~S.~Bhupal Dev, P.~Millington, A.~Pilaftsis and D.~Teresi,
  ``Kadanoff–Baym approach to flavour mixing and oscillations in resonant leptogenesis,''
  Nucl.\ Phys.\ B {\bf 891} (2015) 128
  [arXiv:1410.6434].


\bibitem{Barbieri:1999ma}
R.~Barbieri, P.~Creminelli, A.~Strumia and N.~Tetradis,
``Baryogenesis through leptogenesis,''
Nucl.\ Phys.\ B {\bf 575} (2000) 61
[hep-ph/9911315].

\bibitem{Endoh:2003mz}
  T.~Endoh, T.~Morozumi and Z.~h.~Xiong,
  ``Primordial lepton family asymmetries in seesaw model,''
  Prog.\ Theor.\ Phys.\  {\bf 111} (2004) 123
  [hep-ph/0308276].

\bibitem{Abada:2006fw}
  A.~Abada, S.~Davidson, F.~X.~Josse-Michaux, M.~Losada and A.~Riotto,
  ``Flavor issues in leptogenesis,''
  JCAP {\bf 0604} (2006) 004
  [hep-ph/0601083].

\bibitem{Nardi:2006fx}
  E.~Nardi, Y.~Nir, E.~Roulet and J.~Racker,
  ``The Importance of flavor in leptogenesis,''
  JHEP {\bf 0601} (2006) 164
  [hep-ph/0601084].

\bibitem{Abada:2006ea}
  A.~Abada, S.~Davidson, A.~Ibarra, F.-X.~Josse-Michaux, M.~Losada and A.~Riotto,
  ``Flavour Matters in Leptogenesis,''
  JHEP {\bf 0609} (2006) 010
  [hep-ph/0605281].

	\bibitem{Giudice:2003jh}
	G.~F.~Giudice, A.~Notari, M.~Raidal, A.~Riotto and A.~Strumia,
	``Towards a complete theory of thermal leptogenesis in the SM and MSSM,''
	Nucl.\ Phys.\ B {\bf 685} (2004) 89
	[hep-ph/0310123].
	
	\bibitem{Blanchet:2006dq}
	S.~Blanchet and P.~Di Bari,
	``Leptogenesis beyond the limit of hierarchical heavy neutrino masses,''
	JCAP {\bf 0606} (2006) 023
	[hep-ph/0603107].
	
	\bibitem{Blanchet:2006be}
	S.~Blanchet and P.~Di Bari,
	``Flavor effects on leptogenesis predictions,''
	JCAP {\bf 0703} (2007) 018
	[hep-ph/0607330].

\bibitem{Branco:2011zb}
  G.~C.~Branco, R.~G.~Felipe and F.~R.~Joaquim,
  ``Leptonic CP Violation,''
  Rev.\ Mod.\ Phys.\  {\bf 84} (2012) 515
  [arXiv:1111.5332].


\end{thebibliography}
\end{document}